# International X-ray Observatory (IXO) Assessment Study Report for the ESA Cosmic Vision 2015-2025

X. Barcons, D. Barret, M. Bautz, J. Bookbinder, J. Bregman, T. Dotani, K. Flanagan, R. Fraga, J. Grady, H. Kunieda, D.H. Lumb, K. Mitsuda, K. Nandra, T. Ohashi, L. Piro, N. Rando, L. Strüder, T. Takahashi, T.G. Tsuru , N.E. White, with support of the IXO-Science Definition Team, IXO-Instrument Working Group and IXO-Telescope Working Group.

The International X-Ray Observatory (IXO) will address fundamental questions in astrophysics, including "When did the first SMBH form? How does  large scale structure evolve? What happens close to a black hole? What is the connection between these processes? What is the equation of state of matter at supra-nuclear density?" This report presents an overview of the assessment study phase of the IXO candidate ESA L-class Cosmic Vision mission.

We provide a description of the IXO science objectives, the mission implementation and the payload. The performance will offer more than an order of magnitude improvement in capability compared with *Chandra* and *XMM-Newton* in many different areas of imaging, spectroscopy, timing and polarimetry. This observatory-class facility comprises a telescope with highly nested grazing incidence optics with a performance requirement of 2.5 m² of effective area at 1.25 keV with a 5″ PSF. The instrument complement consists of: 1) an X-ray Microcalorimeter Spectrometer (XMS) offering 2.5 eV energy resolution within a 5 arc min field of view; 2) a combined Wide Field Imager (WFI) and Hard X-ray Imager (HXI) that provides an 18 arc min FOV with CCD-like resolution over an energy range 0.3 to 40 keV; 3) a High Time Resolution Spectrometer (HTRS) for precise spectral-timing even of the brightest X-ray sources; 4) a novel X-ray Polarimeter (XPOL) and 5) an X-ray Grating Spectrometer (XGS) with resolving power ~3000 with effective area 1,000 cm² in the soft X-ray band.

Since earlier submissions to the Astro2010 Decadal Survey, the proposed, substantial technological progress has been made, particularly in the mirror development. Risk reduction measures and important programmatic choices have also been identified. An independent internal Technical and Programmatic Review has also been carried out by ESA, concluding with positive recommendations. Subject to successful conclusion of agreements between the partner space agencies, IXO is fully ready to proceed to further definition, moving towards an eventual launch in 2021-2022.

The full version of the Yellow Book and the Technical Review Board reports are available at:

http://sci.esa.int/science-e/www/object/index.cfm?fobjectid=47796&fbodylongid=2176

# IXO

## Revealing the physics of the hot Universe

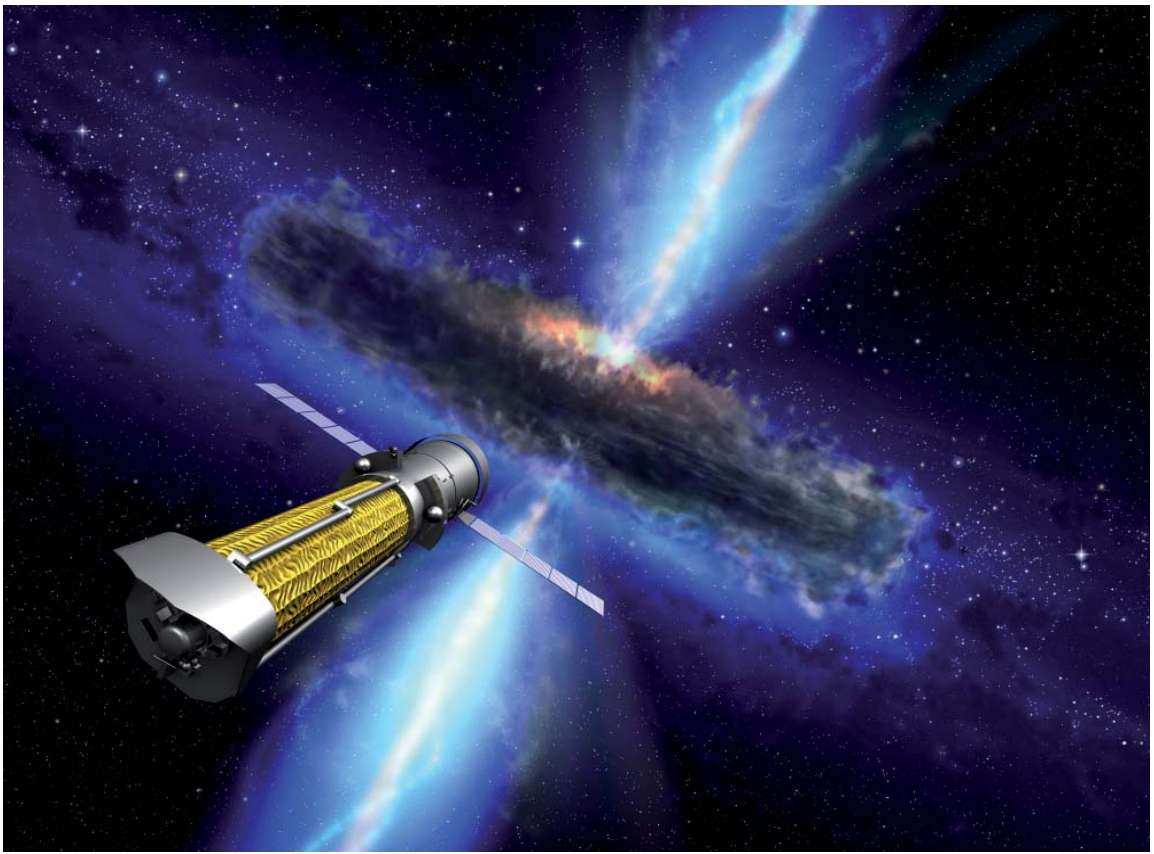

**Assessment Study Report**

**European Space Agency**

# Authorship

## Study Coordination Group
*Co-chairs: H. Kunieda (JP), D.H. Lumb (ESA), N.E. White (USA).*

D. Barret (F), M. Bautz (USA), J. Bookbinder (USA), J. Bregman (USA), T. Dotani (JP), K. Flanagan (USA), J. Grady (USA), K. Mitsuda (JP), K. Nandra (D), T. Ohashi (JP), L. Piro (I), N. Rando (ESA), L. Strüder (D), T. Takahashi (JP), T.G. Tsuru (JP).

## Science Definition Team
*Co-chairs: X. Barcons (E), J. Bregman (USA), T. Ohashi (JP).*

S.W. Allen (USA), M. Arnaud (F), H. Böhringer (D), W.N. Brandt (USA), M. Cappi (I), A. Comastri (I), A.C. Fabian (UK), M. Garcia (USA), A. Hornschemeier (USA), J.P. Hughes (USA), C. Jones (USA), S. Kitamoto (JP), K. Matsushita (JP), M. Mendez (NL), J.M. Miller (USA), R. Osten (USA), F. Paerels (USA), A. Ptak (USA), C.S. Reynolds (USA), S. Sciortino (I), M. Shull (USA), R. Smith (USA), Y. Terashima (JP), Y. Ueda (JP), J. Vink (NL), M.G. Watson (UK), N. Yamasaki (JP).

## Instrument Working Group
*Co-chairs: P. de Korte (NL), J. Nousek (USA), H. Tsunemi (JP).*

D. Barret (F), M. Bautz (USA), R. Bellazzini (I), P. Bodin (F), T. Buckler (USA), D.N. Burrows (USA), W. Cash (USA), J.-W. den Herder (NL), L. Duband (F), E. Figueroa (USA), G.W. Fraser (UK), M. Freeman (USA), R. Fujimoto (JP), K. Hayashida (JP), R. Heilmann (USA), I. Hepburn (UK), A. Holland (UK), K. Irwin (USA), A. Kashani (USA), R. Kelley (USA), C. Kilbourne (USA), M. Kokubun (JP), P. Lechner (D), O. Limousin (F), D. Martin (ESA), R. McEntaffer (USA), K. Mitsuda (JP), K. Nakazawa (JP), C. Pigot (F), L. Piro (I), B. Ramsey (USA), L. Ravera (F), R.E. Rothschild (USA), L. Strüder (D), T. Takahashi (JP), J. Wilms (D).

## Telescope Working Group
*Co-chairs: H. Kunieda (JP), R. Petre (USA), R. Willingale (UK).*

H. Awaki (JP), M. Bavdaz (ESA), J. Bookbinder (USA), F. Christensen (DK), P. Friedrich (D), R. Hudec (CZ), M. Ishida (JP), Y. Maeda (JP), S. O'Dell (USA), G. Pareschi (I), P. Reid (USA), S. Romaine (USA), W. Sanders (USA), M. Schattenburg (USA), W. Zhang (USA).

## ESA Study Team
M. Bavdaz (ESA), D. Lumb (ESA), D. Martin (ESA), T. Oosterbroek (ESA), N. Rando (ESA), P. Verhoeve (ESA).

## Additional Science Contributors
J. Aird (USA), C. Argiroffi (I), D. Ballantyne (USA), M. Begelman (USA), T. Belloni (I), S. Bianchi (I), Th. Boller (D), G. Branduardi-Raymont (UK), L. Brenneman (USA), M. Brusa (D), E. Cackett (USA), M.T. Ceballos (E), F.J. Carrera (E), M. Dadina (I), B. De Marco (I), E.D. Feigelson (USA), N. Grosso (F), G.L. Israel (I), K. Iwasawa (E), J.C. Lee (USA), G. Matt (I), G. Miniutti (E), F. Muleri (I), E. Nardini (I), J.P. Osborne (UK), L.M. Oskinova (D), M.J. Page (UK), D. Patnaude (USA), W.N. Pietsch (D), D. Porquet (F), G. Rauw (Be), J.N. Reeves (UK), G. Risaliti (I), L. Stella (I), B. Stelzer (I), D.K. Strickland (USA), F. Tombesi (USA), M. Türler (CH), P. Uttley (UK), A. Vikhlinin (USA), W.L. Waldron (USA), R. Walter (CH).

The full list of Science Associates can be found under: http://ixo.gsfc.nasa.gov/people/ixoSupporters.html

## Editorial Responsibility
X. Barcons (E), D. Lumb (ESA) & R. Fraga-Encinas (E).

# TABLE OF CONTENTS

# IXO Mission Summary

## Core Science Objectives

| Co-evolution of galaxies and their supermassive black holes | Large scale structure and the creation of chemical elements | Matter under extreme conditions | Life cycles of matter and energy in the Universe |
|---|---|---|---|

## Telescope Module

20 m Focal Length

2.5 m² effective area

5 arcseconds resolution

Silicon pore optics already demonstrated in flight configuration

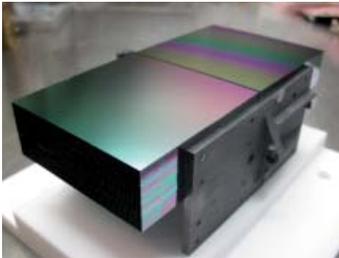

Mirror module assembly 1700 kg

Hierarchical assembly of 8 petals

Includes baffles, thermal control and metrology

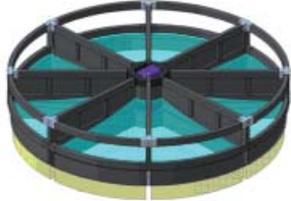

Module designed to be completely compatible with back-up segmented glass optics, NASA development based on NuSTAR

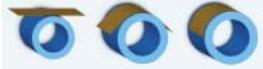

## Instrument Payload

| Instrument | Detector Type | Field of View (arcmin) | Energy Resolution (eV FWHM) | Bandpass (keV) |
|---|---|---|---|---|
| X-ray Microcalorimeter Spectrometer (**XMS**) | Transition edge sensor/bolometer array | 5 x 5 | 2.5 ( @ 6 keV ) | 0.3-12 |
| Wide Field and Hard X-ray Imager (**WFI-HXI**) | Active Pixel Silicon and CdTe imagers | 17 x 17 (WFI) 8 x 8 (HXI) | < 150 <1000 | 0.1-15 10-40 |
| High Time Resolution Spectrometer (**HTRS**) | Fast silicon drift diode array | Point sources | 200 | 0.3-15 |
| X-ray Polarimeter (**XPOL**) | Gas electron multiplier | 2.6 x 2.6 | 1200 | 2-10 |
| X-ray Grating Spectrometer (**XGS**) | Grating segment behind mirror read out with CCDs | Point sources | E /ΔE > 3000 | 0.3-1 |

## Spacecraft and Systems

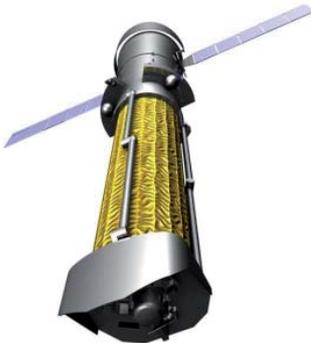

| Launcher | Atlas V/551 | | Orbit | L2 halo, 750,000 km amplitude | Deployable | Articulated booms and shroud |
|---|---|---|---|---|---|---|
| Pointing | 3-axis stabilised- 1.5 arcsecs measurement | | Lifetime | Design 5 yrs consumables 10 yrs | HGA | Steerable X-band |
| Ground Station | Cebreros / DSN | | Data rate | 90 Gbit/day | Observations | ~700/year |

| Mass (kg) | | Power (W) | |
|---|---|---|---|
| Optics/MM | 1100 | Optics - thermal | 1520 |
| Instruments | 645 | Instruments (max) | 1840 |
| Structure/Thermal | 1870 | TT&C (max) | 95 |
| Avionics & Power | 305 | Data Handling | 50 |
| Propulsion | 85 | Propulsion | 65 |
| Spacecraft dry mass | 4010 | AOCS, metrology | 440 |
| Propellant | 420 | Structure, Thermal, Mechanism | 500 |
| Launch Adapter | 240 | Power Subsystem | 245 |
| Total system and maturity margin | 1730 | Total system and maturity margin and harness losses | 950 |
| **Total** | **6400** | **Total** | **5700** |

# 1   Executive Summary

IXO will reveal the entire history of the Hot Universe, from the Cosmic "Dark Ages" to the present day, thereby addressing some of the most important themes posed by ESA's Cosmic Vision 2015-2025 Science Objectives, including:

1) ***The Evolving Violent Universe*** – finding massive black holes growing in the centres of galaxies, and understanding how they influence the formation and growth of the host galaxy;

2) ***The Universe taking shape*** – revealing how the baryonic component of the Universe formed large-scale structures and understanding how and when the Universe was chemically enriched by supernovae;

3) ***Matter under extreme conditions*** – studying how matter behaves in the strongest gravitational fields around black holes and at very high densities in the interiors of neutron stars and accretion disks.

These science questions are described in detail in Section 2.

We show that an observatory with the performance of IXO is necessary to address all these questions: Hot gas, visible primarily by its X-ray emission, dominates the baryonic content of the Universe and permeates and delineates the largest bound cosmic structures. Black holes provide the strongest gravitational fields in nature, and their growth over cosmic time apparently had a remarkable interplay with the development of galaxies and large scale structure in general. IXO will search for supermassive black holes out to redshift $z$=10, observe the process of cosmic feedback where black holes inject energy on galactic and intergalactic scales, find the missing baryons in the cosmic web using background quasars, map bulk motions and turbulence in galaxy clusters, contribute to determining important cosmological parameters by enabling independent measurements of dark matter and dark energy using galaxy clusters, trace orbits close to the event horizon of black holes at all scales (from stellar mass to supermassive black holes), measure black hole spin for several hundred active galactic nuclei (AGN), use spectroscopy to characterize outflows and the environment of AGN during their peak activity, and set constraints on the densest form of observable matter making up the cores of neutron stars.

Being an observatory class mission, IXO will be able to address a large number of additional problems in contemporary astrophysics, such as the origin of cosmic rays in supernovae, studies of the interstellar medium, stellar mass loss and star and planet formation.

To enable these measurements, IXO will employ optics with 10 times more collecting area at 1 keV than any previous X-ray observatory. Furthermore, the focal plane instruments will deliver a 100-fold increase in effective area for high-resolution spectroscopy, deep spectral imaging over a wide field of view, unprecedented polarimetric sensitivity, microsecond spectroscopic timing, and high count rate capability.

The heart of the mission is the X-ray optical system utilising the Silicon Pore Optics (SPO) technology pioneered in Europe. The required 2.5 m² effective collecting area with 5 arcsec angular resolution is achieved with margins, using a 20 m focal length deployable optical bench. The considerable investment in this technology by ESA has already demonstrated in a flight-like assembly, sub-10 arcsecond angular resolution capability, which would already enable a mission with revolutionary capability. A back up technology using segmented glass is also being pursued in Europe and the US, in case of unexpected challenges with the silicon pore approach.

The planned instrument complement will employ the revolutionary X-ray Microcalorimeter Spectrometer (XMS) arrays for high-resolution spectroscopic imaging, active pixel sensor arrays in a wide field imager (WFI), a hard X-ray imager (HXI), a high time resolution spectroscopy (HTRS) instrument for timing of bright sources, a gas pixel imaging X-ray polarimeter (XPOL), and a high-efficiency X-ray grating spectrometer (XGS) to provide high spectral resolution at low energies. All these instruments are on track to reach TRL 5 by the end of the Definition Phase. Details are provided in Section 4.

IXO will be placed in orbit at L2, which provides uninterrupted viewing and an ideal thermal environment. The design assumes a five-year mission lifetime, but has consumables for at least 10 years. Both ESA and NASA studies concluded that the IXO spacecraft could be built with technologies that are fully mature today. All subsystems utilize established hardware with substantial flight heritage. Most components are "off-the-shelf." The IXO spacecraft concept is robust; all IXO resource margins meet or exceed requirements. Substantial redundancy along with failsafe mechanisms for contingency mode operations assure that no credible single failure will degrade the mission. The modular nature of the IXO architecture provides clearly defined interfaces to enable sharing of the development between international partners with minimal risk.

The main IXO requirements are summarised in the mission science requirements, Section 3. The requirements identified per science area, together with an assessment of the observing programme, are provided in the tables in Appendix 1.

The study has confirmed that the technology readiness levels are appropriate to proceed to a definition study in 2011 and will succeed to TRL 5 by 2012. If the IXO Phase A starts in 2011, then a mission implementation schedule allows a launch as soon as 2020. We also consider an alternative later launch date which might be imposed to the mission development by programmatic, rather than technical, aspects.

For the parallel studies, the mission costing has been carried out under the assumption that the mission would be implemented by ESA or NASA alone. The costing has been performed on a system design that has evolved significantly since the report made for the Astro2010 Decadal Survey. This more mature design, and the technology progress since the Astro2010 submission, substantially reduces risk, removing the need for the application of cost threats and penalties included in the Decadal report. The adopted baseline for the ESA workshare is consistent with the assumption for an ESA L-class mission Cost at Completion. An ESA-led scenario would require substantial costs saving to be identified in the Definition Phase. Instrument provision in Europe will be through ESA member states, with substantial contributions from both NASA and JAXA.

The telescope technology developments have made rapid progress to demonstrate <10 arcseconds Half Energy Width. A Technology Development Plan is in place to improve the performance further towards the requirement of 5 arcseconds, and ensure the Technology Readiness Level, TRL>5, will be achieved by the end of 2012.

There are no new technologies to be developed for the spacecraft, while most instruments are at an advanced stage of development. The most innovative (X-ray Microcalorimeter Spectrometer and its associated cryogenic system) is rapidly being brought to a flight ready status through preparations for an, albeit less capable, instrument for the *ASTRO-H* mission (2014), and through a vigorous sensor array development and flight programme. Strong risk mitigation measures have been identified to ensure a robust system design.

Current X-ray observatories, such as *XMM-Newton*, *Chandra* and *Suzaku*, have made major contributions to astrophysics and cosmology in the last decade and raised new questions, requiring at least an order of magnitude improvement in capability (see Figure 1.1). Achieving the main scientific goals that astronomy will face in the next decades requires an X-ray observatory class mission like IXO.

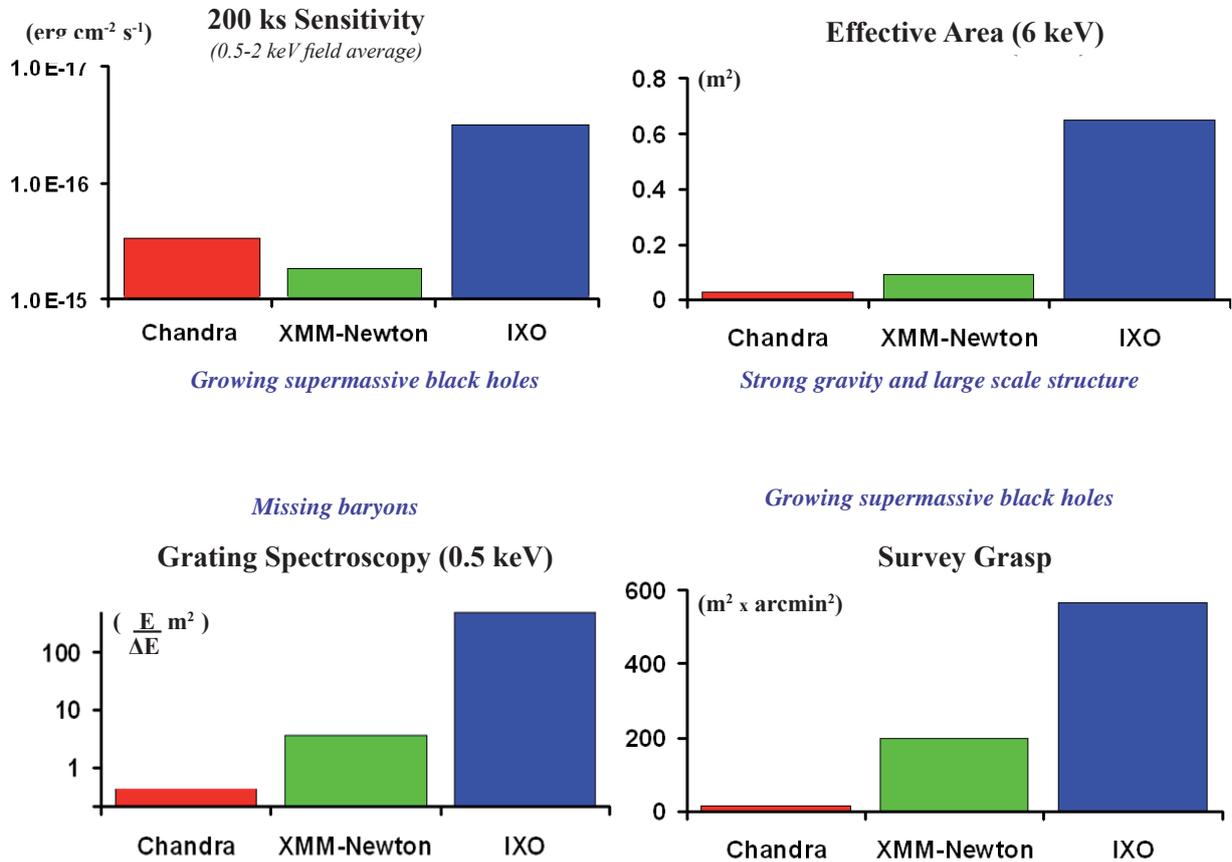

**Figure 1.1**

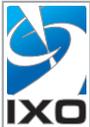

IXO will provide an enormous leap forward with respect to preceding X-ray observatories, in terms of effective area, point source sensitivity, survey grasp, spectroscopic grasp and maximum observable flux. These order of magnitude (or more) improvements offer an enormous potential for unforeseen discoveries in many areas of Astrophysics that no other mission can offer.

# 2   IXO Science Objectives: The hot Universe

IXO is an observatory class Astrophysics mission that has been conceived to study the hot Universe, all the way from the earliest galaxies where supermassive black holes trapped radiating hot matter, through the largest gravitationally bound hot gas reservoirs, and down to our immediate cosmic environment where hot matter surrounds stars and black holes. IXO aims to provide direct insight into some of the most important themes posed by ESA's Cosmic Vision 2015-2025 science objectives (*Cosmic Vision: Space Science for Europe* 2015-2025, ESA BR-247):

- (Q4.3) The Evolving Violent Universe, by finding massive black holes growing in the centres of galaxies, from early times to the present, and understanding how they influence the formation and growth of the host galaxy through feedback processes.

- (Q4.2) The Universe taking shape, by studying how the baryonic component of the Universe formed large-scale structures and by finding the large fraction of baryons that are still missing; and understanding how and when the Universe was chemically enriched by supernovae.

- (Q3.3) Matter Under Extreme Conditions, by revealing how matter behaves in very strong gravity, such as occurs around black holes and compact objects, where General Relativity predicts many effects; and how matter behaves at densities higher than in atomic nuclei in the interiors of neutron stars.

IXO will also help to provide answers to other Cosmic Vision themes, most notably to *(Q4.1) The Early Universe* (which includes the investigation of the Dark Energy that dominates the content of today's Universe), by enabling independent measurements of dark matter and dark energy using galaxy clusters. Being an observatory class mission, IXO will also be able to address a large number of additional problems in contemporary astrophysics, a few of which are highlighted here: the origin of cosmic rays in supernovae, studies of the interstellar medium, stellar mass loss, and star and planet formation. Likewise, IXO will provide responses to a very large fraction of the questions posed by the Astronet Science Vision exercise (de Zeeuw & Molster 2007), which proposes a vision for the science goals of European astronomy for the next two decades, and features prominently in the Astronet roadmap. **IXO will undoubtedly play a significant role in answering a wide range of the science questions listed in both the Cosmic Vision 2015-2025 and Astronet documents.**

X-ray observations conducted by the suite of IXO instruments will deliver astronomical data for each single photon detected which include its position, energy, arrival time and even polarization. This will enable astronomers to obtain not only X-ray images of the sky, but a spectrum and a time series to study variable sources in every point of the sky covered by the IXO instrument in operation, thus offering an enormous potential for astrophysical investigations (see, e.g. Figure 2.1).

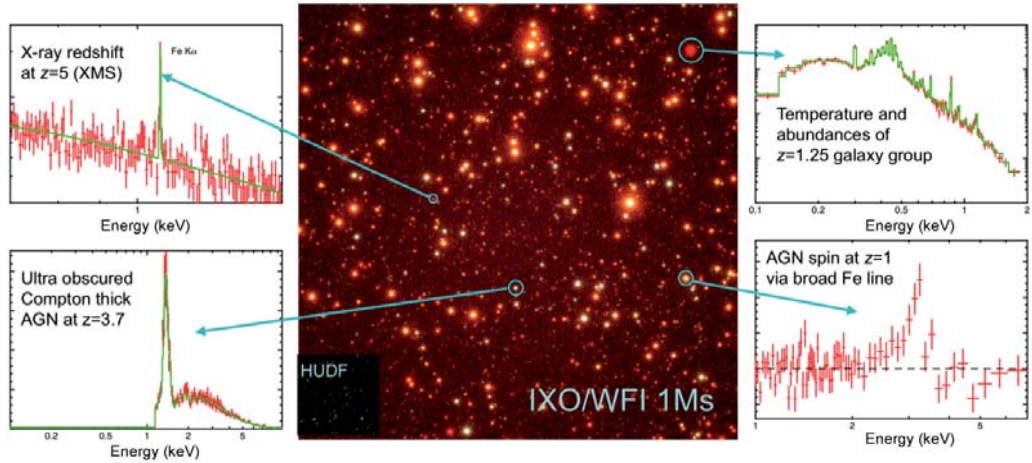

**Figure 2.1.** *IXO Wide Field Imager (IXO-WFI) simulation of the Chandra Deep Field South with Hubble Ultra Deep Field (HUDF) in inset. Simulated spectra of various sources are shown, illustrating IXO's ability to determine redshift autonomously in the X-ray band (**top left panel**), determine temperatures and abundances even for low luminosity groups to z>1 (**top right panel**), make spin measurements of AGN to a similar redshift (**bottom right panel**) and uncover the most heavily obscured, Compton-thick AGN (**bottom left panel**).*

IXO is being designed to be part of the coherent set of large telescopes that Europe, often with international partners, is building or planning  (including ALMA, LOFAR, JWST, E-ELT and SKA). The sensitivity of IXO, which stems from a technological effort to have very good angular resolution (5 arcsec) combined with a large effective area, matches that of the other large facilities. To illustrate how these facilities will complement each other, for example, in the study of the first galaxies in cosmic history, JWST and the E-ELT will see their starlight, ALMA their cold interstellar medium and IXO their growing massive black holes (see Figure 2.2). An extraordinary recent discovery in astrophysics is that these three components (stars, black holes, and gas) are locked together in a process dubbed 'feedback' which can only be understood by combining studies from the whole of these future observatories.

**Figure 2.2.** *Sensitivity of future key observatories, along with the spectrum of a galaxy with strong star formation and a merging massive binary black hole, like NGC 6240 at z=7. Below $10^{16}$ Hz emission is ultimately due to nuclear fusion and above to gravity. Only IXO is able to detect unambiguously the direct signal from an accreting binary super-massive black hole.*

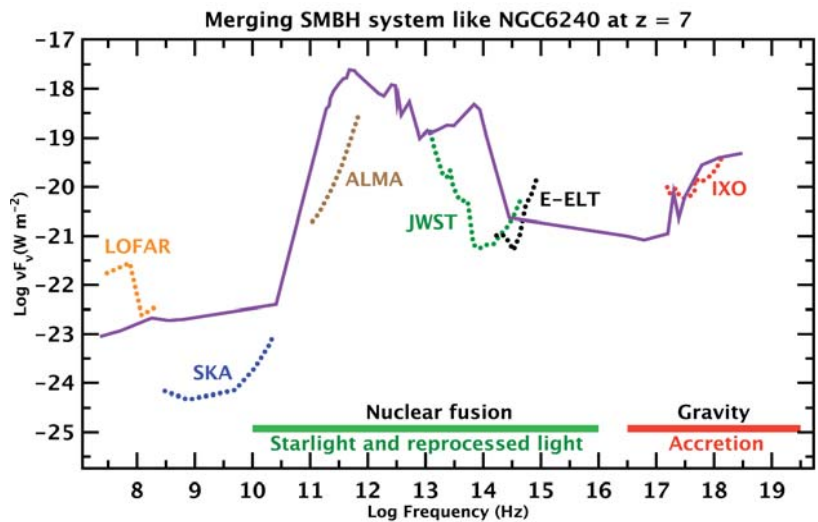

IXO will address the high-energy phenomena that happen in the same sources that other major facilities at longer wavelengths like ALMA, JWST and E-ELT will be targeting, notably in the high-*z* Universe.

| THEME/QUESTION | WHAT IXO WILL DO | INSTRUMENT |
| --- | --- | --- |
| **CO-EVOLUTION OF GALAXIES AND THEIR SUPERMASSIVE BLACK HOLES (SMBH)** | | |
| The first SMBH | Find young growing massive black holes at the dawn of the Universe (z~6-10), by conducting multi-tiered surveys, in conjunction with observations at longer wavelengths. To test the growth mode of SMBH by measuring their spin. | WFI |
| Obscured growth of SMBH | Find and characterise SMBH growing in an obscured phase at z~1-3, including Compton-thick AGN, and measure their energetics. | WFI, HXI, XMS |
| Cosmic feedback from SMBH | Measure feedback from growing SMBH in galaxies, groups and clusters, by relating AGN activity to star formation at high-z, and measure gas motions in galaxies and clusters induced by AGN . | XMS, WFI |
| **LARGE-SCALE STRUCTURE AND THE CREATION OF CHEMICAL ELEMENTS** | | |
| Missing baryons and the Intergalactic Medium | Find the remaining missing baryons in the current Universe, likely in a Warm-Hot Intergalactic Medium phase, by detecting hundreds of intervening absorption systems towards bright background targets. | XGS, XMS |
| Cluster Physics and Evolution | Find out how gas dynamically evolves in the cluster dark matter potentials, by measuring gas motions and turbulence. Determine the physical processes behind the production of cosmic rays in clusters. Also find when and how the excess entropy was generated in clusters by studying their precursors at early epochs. | XMS, HXI |
| Galaxy cluster cosmology | Provide an independent measurement of Dark Energy density and its equation of state by counting clusters at various epochs and measuring their gas fraction. | WFI |
| Chemical evolution along cosmic time | Measure chemical abundances of the various families of elements in intracluster gas, find at which epoch they were dispersed and derive what is the main production mechanism. . | XMS |
| **MATTER UNDER EXTREME CONDITIONS** | | |
| Strong gravity and accretion physics | Measure black hole spin of stellar and supermassive black holes via time-averaged spectroscopy, reverberation mapping, timing and polarimetry; probe General Relativity in the strong field regime by mapping the emission of the innermost regions of accretion disks; measure the kinetic energy of outflows produced by matter falling onto black holes. | WFI, HXI, HTRS, XPOL, XMS |
| Neutron Star Equation of State | Measure Mass and Radius of tens of neutron stars via a number of timing and spectroscopic techniques, and constrain the equation of state at supra-nuclear density. Detect vacuum polarisation effects in highly magnetised neutron stars. | HTRS, XMS, XPOL |
| **LIFE CYCLES OF MATTER AND ENERGY IN THE UNIVERSE    (Additional Science Goals)** | | |
| Supernovae : explosion mechanisms | Probe the Supernova explosion material and its environment, by measuring temperatures, velocities, turbulences in Supernova Remnants. | XMS |
| Supernovae : nucleosynthesis | Measure nucleosynthesis products for the various types/subtypes of Supernovae, by detecting weak emission lines from radioactive elements in Supernova Remnants. | XMS, HXI |
| Cosmic ray acceleration in Supernova Remnants | Measure hard X-ray synchrotron emission due to fluctuating magnetic fields. | WFI, HXI, XPOL |
| Characterising the Interstellar Medium in the Galaxy | Determine the aggregation state of elements in the Inter-stellar Medium in the Galaxy (solid state, molecular, atomic, etc), by precise measurements of the edge energy of bound-free transitions . | XGS |
| Galactic Centre | Study the energetics and geometry of the various components around Sgr A*. | WFI, XPOL |
| Accretion in Young Stellar Objects | Characterise the geometry of accretion disks in Young Stellar Objects, via time resolved high-resolution spectroscopy. | XGS |
| The atmospheres of Solar System planets | Study particle populations and their acceleration mechanisms occurring in the outer layers of the atmospheres of these planets and show their link to solar activity. | XMS |

**Table 2.1.** *Summary of IXO's Science Goals.*

The science case of IXO is presented along the main Cosmic Vision themes that it will address in the following 3 sections: ***Co-evolution of galaxies and their supermassive black holes***, ***Large scale structure and the creation of chemical elements;*** and ***Matter under extreme conditions***. A summary of what IXO will actually do to address these themes, along with the IXO instruments in use, is shown in Table 2.1. Finally, under ***Life cycles of matter and energy***, the breadth of fundamental astrophysics enabled by the dramatic increase in capabilities IXO are discussed.

Beyond the science goals described in this document, IXO will have an unrivalled potential for breakthrough discoveries that we cannot forecast, given our current knowledge of the X-ray Universe. The very high throughput (~10 times larger than *XMM-Newton*), coupled to the excellent angular resolution, the capability of performing spatially resolved spectroscopy of spectral resolution ~330-4800, timing studies of the brightest sources in the sky, spatially resolved polarimetry and sensitive spectroscopy of resolution >3000 for single objects will open new avenues in the investigation of X-ray sources and especially in unveiling new physical phenomena at work in these sources. The typically order-of-magnitude leap forward that IXO will feature in many of these parameters with respect to current or foreseen facilities will open unsuspected insight into the physics of the hot evolving Universe. The questions posed in this science case will indeed find answers with IXO, but previous experience has shown that order-of-magnitude leaps forward in capability will reveal unforeseen breakthroughs which may surpass the questions posed here.

## 2.1 Co-evolution of galaxies and their supermassive black holes

### 2.1.1 The first supermassive black holes

One of the main themes in extragalactic astronomy for the next decade and beyond will be the formation and evolution of the first galaxies, their stars, gas and black holes. While most of the interstellar medium of the forming galaxies likely resides in a cold phase best seen in sub-millimetre observations, and starlight is optimally seen through optical and infrared observations, matter funneled onto the forming supermassive black holes in galaxy centres constitutes an important ingredient of the hot Universe, which is essential to understand how galaxies formed. Studying the behaviour of the hot matter in the most immediate environment of black holes, where General Relativity is no longer a small perturbation of Newtonian mechanics, is a key science goal of IXO, discussed in Section 2.3. In this section we show how IXO will reveal the way that supermassive black holes grew out of hot infalling material, in an interplay with the stars and the cold gas. Many future observatories, including JWST, ALMA, and the ELTs will intensively observe starlight and cold gas with a broad range of redshifts, out to the dawn of the Universe when the first galaxies formed. It has, however, become clear that the properties and evolution of galaxies are intimately linked to the growth of their central black holes. Any understanding of the formation of galaxies, and their subsequent evolution, will therefore be incomplete without similarly intensive observations of the accretion light from supermassive black holes (SMBHs) in galactic nuclei. To make significant progress, the unique deep X-ray survey capabilities of IXO are needed to chart the formation of typical SMBHs at *z*>6 (when the Universe was <5% of its current age), and their subsequent growth over cosmic time, which is most effectively achieved with X-ray observations.

#### Black hole accretion and galaxy evolution: a fundamental connection

A major recent development in astrophysics has been the discovery of the relationship in the local Universe between the properties of galaxy bulges, and dormant black holes in their centres (e.g. the M-σ relation between the black hole mass and bulge velocity dispersion; Ferrarese & Merrit 2000; Gebhardt et al. 2000). These tight relationships represent definitive evidence for the co-evolution of galaxies and Active Galactic Nuclei (AGN). The remarkable implication of this is that some consequence of the accretion process on the scale of the black hole event horizon is able to influence the evolution of an entire galaxy. This is possible because the gravitational potential energy acquired by an object approaching a black hole is a million times larger than the energy of an object orbiting in the potential of a typical galaxy. As a black hole grows to 0.2% of the bulge mass through accreting matter, it releases nearly 100 times the gravitational binding energy of its host bulge.

The main idea is that radiative and mechanical energy from the AGN regulates both star formation and BH accretion during periods of galaxy growth. Some broad evidence for this type of feedback has been established by the tight coupling between the energy released by supermassive black holes and the gaseous structures surrounding them. Despite the intense current interest in this topic, and its great importance, the physical processes involved are only known to first order, and direct evidence for this kind of AGN feedback is scarce (see, however, Section 2.1.3, where such evidence is reviewed in more detail).

Black hole driven feedback is thought to be essential in shaping the first galaxies. Current models propose that mergers of small gas-rich proto-galaxies in deep potential wells at high redshift drive star formation and black hole growth (in proto-quasar active galaxies) until a luminous quasar forms. At this point, a black hole driven wind evacuates gas from the nascent galaxy, limiting additional star formation and further black hole growth (Silk & Rees 1998). Further episodes of merger-driven star formation, accretion, and feedback are expected to proceed through cosmic time. This provides a plausible origin for the M-σ relation (e.g. di Matteo et al. 2005). Several puzzling aspects of galaxy formation, including the early quenching of star formation in galactic bulges and the galaxy mass function at both high and low ends, have been attributed to black hole "feedback" (Croton 2006). This topic requires an X-ray observatory with the deep survey capabilities of IXO to trace the evolution of super massive black holes in the Universe.

## Black hole growth in the early Universe

The very first stars were formed a few hundred million years after the Big Bang in primordial structures, where gravity was able to overpower the pressure of the ambient baryons (Loeb 2010). The first seed black holes were left behind as remnants of the most massive stars. Radiation from massive stars and from these first black holes in the cores of primeval galaxies was responsible for reionizing the Universe at a redshift that WMAP has measured to be around $z\sim11$ (Dunkley et al. 2009). Identifying and studying AGN in their formation epoch ($z\sim6$-10) is essential to better constrain the AGN contribution to re-ionisation, bridging the gap between the few known sources at this redshift, and the information we can extract from the microwave background. Still, the highest redshift galaxies and quasars currently known are all in the range $z$=6-8 (several of which were signaled by Gamma-ray bursts), with tentative candidates at $z\sim10$ just emerging (Bouwens et al. 2010).

The known AGN population at $z$=6-8 currently consists of luminous optical quasars (e.g. Fan et al. 2003). How powerful SMBH formed at such early epochs is a major unknown in contemporary astrophysical cosmology. Models range from growth by Eddington-limited accretion from modest mass black hole seeds (of a few $10\,M_\odot$), to monolithic collapse of hot, dense gas clumps (Bromm & Loeb 2003; Begelman et al. 2006) or "quasistars" (Begelman et al. 2008). Gas-dynamical cosmological simulations are able to produce quasars with $\sim 10^9\,M_\odot$ at $z$=6.5 through a rapid sequence of mergers in small groups of proto-galaxies (Li et al. 2007). The growth is likely to proceed in a self-regulated manner owing to feedback with the progenitor host, with a period of intense star formation and obscured accretion preceding the optically bright quasar phase. The complex physics involved in such a scenario is, however, poorly understood. Furthermore, evidence for widespread merger-driven AGN activity and feedback at high redshift is scarce, and other models are equally valid as far as current observational constraints are concerned.

It must also be borne in mind that luminous, optical quasi-stellar objects, hosting among the most massive black holes ($>10^9\,M_\odot$) in the Universe, are extremely rare. Typical AGN, which are of lower luminosity and often obscured, remain largely undiscovered. Uncovering such objects at $z$=6-7 (and even higher redshifts) holds the key to our understanding of this crucial phase in the development of the Universe. It is very likely that SMBHs as massive as $10^6\,M$, possibly hosted by vigorously star forming galaxies, existed as early as $z\sim10$. X-ray observations offer a unique tool to discover and study the accretion light from moderate luminosity AGN at $z$=6-10, which are rendered invisible in other wavebands due to intergalactic absorption and dilution by their host galaxy. Testing the various competing scenarios for the evolution of early SMBH will be rendered possible by IXO, which will reveal the AGN population at the highest redshifts for the first time. Such surveys will normally be performed in sky regions intensively studied at longer wavelengths by E-ELT, JWST or ALMA, and they will therefore benefit from the redshift measurements obtained by these facilities. Massive survey facilities, like LSST, will also be a prime handle to identify the distant growing SMBH, through either spectroscopic or photometric redshifts.

In addition to constraining the demographics of high-redshift active galaxies via its surveys, IXO will also deliver some information on how SMBH grew so rapidly in the early Universe. Though theoretically challenging, there appears to be (just) enough time to grow the most massive black holes currently found at $z$=4-6.5, observed as luminous quasars. Hints on mass accretion rate of these objects will be obtained by X-ray spectroscopy, to a much better sensitivity than what is available today. IXO measurements of the X-ray continuum shape, X-ray variability, and discrete spectral features such as Fe K lines of growing SMBH at $z$~4 and above will all be brought to bear to learn about the mode(s) of accretion for the SMBH. These properties depend upon, and thus will be used to constrain, the black-hole mass, mass accretion rate, and the physical conditions in the inner accretion flow. Furthermore, the incidence of winds and outflows in SMBH at those redshifts, which some models predict to be high, will also need the high throughput spectroscopic capabilities of IXO (see Section 2.3.1.3).

## IXO: Breaking through to typical AGN at "first light"

The highest redshift quasi-stellar objects known have been discovered in optical/infrared surveys covering large areas of the sky (e.g. SDSS; Fan et al. 2001). These are fascinating sources, but it is crucial to remember that they are among the most extreme and unusual objects in the Universe. Large area optical and NIR surveys will be continued with, e.g. Pan-STARRS and VISTA, and perhaps with LSST and *WFIRST/Euclid*, which will discover many more high-$z$ quasi-stellar objects. The optical/infrared surveys are, however, fundamentally limited to objects in which the AGN light is a significant fraction of that of the host galaxy. For this reason, X-ray observations can probe much lower AGN bolometric luminosities than the optical. Current deep X-ray surveys probe factors of 100-1000 fainter down the luminosity function than SDSS, at or around L* (the typical galaxy luminosity), where the bulk of the accretion power is produced. We furthermore expect the majority of AGN at high redshift to be heavily obscured by gas and dust, where the accretion light is rendered invisible in the optical but detectable at X-ray energies. X-ray observations are thus absolutely essential in both the discovery and characterisation of typical accreting black holes at high redshift.

**Figure 2.3.** *Schematic view of the innermost part of an Active Galactic Nucleus. X-rays are emitted from an electron corona around the accretion disk, and therefore provide a direct view of the immediate vicinity of the growing SMBH. They also carry information about the kinetic power deposited in the winds, through absorption features in the spectrum. Ultraviolet and optical emission lines come from the more distant Broad Line Region clouds. Infrared radiation is emitted even further out from the molecular torus.*

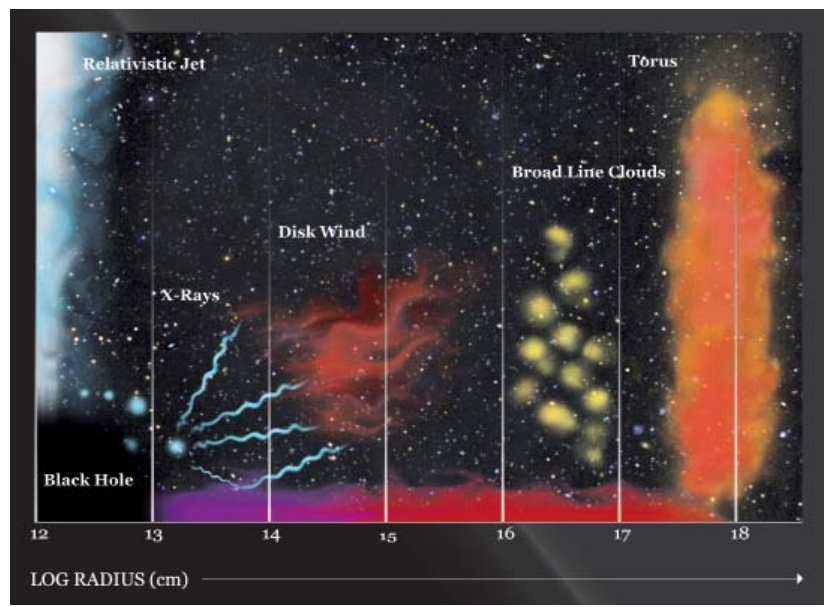

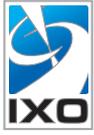

There are three reasons why X-rays are unique tracers of growing supermassive black holes: they come from very close to the event horizon, they can escape through obscured zones and they are uncontaminated by the host galaxy (Figure 2.3). IXO will be a unique tracer of growing supermassive black holes in the era that ALMA, JWST and the E-ELT will open.

The current deepest surveys with *Chandra* reach great depths near the central aim point (i.e. the telescope optical axis). These have yielded a handful of AGN candidates at z>5 (Luo et al. 2010), but none is yet confirmed at z>6. To harvest significant samples of moderate luminosity AGN at z=6-10 we need to reach up to and beyond these depths over much larger areas. IXO's combination of huge collecting area and excellent spatial resolution over a large field of view fulfils this requirement. These observations will have enormous discovery and constraining power. Current semi-analytic models (e.g. Rhook & Haehnelt 2008; Salvaterra et al. 2007; Marulli et al. 2008) or extrapolations of the measured evolution of the luminosity function at z<~4 (Figure 2.4, e.g. Silverman et al. 2008; Ebrero et al. 2009; Brusa et al. 2009; Aird et al. 2010) predict vastly different amounts of black hole growth in the early Universe. A suitably designed "multiple-cone" IXO survey with the IXO Wide Field Imager (WFI), consisting of ultra-deep, deep and shallower wide area components will yield several hundred X-ray selected AGN at z~6, and as many as a few 10s at z~10, depending on model extrapolations, and thus place strong constraints on AGN accretion in the very early Universe. Additional very high redshift AGN will be identifiable via serendipitous observations, which will yield a rich archive, but dedicated surveys are needed as they target the same areas as the complementary multi-wavelength facilities. Coordinating the selection of IXO deep survey areas with well-explored regions at other wavelengths is necessary for the identification and redshift determination of faint X-ray sources pinpointed to high accuracy with IXO's excellent angular resolution. Facilities like EVLA, ALMA, JWST and E-ELT will also yield the host galaxy properties of these early AGN (in particular redshifts, but also stellar and gas masses, star formation rates), which will be crucial in discriminating between various SMBH formation models.

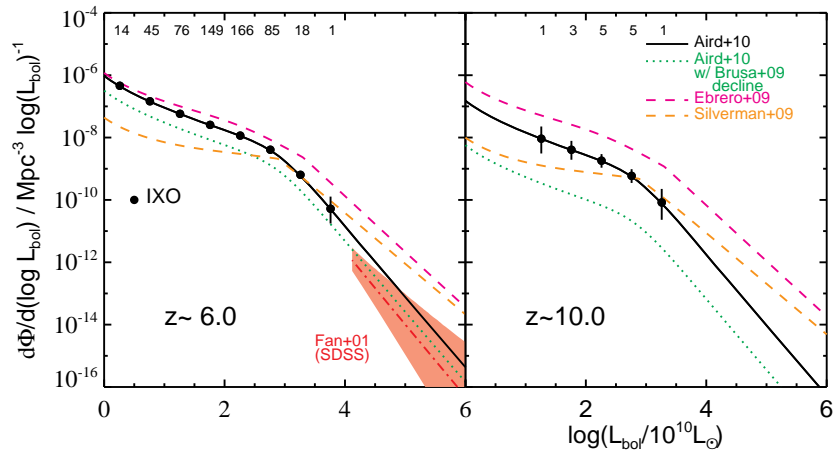

**Figure 2.4.** *Predicted bolometric luminosity functions at z~6 and z~10, for a variety of current models. The red dot-dashed line and shaded area show current observational constraints at z~6 based on large area optical surveys (SDDS: Fan et al. 2001), which probe the high luminosity tail of the population. Black dots show the predictions for one specific model, with a specifically designed multi-cone IXO survey; the predicted number of sources in each luminosity bin are also given at the top. IXO will enable extremely accurate measurements of the AGN luminosity function at z~6 and will place constraints out to z~10, allowing discrimination between the different evolutionary models, and revealing the crucial role of SMBH accretion in the very early Universe.*

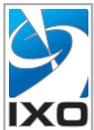

IXO surveys will find large numbers of moderate luminosity AGN at z~6-10, right after the Universe was re-ionised. ALMA, E-ELT and JWST observations in the same survey fields will deliver the redshifts of these galaxies, as well as other properties of the host galaxy.

## 2.1.2 Obscured growth of supermassive black holes

Since Active Galactic Nuclei are the very phenomenon where black holes grow by mass accretion, we need to uncover all populations of AGN over a broad range of redshifts, luminosities, and obscuration, to understand the whole growth history of SMBH and their relation to galaxy formation. Theories suggest that a significant portion of SMBH growth in massive galaxies occurred during a phase of heavy obscuration, accompanied by intense star formation (e.g., Hopkins et al. 2006). Observations indicate that unobscured and heavily obscured AGN represent different phases after major galaxy merger events, with most of the SMBH growth occurring in the obscured systems (Treister et al. 2010). Thus, it is a key issue to reveal the evolution of obscured AGN, in particular at redshift range of 1-3, where both mass accretion and star formation had peak activities in the cosmic history.

The extreme "Compton-thick" AGN, with absorbing columns above $N_H > 10^{24.5}$ cm$^{-2}$ are of particular importance, as they may represent a key phase of the SMBH-host galaxy co-evolution and harbour a large fraction of the SMBH. Unveiling such heavily obscured AGN is being done nowadays largely via selection of mid-IR and near-IR colours (Daddi et al. 2007, Fiore et al. 2008), but this is highly uncertain due to stellar contamination and only a signature of X-ray emission towards the highest energies certifies the presence of a deeply buried AGN. IXO will not only have a better sensitivity at energies above 10 keV than any of its predecessors, but will in addition bring the most powerful spectral sensitivity below 10 keV, which will be instrumental in finding other features of the Compton-thick AGN, e.g., a strong Fe line and/or weak reflected components.

Due to difficulties in detecting heavily obscured AGN, we are far from having a complete census of the AGN population even in the local Universe. In these objects, the direct emission from the nucleus in the UV, optical, and near IR bands as well as at $E$<10 keV X-ray is blocked by obscuring matter, making it difficult to probe the central engine. Even in the deepest *Chandra* and *XMM-Newton* surveys now available, ~40% of the X-ray background above 6-8 keV is still unresolved (Worsley et al. 2005), where the Compton-thick AGN that IXO will find are likely to provide a substantial contribution. Although indirect estimates on the number density of Compton-thick AGN can be obtained in the framework of population synthesis models (e.g., Gilli et al. 2007), these are highly uncertain because their contribution to the X-ray background is inevitably coupled to parameters such as the average shape of the broad band spectra of Compton-thin AGN, the contribution of minor populations like blazars, and the precise (<10%) level of the absolute intensity of the X-ray background. More importantly, it is not clear whether Compton-thick AGN follow the same cosmological evolution as found for Compton-thin AGN, which exhibit a "down-sizing" behaviour. The only way to reliably determine the evolution of Compton-thick AGN is to directly detect them in large numbers and characterise their entire population.

Many observations suggest the presence of a large (but uncertain) number of Compton-thick AGN in the local Universe. Evidence for heavily obscured AGN is found from optical narrow emission line galaxies (Risaliti et al. 1999) and from infrared selected galaxies (e.g., Maiolino et al. 2001; Imanishi et al. 2007). Although current estimates are highly uncertain, their local number density is estimated to be comparable to that of known, Compton-thin type-2 Seyfert galaxies (Ueda et al. 2007), and most of them are possibly missed even by the deepest current X-ray surveys.

At higher redshifts, it has been suggested from deep multi-wavelength surveys that Compton-thick AGN at $z$~2 may be hiding among infrared bright, optically faint galaxies, although each object is generally too faint in X-rays to be detected individually even with the deepest *Chandra* exposures. Stacking analyses assign to them an average intrinsic X-ray luminosity > $10^{43}$ erg s$^{-1}$, and the estimated number density is of the order of and might be even higher than that of Compton-thin AGN at the same luminosity and redshift range (Fiore et al. 2008). If these numbers are confirmed, then Compton-thick AGN might provide a major contribution to the total accretion power in the Universe (Fabian & Iwasawa 1999). This could change our current view of the accretion history of the Universe based on Compton-thin AGN, where the mass function of SMBHs in the local Universe can be well explained by smooth prolonged accretion, based on Soltan's argument (Marconi et al. 2004). The inferred very high number density of Compton-thick AGN at $z\sim2$ is qualitatively consistent with the evolution of the absorbed fraction seen in Compton-thin AGN (i.e., more obscuration at higher redshifts). However, a simple AGN unification scheme may not hold since the evolutionary phase of Compton-thick AGN in galaxy evolution could be different from that of Compton-thin AGN; models predict that Compton-thick AGN correspond to the stage where the AGN blows out gas from the galaxy and terminates star formation (Hopkins et al. 2006).

While X-ray selection has provided the most robust AGN samples to date, finding the most obscured objects has proved difficult with current X-ray missions. Mid-IR selection is promising, but as stated above has not yet yielded samples that are both reliable and complete, due to the contamination from starburst activities in separating AGN components. Observations of galaxies in the hard X-ray band are the only way to select accreting black holes in an unbiased fashion, and to probe how AGN work near the black hole's gravitational radius.

Future hard X-ray (>10 keV) imaging missions (*NuSTAR*, *ASTRO-H*) will provide a step forward in revealing AGN with column densities of $N_H \sim$ a few $10^{24}$ cm$^{-2}$. At the highest column densities, even the 10-40 keV light is suppressed (by a factor $\sim$10 at $N_H = 10^{25}$ cm$^{-2}$), leaving the AGN visible only in scattered X-rays. The spectral sensitivity of IXO in the 2-10 keV band will reveal the telltale intense iron K emission characteristic of a Compton reflection dominated source and can be combined simultaneously with hard X-ray data of unprecedented sensitivity. From IXO deep surveys using simultaneously the WFI and the Hard X-ray Imager (HXI), it is expected that more than 50% of the 10-40 keV X-ray background can be resolved into discrete AGN including Compton-thick ones. Identifying and investigating the nature of these objects (e.g., star formation rate) will be possible thanks to contemporary observatories at longer wavelengths such as ALMA, JWST and the E-ELT.

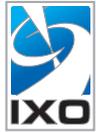

> The spectral capabilities of IXO across the whole X-ray band will be a key to unveil how much of the growth of supermassive black holes occurs in obscured objects, including the Compton-thick AGN.

## Cosmological spin evolution

IXO will be able to measure the spin of supermassive black holes for distant AGN, using their Fe emission line profile (see Section 2.3.1.2). Current X-ray observatories are beginning to measure spin parameters in a small number of sources. The variety of techniques open to IXO observers will provide the best means of obtaining the largest possible number of spin measurements. IXO will increase the number of current spin measurements by at least an order of magnitude, revealing the nature of the first black holes to inhabit young galaxies at high redshift.

The spin distribution of AGN is a powerful discriminator between growth histories of supermassive black holes that may form identical mass-functions. Growth by smooth accretion between major mergers favours rapidly spinning SMBH, while short-lived episodic accretion of small fragments (compared to the SMBH) favours low spin parameters (see Figure 2.5). With IXO we will measure the spin of a few hundred SMBH and therefore determine which process dominated their growth.

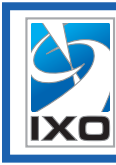

> IXO will measure the spin distribution of supermassive black holes as a direct test of the dominant growth mode of the black holes.

**Figure 2.5.** *Spin as a probe of SMBH growth history. Simulations of the distribution of supermassive black hole spin of AGN depending on the dominant mode for black hole growth: growth by major mergers only (red), by mergers and standard prolonged accretion (blue) and from chaotic accretion, i.e. short lived accretion episodes of small fragments (green). Figure adapted from Berti & Volonteri 2008.*

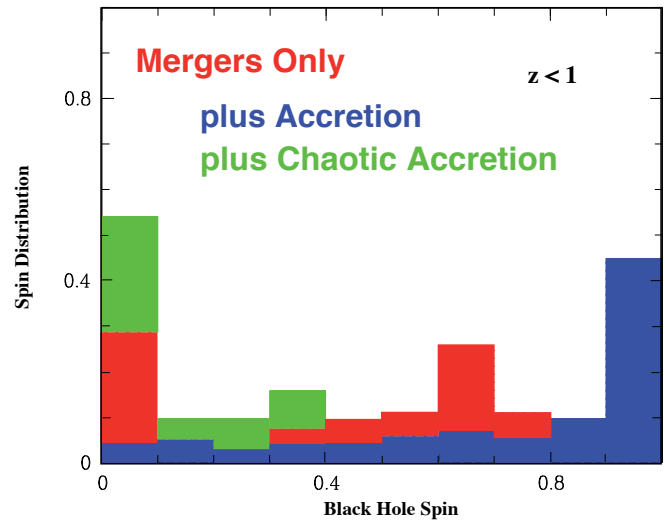

### 2.1.3 Cosmic feedback from supermassive black holes

#### Feedback and galaxy formation

Galaxy formation models often invoke feedback from growing SMBHs as the process that drives galaxy "downsizing" (i.e., massive galaxies forming first, smaller ones forming later), by preventing the largest galaxies to continue forming stars. Regulation of star formation by AGN feedback is also invoked to explain the rapid move of forming galaxies from the blue to the red sequence and in establishing the M-sigma relation between SMBH mass and galaxy bulge velocity dispersion. In conjunction with other major facilities capable of mapping star formation out to very early epochs, IXO will enable us to test whether AGN activity has a direct impact in the stellar population of distant galaxies.

There is no question that a growing black hole *could* drastically affect its host galaxy. Whether and how it *does* so, however, is an open question that depends on how much of the energy released actually interacts with the matter in the galaxy. If the energy is in electromagnetic radiation and the matter largely stars, then very little interaction is expected. If the matter consists of gas, perhaps with embedded dust, the radiative output of the black hole can both heat the gas, and drive it via radiation pressure. Alternatively, if significant AGN power emerges in winds or jets (see Figure 2.6), mechanical heating and pressure provide the link. Either form of interaction can be sufficiently strong that gas can be driven out of the galaxy entirely (Silk & Rees 1998).

The radiative form of feedback is most effective when the black hole is accreting close to its Eddington limit. The mechanical form associated with jets, on the other hand, operates at accretion rates below the Eddington limit. X-ray observations are essential for studying both forms of feedback. The mechanical forms of feedback rely on dynamical (ram) pressure to accelerate gas to high speeds (Figure 2.7). If this gas is initially of moderate temperature, the interaction will shock it to high temperatures where it can only be detected in X-rays. Since the efficiency of ram pressure acceleration is roughly proportional to the volume filling factor of the accelerated gas, most of the energy is probably absorbed by the hot component of the interstellar medium. Gas accelerated by radiation pressure or radiative heating is likely to be cold and dusty. The interaction is therefore much more difficult to observe directly. X-ray and far infrared emission can escape from the inner regions where the interaction occurs, revealing the AGN itself.

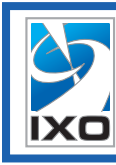 Direct feedback from AGN in quenching star formation in high-redshift galaxies will be probed by observing with IXO the same deep galaxy survey fields that E-ELT or JWST will be observing to map their starlight at infrared wavelengths.

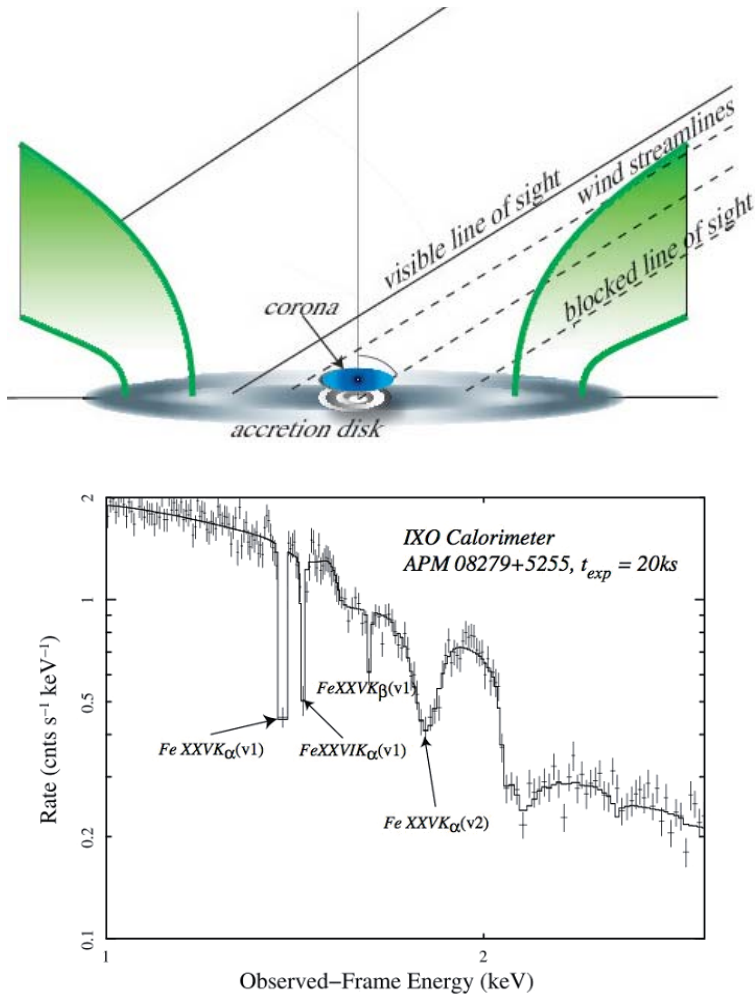

**Figure 2.6. Top panel**: *Schematic view of the Compton-thick outflowing material around the Cloverleaf Quasar, which blocks the direct view to the SMBH and accretion disk, although the optically active Broad Line Region can be seen (Chartas et al. 2007). The Compton-thick wind blocks the view of the near side of the accretion disk, but scattered and fluorescent emission from the far side and the outflow can reach the observer. **Bottom panel**: The characteristic X-ray Broad Absorption Line signatures of the fast (~0.2c) outflow in APM 08279+5255 (Chartas et al. 2002), which may be in the "blowout phase", as seen by the IXO-XMS calorimeter in just 20ks. Such powerful winds might actually pinpoint a key phase in the post-merger galaxy evolution, and their study in more typical moderate luminosity AGN will be possible with IXO in reasonable integration times.*

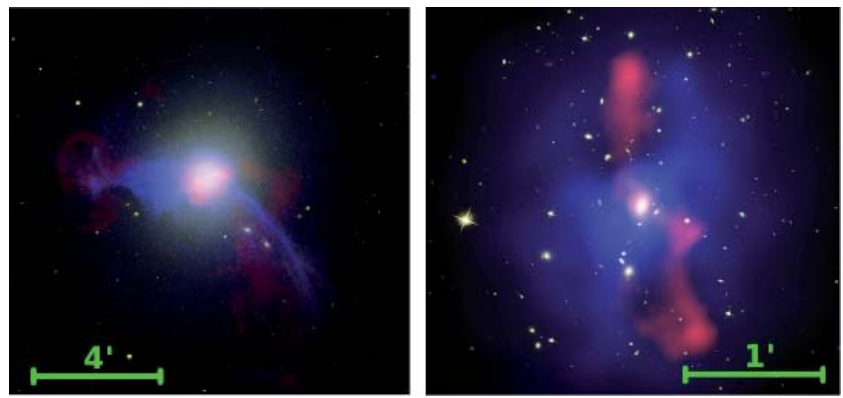

**Figure 2.7. Left**: *X-ray emission (blue), radio emission (red) superposed on an optical image of M87. The X-ray structure was induced by ~$10^{58}$ erg outburst that began 10 Myr ago (Forman et al. 2005). The persistence of the delicate, straight-edge X-ray feature (part of the South West filament) indicates a lack of strong turbulence. **Right**: X-ray emission (blue), 320 MHz radio emission (red) superposed on an HST optical image of the z=0.21 cluster MS0735.6+7421 (McNamara et al. 2005). Giant cavities each 200 kpc (1 arcmin) in diameter were excavated by the AGN. With a mechanical energy of $10^{62}$ erg, MS0735 is amongst the most energetic AGN known. This figure shows that AGN can affect structures on galaxy and cluster scales.*

Radiative acceleration appears to be particularly dramatic in the outflows from some luminous quasars. UV observations indicate that outflows reaching 0.1-0.2c may be present in most quasars. X-ray observations are required to determine the total column density and hence the kinetic energy flux. Current work on a small number of objects implies that this can be comparable to the radiative luminosity (see, e.g., Figure 2.4). To diagnose the energetics of quasars we need large samples of quasars, comparing them with the less energetic (but still substantial) outflows from lower luminosity AGN, which can reach speeds of several thousand km/s.

IXO's huge spectroscopic throughput will extend detailed evidence on how both radiative and mechanical AGN feedback operates to redshifts $z$=1-3, where the majority of galaxy growth is occurring. IXO will be sensitive to all ionization states from Fe I – Fe XXVI, allowing us to study how feedback affects all phases of interstellar and intergalactic gas, from million-degree collisionally ionized plasmas to ten-thousand degree photoionized clouds, and to measure the velocities and energetics of galactic outflows. These measurements will probe over 10 decades in radial scale, from the inner accretion flow where the outflows are generated ($\sim 10^{-4}$ pc, see Figure 2.3) to the halos of galaxies and clusters ($\sim 1$ Mpc), where the outflows deposit their energy.

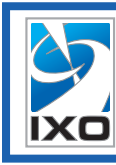

AGN feedback in the form of winds and outflows, which might overpower radiative feedback, will be probed by IXO out to $z\sim$1-3 thanks to its large throughout and spectral resolution.

## Thermal regulation of gas in bulges and cluster cores

Mechanical feedback dominates in galaxies, groups, and clusters at late times, as shown by X-ray observations of gas in the bulges of massive galaxies and the cores of galaxy clusters (e.g. Figure 2.7). The energy transfer process is surprisingly subtle. The radiative cooling time of the hot gas in these regions is often much shorter than the age of the system, so that without any additional heating, the gas would cool and flow into the centre. For giant ellipticals the resulting mass cooling rates would be of order 1 solar mass per year. At the centres of clusters and groups, cooling rates range between a few to thousands of solar masses per year. Spectroscopic evidence from *Chandra* and *XMM-Newton* show that some cooling occurs, but not to the extent predicted by simple cooling (Peterson & Fabian 2006). Limits on cool gas and star formation rates confirm this. Mechanical power from the central AGN acting through jets must be compensating for the energy lost by cooling across scales of tens to hundreds of kpc (McNamara & Nulsen 2007).

The gross energetics of AGN feedback in clusters is reasonably well established. This is directly relevant to feedback in galaxies as well, as the most massive galaxies live and evolve in these cluster hot halo environments. Remarkably, relatively weak radio sources at the centres of clusters often have mechanical power comparable to the radiative output of a quasar, which is sufficient to prevent hot atmospheres from cooling (McNamara & Nulsen 2007). The coupling between the mechanical power and the surrounding medium are, however, poorly understood. Moreover, it is extremely hard to understand how such a fine balance can be established and maintained.

The heat source – the accreting black hole – is roughly the size of the Solar System, yet the heating rate must be tuned to conditions operating over scales 10 decades larger. The short radiative cooling time of the gas means that the feedback must be more or less continuous. How the jet power, which is highly collimated to begin with, is isotropically spread to the surrounding gas is not clear. The obvious signs of heating include bubbles blown in the intracluster gas by the jets (Figure 2.7 and Figure 2.8) and nearly quasi-spherical ripples in the X-ray emission that are interpreted as sound waves and weak shocks. Future low-frequency radio observations of the bubbles and cavities are of great importance in determining the scale of the energy input. The disturbances found in the hot gas carry enough energy flux to offset cooling, but the microphysics of how such energy is dissipated in the gas is not understood.

The persistence of steep abundance gradients in the cluster gas, imprinted by supernovae in the central galaxy, means that the feedback is gentle, in the sense that it does not rely on violent shock heating or supersonic turbulence. Long filaments of optical line-emitting gas in some objects also suggest low levels of turbulence. Yet the continuous streams of radio bubbles made by the jets, the movement of member galaxies and occasional infall of subclusters must make for a complex velocity field.

With high resolution imaging and moderate resolution spectroscopy the *Chandra* and *XMM-Newton* observatories have established AGN feedback as a fundamental astrophysical process in nature. Evidence has accumulated that gas in cluster centres cools down to temperatures of up to a factor 10 below that of the outer parts, but not beyond (e.g., Sanders et al. 2008). This is due to the AGN feedback depositing energy into the cluster gas as shown in Figure 2.8. IXO will, for example, determine the amount of gas at 0.25 keV through its OVII emission at 21.6 Å. However, the dynamics of these powerful outflows are not yet understood.

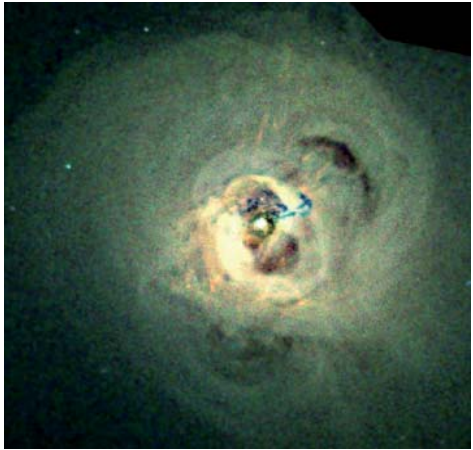

**Figure 2.8.** *How did feedback from black holes influence galaxy growth? Chandra X-ray observations of the Perseus cluster (Fabian et al. 2003) and other nearby clusters have revealed the indelible imprint of the AGN on the hot gas in the core. The X-ray image shows two inner cavities containing the active radio lobes as well as two outer "ghost" cavities. A number of ripples (actually sound waves), which result from relativistic radio bubbles injected by the AGN, are also evident. Radiative and mechanical heating and pressure from black holes has a profound influence not only on the hot baryons, but on the evolution of all galaxies whether or not they are in clusters.*

Understanding the dynamics demands a leap in spectral resolution by more than one order of magnitude above that of *Chandra* and *XMM-Newton*. The IXO spectral resolution and sensitivity is needed to understand how the bulk kinetic energy is converted to heat. Its capabilities are essential in order to measure and map the gas velocity to an accuracy of tens of km/s, revealing how the mechanical energy is spread and dissipated (Figure 2.9). From accurate measurements of line profiles and from the variations of the line centroid over the image it is possible to deduce the characteristic spatial scales and the velocity amplitude of large (> kpc) turbulent eddies, while the total width of the line provides a measure of the total kinetic energy stored in the stochastic gas motions at all spatial scales. Such data will provide crucial insight into the Intra-cluster Medium heating mechanisms. Observations of the kinematics of the hot gas phase, which contains the bulk of the gaseous mass, and absorbs the bulk of the mechanical energy in massive elliptical galaxies, are only possible at X-ray wavelengths.

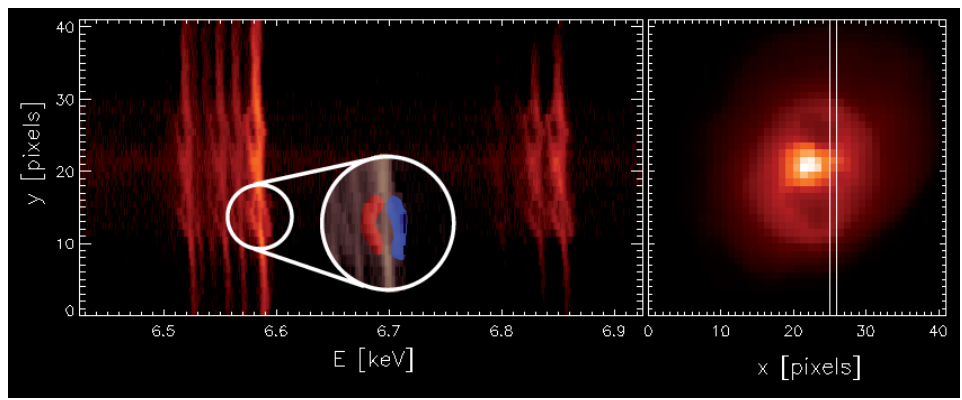

**Figure 2.9.** *Simulated high-resolution X-ray spectra from the shells and X-ray cavities in the Perseus cluster (see Figure 2.8). The right panel shows the X-ray image and the chosen spectral slit, while the left shows the spectrum of the K-alpha lines from Fe XXV and Fe XXVI (both lines are multiplets), as will be observed by the IXO-XMS in an exposure of 250 ks. At the location of the cavities, each of the lines splits into three components (approaching, rest-frame and receding), from which the velocity (and age) and therefore the jet power can be derived. Hydrodynamic simulations of jets in galaxy clusters with parameters appropriate for Perseus (jet power $10^{45}$ erg s$^{-1}$; Heinz, Brügen & Morsony, 2010) have been used. See http://www.astro.wisc.edu/~heinzs/ for a movie.*

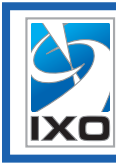

> IXO will measure gas velocities and turbulence in the intracluster gas, injected by the Active
> Galactic Nucleus, and therefore, directly measure the kinetic energy deposited on large-scales.

## Feedback from starburst driven superwinds

In addition to AGN feedback, star formation itself, through supernovae and winds, also influences the
ability of galaxies to form further stars. This is specially true in starburst galaxies, which drive galactic-
scale outflows or superwinds that may be responsible for removing metals from galaxies and polluting the
Inter-galactic Medium. Superwinds are powered by massive star winds and by core collapse supernovae
which collectively create $T<10^8$ K bubbles of metal-enriched plasma within star forming regions. These
over-pressured bubbles expand, sweep up cooler ambient gas, and eventually blow out of the disk into
the halo. Tremendous progress has been made in mapping cool entrained gas in outflows through UV/
optical imaging and absorption line spectroscopy, demonstrating that superwinds are ubiquitous in galaxies
forming stars at high surface densities and that the most powerful starbursts can drive outflows near escape
velocity. Theoretical models of galaxy evolution have begun to incorporate superwinds, using ad-hoc
prescriptions based on our knowledge of the cool gas. However, these efforts are fundamentally impeded by
our lack of information about the hot phase of these outflows. The hot X-ray emitting phase of a superwind
contains the majority of its energy and newly-synthesized metals, and given its high specific energy and
inefficient cooling it is also the component most likely to escape from the galaxy's gravitational potential
well. Knowledge of the chemical composition and velocity of the hot gas are crucial to assess the energy
and chemical feedback from a starburst. These processes may be responsible for Inter-galactic Medium
enrichment and the galaxy mass-metallicity relationship. The IXO-XMS will enable direct measurements
to be made of the rates at which starburst galaxies of all masses eject gas, metals, and energy into the Inter-
galactic Medium, through the determination of the composition, energetics and flow rates of the hot gas
(Figure 2.10).

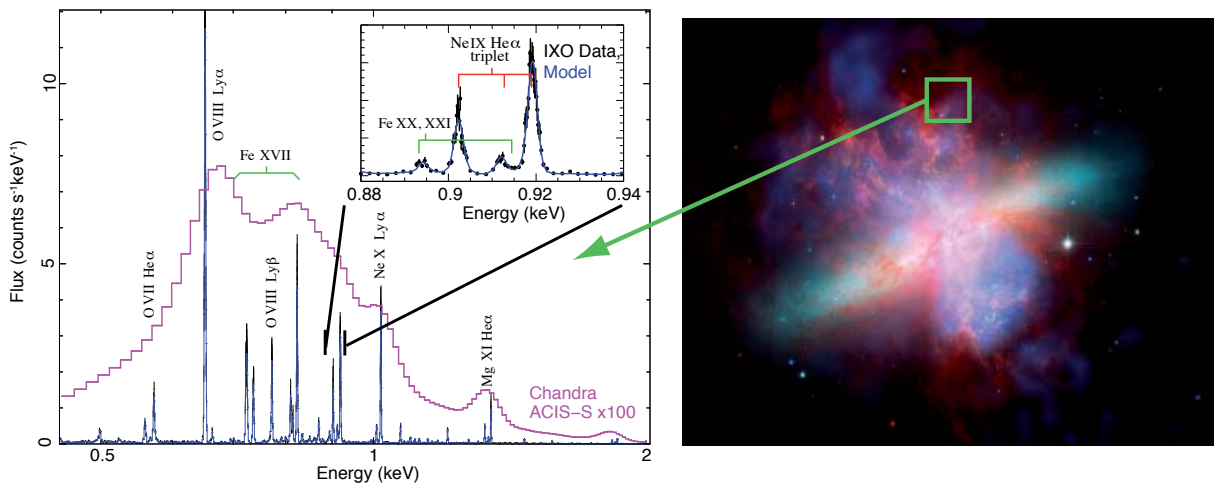

**Figure 2.10.** *Left: IXO high-resolution X-ray spectra (blue) show the metal-enriched hot gas outflowing from star-
burst galaxy Messier 82, a part of the feedback process unresolvable with current X-ray CCD data (magenta). The
insert shows that the He-like emission line triplet of NeIX can be resolved, and that not only velocities can be meas-
ured, but the plasma temperature and ionisation state can be diagnosed. Right: Superwinds in Messier 82, exhibiting
a starburst-driven superwind. Diffuse thermal X-ray emission as seen by Chandra is shown in blue. Hydrocarbon
emission at 8μm from SPITZER is shown in red. Optical starlight (cyan) and Hα+[NII] (yellow) are from HST-ACS
observations.*

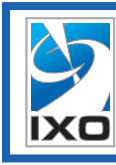 IXO's spatially resolved high-resolution spectroscopy around starburst galaxies will measure the amount of gas, metals and energy deposited by galactic superwinds into the Inter-galactic Medium.

## 2.2    Large scale structure and the creation of chemical elements

### 2.2.1    The hot diffuse components of the Universe

The formation and evolution of the large-scale structure of the Universe is a central issue of cosmology. Recent observational progress, showing that 95% of the total mass-energy of the Universe is contained in cold dark matter and dark energy, has allowed us to robustly define the cosmological framework in which structures form. Much progress has also been made in reconstructing the evolution of the dark matter distribution from its initial density fluctuations. In contrast, *we still do not understand the evolution of the baryonic 'visible' component of the Universe*, collected in the dark matter potential wells. Present observations, as well as theoretical work, indicate that this is fundamentally a multi-scale problem, involving a complex interplay between gravitational and poorly understood non-gravitational processes. Galaxy formation depends on the large scale environment and on the physical and chemical properties of the intergalactic gas from which they form, which in turn is affected by galaxy feedback through energy released from star explosions and active nuclei accreting matter from their environment (see Section 2.1, and specifically 2.1.3).

The vast majority of the baryons that we see reside in clusters of galaxies and are in a hot phase (see Section 2.2.3). Clusters are, therefore, the largest bound reservoirs of ordinary matter, which sits in their potential wells at temperatures of hundreds of millions of degrees. This hot gas has been chemically enriched in cluster galaxies and can be therefore used to trace the chemical evolution of the Universe (see Section 2.2.4). But still, an even larger fraction of the baryon content of the Universe is still missing, in a yet undetected hot phase of the intergalactic medium, which might host well over 1/3 of the total baryons of the Universe (Section 2.2.2).

### 2.2.2    Missing baryons and the Warm-Hot Intergalactic Medium

Ordinary matter (baryons) represents 4.6% of the total mass/energy density of the Universe, but less than 10% of this baryonic matter appears in collapsed objects (stars, galaxies, groups; Fukugita & Peebles 2004). Theory predicts that most of the baryons reside in vast unvirialized filaments that connect galaxy groups and clusters (the "Cosmic Web"). After reionization, the dominant heating mechanism is through the shocks that develop when large-scale structures collapse in the dark matter potential wells. As large-scale structure becomes more pronounced with cosmological time, the gas is increasingly shock-heated, reaching temperatures of $10^{5.5}$-$10^7$ K for z < 1 (Figure 2.11, left). Additional heating occurs through star-formation driven galactic winds and AGN, processes that enrich the surroundings with metals.

Ly $\alpha$ studies and OVI absorption line studies detect warm baryons, but 30-50% of the baryons remain unaccounted for (Danforth & Shull 2008). These "missing baryons" can only be observed through X-ray studies, but even detecting the very first of them through X-ray absorption spectroscopy of distant sources is likely at the edge (if not beyond) the capabilities of current X-ray observatories like *Chandra* and *XMM-Newton*. Therefore, a basic goal is to determine if the missing baryons exist in the predicted hot phase and this requires a higher throughput and higher spectral resolution X-ray observatory like IXO. In the standard cosmology, chemical enrichment of the Inter-galactic Medium occurs through galactic superwinds, a powerful feedback mechanism that also heats the gas (see Section 2.1.3). The shocks and superwinds leave distinctive absorption features, such as pairs of absorption lines (for a line of sight passing through a galactic superwind shell) and turbulent broadening. This feedback mechanism not only extends the cross section of the metal-enhanced regions, its effects are ion-dependent.

These observational diagnostics allow us to pursue another key science goal, which is to test the large-scale structure and galactic superwind heating of the Cosmic Web. The extent of the superwinds and the elemental mixing can be determined by studying the spatial relationship between hot gas seen through X-ray absorption and the location of galaxies (Stocke et al. 2006). By studying the same sight lines with X-rays and UV-optical bands, we will discover the relationship of all temperature components to the galactic environment. Therefore, X-ray studies will measure the extent of galactic superwinds and the chemical mixing process. Finally, we shall measure the connections of Cosmic Web filaments to galaxy groups and clusters.

These goals are achievable by measuring the He-like and H-like X-ray resonance lines of carbon (C V, C VI), nitrogen (N VI, N VII) and oxygen (O VII, OVIII) toward background AGN (other ions may be detected in a few cases, such as Ne IX, Ne X, Fe XVII, and Fe XVIII). Existing measurements of intergalactic OVII and OVIII are near current instrumental detection thresholds and, therefore, need confirmation (Nicastro et al. 2005; Kaastra et al. 2006; Bregman 2007; Buote et al. 2009), but the adjacent ion, intergalactic OVI, is detected in the UV along many sightlines and there are clear detections of OVII and OVIII within the Local Group. A conservative estimate of the equivalent width distribution dN/dz is obtained from models normalized to the OVI measurements (Cen and Fang 2006, Figure 2.11, right; the quality of the OVI normalization will improve significantly with upcoming HST-COS observations). These show that we need an order of magnitude improvement over current sensitivities to conduct an X-ray survey of intergalactic absorption lines from the above elements. In addition, to study the velocity structures of lines, we need resolution that approaches the Doppler width of a line, typically 50-100 km/s, and that is able to separate the two absorption lines from an expanding superwind shell occurring near a typical escape velocity of 200-1000 km/s.

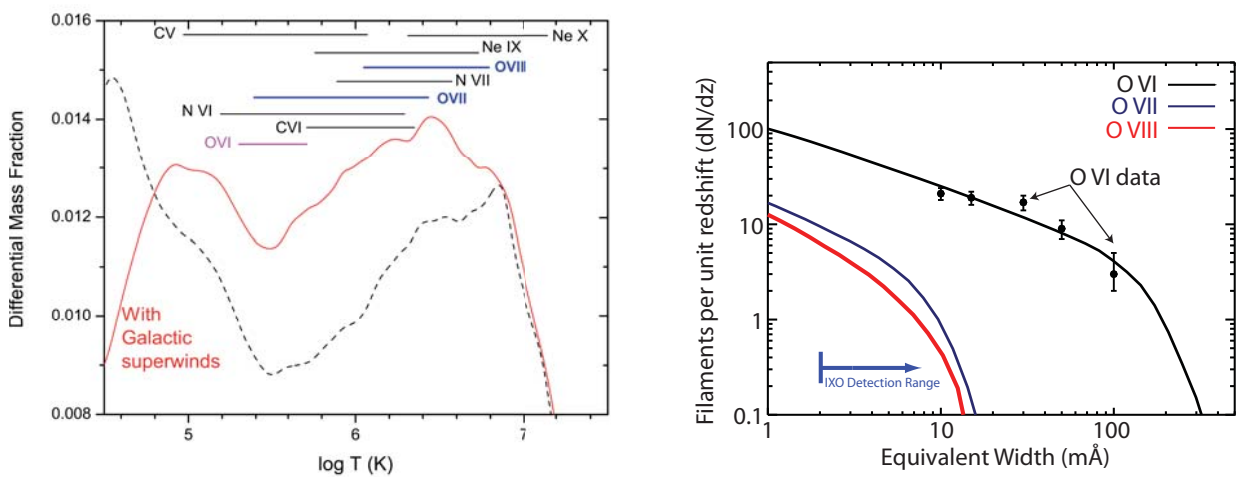

**Figure 2.11. Left:** *The differential gas mass fraction as a function of temperature at low redshift for the ΛCDM cosmological simulation of Cen & Ostriker (2006). This distribution is sensitive to the presence of galactic superwinds (solid red line; dashed line is without superwinds). The ions with the strongest resonance lines in the $10^5$-$10^7$ K range are shown, and except for OVI (UV line; 1035 Å), the other lines lie in the X-ray band. In all cases, models predict that most of the baryonic mass resides in gas that can only be seen through X-rays.* **Right:** *The differential number of absorbers as a function of equivalent width for OVI (λ1035), OVII (Kα), and OVIII (Kα), based on the model of Cen & Fang (2006); the OVI data are from Danforth and Shull (2005). Despite the smaller number of OVII/OVIII absorbers expected than those seen in OVI, the former contain much more mass than the latter.*

IXO has two instruments - XMS and XGS - that can detect these absorption lines, although the grating spectrometer has a sensitivity advantage and it has the ability to study velocity structure. The sensitivity to line detection is 15 times better than with *XMM-Newton* or *Chandra,* with lines of equivalent width as weak as 2 mÅ (0.05 eV, see Figure 2.12). The IXO-XGS grating spectrometer will use several dozen of the brightest sources in the sky, providing a dataset with which to obtain at least 100 absorption line systems, each having one or more lines from the He-like and H-like species of O, C, and N, which cover the temperature range $10^5$-$10^7$ K. Simulations show that the OVII line will be the strongest, but that multiple lines will be detected in about 80% of the systems. For each line, we obtain the column density, velocity and the velocity width. This data set will revolutionize the field by answering the most fundamental questions in the study of the missing baryons.

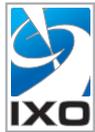

IXO will detect hundreds of intervening absorption lines towards bright distant sources, which are produced by the highly ionized baryonic gas thought to reside as filaments in the inter-galactic medium. Those are thought to be the missing baryons in today's Universe. The role of galactic superwinds in heating and chemically enriching the Inter-galactic Medium will also be tested by IXO.

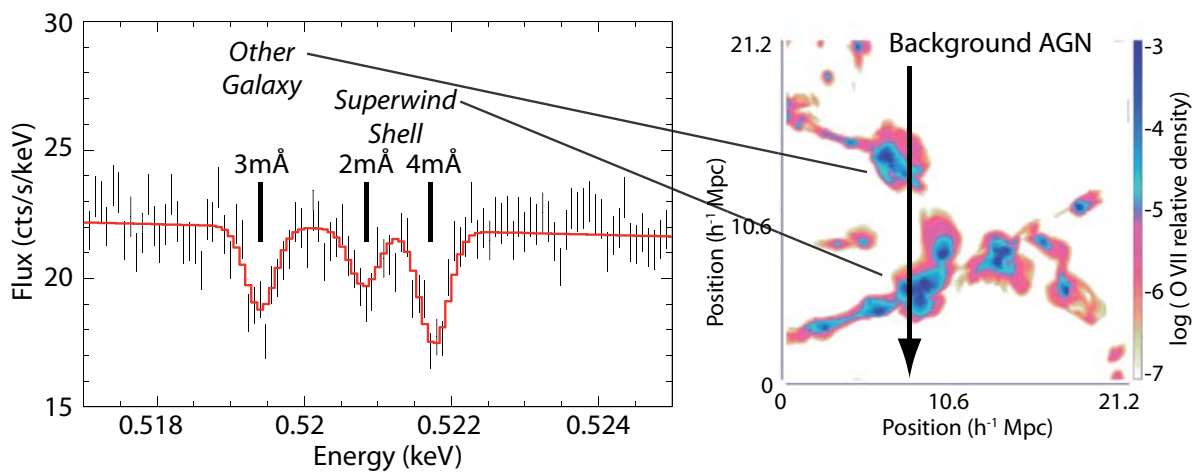

**Figure 2.12.** *Left: IXO-XGS simulated spectrum (500 ks) of a background AGN (0.5-2 keV flux of 1.5 x 10⁻¹¹ erg/ cm²/s) going through a superwind shell ejected by a nearby galaxy (z=0.1) and through the halo of another galaxy 1200 km/s farther. The three OVII absorption lines are seen and resolved by the IXO gratings (the two superwind shell lines are 600 km/s apart). Adapted from the simulation by Cen & Ostriker (2006) which is shown on the* **right**.

### 2.2.3 Cluster physics and evolution

Galaxy clusters, the largest collapsed structures defining the nodes of the Cosmic Web and delineating the large-scale structure of the Universe, are privileged sites to reveal the complex physics of structure formation. Over 80% of their mass (up to $10^{15}$ M$_\odot$) resides in the form of dark matter. The remaining mass is composed of baryons, most of which (about 85%) are in the form of a diffuse, hot T > $10^7$ K plasma (the Intra-cluster Medium ) that radiates primarily in the X-ray band. Thus, in galaxy clusters, through the radiation from the hot gas and the galaxies, we can observe and study the interplay between the hot and cold components of the baryonic matter and the dark matter. X-ray observations of the evolving cluster population provide a unique opportunity to address such open and fundamental questions as: **(1)** How do hot diffuse baryons dynamically evolve in dark matter potentials? **(2)** How and when was the excess energy that we observe in the Intra-cluster Medium generated?

To achieve these goals, we need to extend the detailed studies currently limited to the relatively nearby Universe (*z*<0.5) to the more distant Universe where the first low-mass clusters (few $10^{13}$ M$_\odot$) began to emerge (*z*~2). The thermo-dynamics and the chemical properties of these early systems needs to be studied and their evolution traced into today's massive clusters.

## How do hot baryons dynamically evolve in dark matter potentials?

Clusters grow via accretion of dark and luminous matter along filaments and the merger of smaller clusters and groups. X-ray observations show that many present epoch clusters are indeed not relaxed systems, but are scarred by shock fronts and contact discontinuities (Markevitch & Vikhlinin 2007), and that the fraction of un-relaxed clusters likely increases with redshift. Although the gas evolves in concert with the dark matter potential, this gravitational assembly process is complex, as illustrated by the temporary separations of dark and X-ray luminous matter in massive merging clusters such as the "Bullet Cluster" (Clowe et al. 2006).

There are important questions to be answered; both to understand the complete story of galaxy and cluster formation from first principles and, through a better understanding of cluster physics, to increase the reliability of the constraints on cosmological models derived from cluster observations (see Section 2.2.4). These include: **(1)** How is the gravitational energy that is released during cluster hierarchical formation dissipated in the intra-cluster gas, thus heating the Intra-cluster Medium, generating gas turbulence, and producing significant bulk motions? **(2)** What is the origin and acceleration mechanism of the relativistic particles observed in the Intra-cluster Medium? **(3)** What is the total level of non-thermal pressure support, which should be accounted for in the cluster mass measurements, and how does it evolve with time? To answer these questions, we need to map velocities and turbulence which requires more than an order of magnitude improvement in spectral resolution, as compared to currently available CCD resolution, while keeping good imaging capabilities.

IXO will bring a new key observational capability - spatially resolved high-resolution spectroscopy - which will enable gas flow velocities and gas turbulence being locally measured in clusters of galaxies and other extended sources. In particular, sub-cluster velocities and directions of motions will be measured by combining redshifts measured from X-ray spectra (which give relative line-of-sight velocities) and total sub-cluster velocities deduced from temperature and density jumps across merger shocks or cold fronts (Markevitch & Vikhlinin 2007). These measurements, combined with high quality lensing observations from future instruments (e.g. LSST and *Euclid*), will probe how the hot gas reacts in the evolving dark matter potential. X-ray line width measurements will allow the level of gas turbulence to be mapped in detail for the first time. As an example, Figure 2.13 shows that the 2.5 eV resolution of the IXO-XMS can measure turbulent line widths down to about 100 km/s in a small (1 arcmin²) region of the Hydra A cluster.

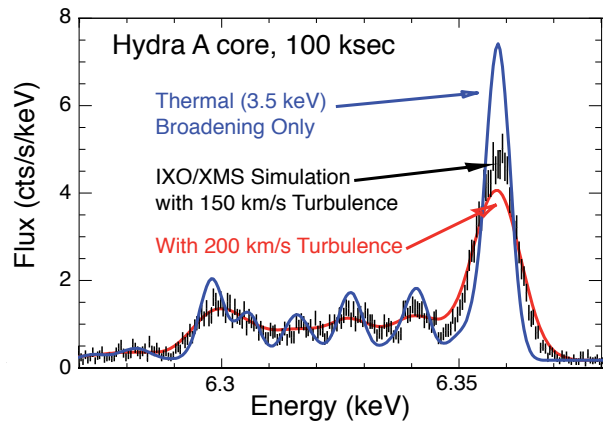

**Figure 2.13.** *IXO spectrum of Fe XXV lines shows that turbulence of ~150 km/s or ~200 km/s may be distinguished from thermal broadening alone. This cannot be done at CCD resolution. Simulated IXO-XMS data in black, models in colour.*

IXO-HXI will provide the high sensitivity and spatial resolution in the hard energy band (10-40 keV) to detect and map the inverse Compton emission that has so far not even clearly been detected in clusters. This promises unique information on the energy density of the relativistic particles, and when combined with next generation radio observatories like SKA, would probe the history of magnetic fields in clusters. Capabilities like those of IXO are needed to understand these observationally elusive, but important components of the Intra-cluster Medium.

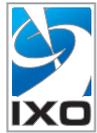

> Through spatially resolved high-resolution spectroscopy IXO will measure intra-cluster gas motions and turbulence, showing how baryonic gas evolves in the dark matter cluster potential wells.

## How and when was the excess energy in the intergalactic medium generated?

As discussed earlier, one of the most important revelations from recent observations is that non-gravitational processes, particularly galaxy feedback from outflows created by supernovae and super-massive black holes, must play a fundamental role in the history of all massive galaxies as well as the evolution of groups and clusters as a whole. Galaxy feedback is likely to provide the extra energy required to keep large quantities of gas in cluster cores from cooling all the way down to molecular clouds, to account for the energy (i.e. entropy) excess observed in the gas of groups and clusters, to regulate galaxy and star formation, and to produce the galaxy red sequence (see also Section 2.1.3 discussing the physics of AGN feedback mechanism).

It is now well established from *XMM-Newton* and *Chandra* observations of local clusters and groups that their hot atmospheres have much more entropy than expected from gravitational heating alone (Pratt et al. 2010; Sun et al. 2009). Determining when and how this non-gravitational excess energy was acquired is an essential goal of IXO. Galaxy feedback is a suspected source, but understanding whether the energy was introduced early in the formation of the first halos (with further consequence on galaxy formation history), or gradually over time by AGN feedback, SN driven galactic winds, or an as-yet unknown physical process, is crucial to our understanding of how the Universe evolved.

The various feedback processes, as well as cooling, affect the intergalactic gas differently, both in terms of the amount of energy modification and of the time-scale over which this occurs. According to current evidence on the evolution of star formation with cosmic time, before the epoch of cluster formation ($z\sim2$) little star formation (and therefore little supernovae) had occurred. Measuring the gas entropy and metallicity (a direct probe of SN feedback) at that epoch and comparing it to that of clusters in the local Universe is a key to disentangle and understand the respective role for each process. Since non-gravitational effects are most noticeable in groups and poor clusters, the building blocks of today's massive clusters, these systems are of particular interest.

A major goal for IXO is therefore to study the properties of the first small clusters emerging at $z\sim2$ and directly trace their thermodynamic and energetic evolution to the present epoch. Future wide-field Sunyaev-Zel'dovich, X-ray (e.g. *eROSITA*) and optical-IR surveys (e.g *WFIRST/Euclid*) will discover many thousands of clusters with $z<2$, but will provide only limited information on their individual properties, especially at high $z$. These surveys will provide excellent samples of clusters for follow-up studies with IXO at the high sensitivity and resolution required to determine the X-ray properties of these low mass systems. In addition, $\sim 4$ low mass clusters per deg$^2$ with M$>10^{13}$M$_\odot$, will be detected serendipitously within the 18 arc min diameter field of the IXO Wide Field Imager. The power of IXO is illustrated in Figure 2.14 that shows simulated, deep spectra for high redshift clusters as will be obtained with the IXO-XMS calorimeter. These spectra will provide gas density and temperature profiles, and thus, entropy and mass profiles to $z\sim1$ for low mass clusters (kT$\sim2$ keV, Figure 2.14, left) with a precision currently achieved only for local systems. Measurements of the global thermal properties of the first poor clusters in the essentially unexplored range $z$=1.5-2 also will become possible (Figure 2.14, right).

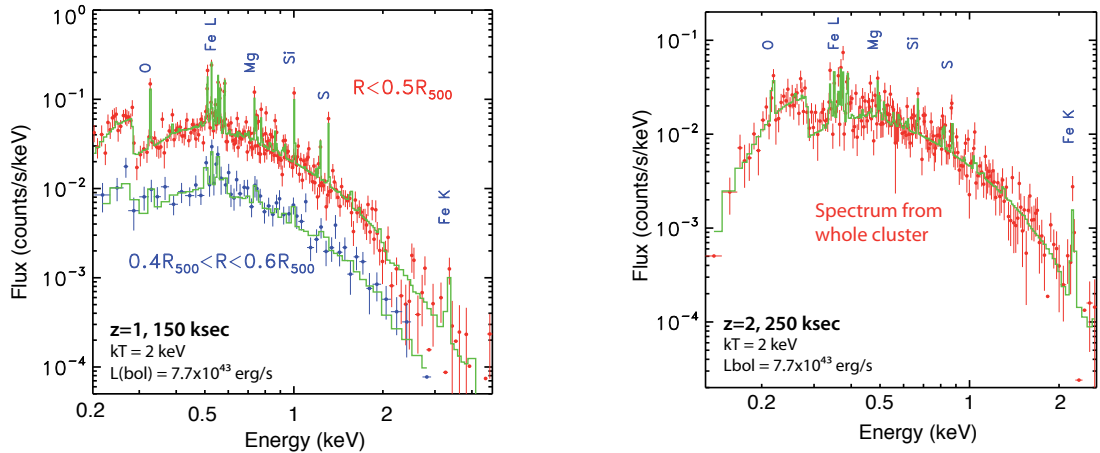

**Figure 2.14.** *IXO will measure local gas densities, temperatures and metallicities in all clusters out to z=2, a goal only currently achievable for massive clusters in the low-redshift Universe. The figure shows IXO-XMS simulations of low-mass galaxy clusters. **Left**: At z=1, spectra can be extracted in annuli at different radii. In the 0.4R$_{500}$-0.6R$_{500}$ region (R$_{500}$ is where the mean cluster mass density is 500 times the cosmic critical density), the temperature and iron abundances are measured with an accuracy of ±5% and ±20% respectively. **Right**: At z=2, the global temperature and abundances can be measured to an accuracy of ±3% for kT, and to ±3.5% for O and Mg, ±25% for Si and S, and ±15% for Fe.*

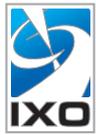
> By measuring gas density and temperature profiles in groups and clusters out to z~2, IXO will reveal the evolution of these objects and, in particular, the epoch when their entropy excess was generated.

In conclusion, numerical simulations have reached a stage where modelling, including all hydro-dynamical and galaxy formation feedback processes, is becoming feasible, although AGN feedback modelling is still in its infancy. The appropriate physics of these processes is not well understood, and advances are largely driven by observations. Thus, constant confrontation between numerical simulations of galaxy cluster formation and observations is essential for making progress in the field. IXO observations of the hot baryons, the most significant baryonic mass component of clusters, combined with radio observations (e.g. SKA), observations of the cold baryons in galaxies (e.g. from JWST, ALMA, and E-ELT) and of the dark matter via lensing data (LSST, *WFIRST/Euclid*) will provide, for the first time, the details for a sufficiently critical comparison. We expect that the major breakthrough of a detailed understanding of structure formation and evolution on cluster scales will come from simulation-assisted interpretation and modelling of these new generation observational data.

### 2.2.4 Galaxy cluster cosmology

The largest well defined objects in the Universe are galaxy clusters which form an integral part of the cosmic Large-scale Structure. They are, therefore, very promising probes to assess the statistics of the Large-scale Structure, its growth over cosmic time, and lend themselves for tests of cosmological models. Systematics in characterising galaxy clusters by their mass are the most critical issue in using clusters for precision cosmology. X-ray observations provide the most robust and detailed observables for cluster characterisation, and systematics can be further minimized by combination with observations at other wavelengths. In the last 10-20 years, galaxy clusters played an important role in establishing the current cosmological paradigm based on measurements of the cluster number density (Henry & Arnaud 1991). These provided evidence for a (now fully confirmed) low amplitude of matter fluctuations. Also, based on the baryon mass fraction compared to Big Bang nucleosynthesis (White et al. 1993), distant cluster studies point to a low (dark + baryonic) matter density.

Two crucial observational approaches are used to test for the appropriate cosmological model and the effect of Dark Matter and Dark Energy: **(1)** the measurement of the cosmic expansion history (geometric method) applying e.g. to the observations of distant SN Ia and Baryonic Acoustic Oscillations and **(2)** the assessment of the growth of Large-scale Structure via a time dependent growth factor used in connection with observations of the galaxy distribution, cosmic gravitational lensing shear, and the abundance and spatial distribution of galaxy clusters. The ESA/ESO Working Group Report on Fundamental Cosmology highlights the complementarity of these methods and points out the important role of galaxy cluster cosmology (once systematic uncertainties can be overcome). Both types of cosmological tests are required to distinguish without major degeneracies between quintessence type models and modified General Relativity for the explanation of the accelerated cosmic expansion (Frieman et al. 2008; Sapone & Amendola 2007). Galaxy clusters are particularly sensitive probes of structure growth through observations of their abundance (for a given mass limit) as a function of redshift (Jenkins et al. 2001; Tinker et al. 2008). Clusters allow us to probe the redshift range $z$=0-2 where the effect of Dark Energy is most significant. Therefore, galaxy clusters studies will form an indispensable part of the future precision cosmology efforts.

A number of galaxy cluster surveys ongoing and in preparation at different wavelengths (Sunyaev-Zel'dovich Effect): e.g. SPT, ACT and *Planck*; Optical/NIR: e.g. DES, PanSTARRS, *WFIRST/Euclid* and LSST; and X-rays: *eROSITA*) will produce large catalogs of galaxy clusters stretching out to redshifts of 2, with very sparse characterisation of the more distant clusters. Next generation Sunyaev-Zel'dovich Effect experiments and the NIR sky survey by *WFIRST* and *Euclid* will be most important to detect clusters at the highest $z$. X-ray observations will still provide the essential information on the structure and mass of clusters from detailed images and properties of the most massive baryonic cluster component (~85%), the Intra-cluster Medium. A high throughput telescope with the specifications of IXO is required to obtain precision data on cluster at redshifts beyond $z$=1.2. It will be important in two ways: **(1)** improving the cosmological constraints of the mentioned surveys by significantly better calibrating their cluster mass modelling and **(2)** by using IXO follow-up to obtain precision X-ray parameters such as Intra-cluster Medium temperature, gas mass, and $Y_X$ (= gas mass times temperature) for well selected cluster samples.

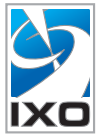
IXO will provide cluster mass estimates out to $z$~2, which are strictly needed to obtain cluster mass functions and thence derive the Dark Energy density and equation of state out to that redshift. IXO will also deliver independent measurements of Dark Energy by measuring the cluster gas fraction.

## Measuring the Growth of Structure

The mass function of galaxy clusters is an exponentially sensitive probe of the linear density perturbation amplitude in the Dark Matter distribution and can be theoretically predicted with high precision for different cosmological models. Given precise cluster masses, the perturbation growth factor can be recovered to ~ 1% accuracy from a sample of 100 massive clusters at a given redshift (Vikhlinin et al. 2010). If the cluster mass function can be observationally measured to 1-2% accuracy at $z$~1, the Dark Energy equation of state parameter, w (the ratio of Dark Energy pressure to its density), can be determined with 3-6% uncertainty. Therefore, observations of the cluster mass function evolution provide an arguably more advanced assessment of cosmic structure growth than the two other most promising methods, weak lensing tomography (Amara & Refregier 2008) and redshift-space distortions in galaxy clustering analysis (Guzzo et al. 2008). The most critical link between theory and observational test is the relation between directly observable cluster parameters and the cluster mass.

Recent studies have shown that X-ray observations can provide tight relations for the parameters X-ray luminosity (central region excised), temperature, and $Y_X$ with a scatter of 17%, 10-12%, and 8-10% , respectively (Kravtsov et al. 2006; Allen et al. 2008; Pratt et al. 2009; Vikhlinin et al. 2006; Mantz et al. 2009). Such tight relations along with *Chandra* and *XMM-Newton* follow-up studies of ROSAT detected clusters has enabled us to gain interesting constraints on w, to provide evidence that the growth of cosmic structure has slowed down at $z$<1 (Mantz et al. 2008; Vikhlinin et al. 2009) and to place first constraints on possible departures from General Relativity (Rapetti et al. 2009). Further important results come from the mass determination comparison with lensing data and detailed hydrodynamical simulations which both point towards a low bias of X-ray cluster masses of few – 20%. Improving the mass estimates is the key to progress. For instance, while the *eROSITA* survey is expected to provide constraints on w (dw/dz) of the order 10% (50%) with present calibration, the results can be improved to 2% (12%) if an independent mass calibration at the 1% accuracy level is available (Haiman et al. 2005). Combining lensing and X-ray observations is crucial since X-rays yield the most certain detections of real clusters and lensing provides a statistically unbiased mass measurement, but with large (~30%) uncertainty. IXO observations combined with lensing data for thousands of galaxy clusters detected by *WFIRST* or *Euclid* will provide a large synergy for precise cluster mass estimation. In this way an uncertainty of 1-2% is expected to be achievable in the normalization of the mass scaling relation. IXO will be crucial in reducing the intrinsic scatter in the relation, and thus the systematics in mass estimate from mass proxies, by allowing us to include further structure parameters in the scaling relations, like the X-ray spectral line broadening as a measure of the dynamical distortion of the clusters (see Section 2.2.3), which affects the mass measurement. A major follow-up programme (~ 5-10 Msec) to get good observational mass proxy parameters (like $Y_X$) for ~ 2000 galaxy clusters with IXO spanning up to *z*=2 can provide the constraints shown in Figure 2.15 (Vikhlinin et al. 2009). In both cases IXO can help to significantly tighten the cosmological parameter constraints, even though IXO is not an observatory focused primarily on cosmology. In addition, IXO will open a window for the discovery of even more distant clusters (*z*> ~ 2) in serendipitous cluster searches in IXO-WFI pointings.

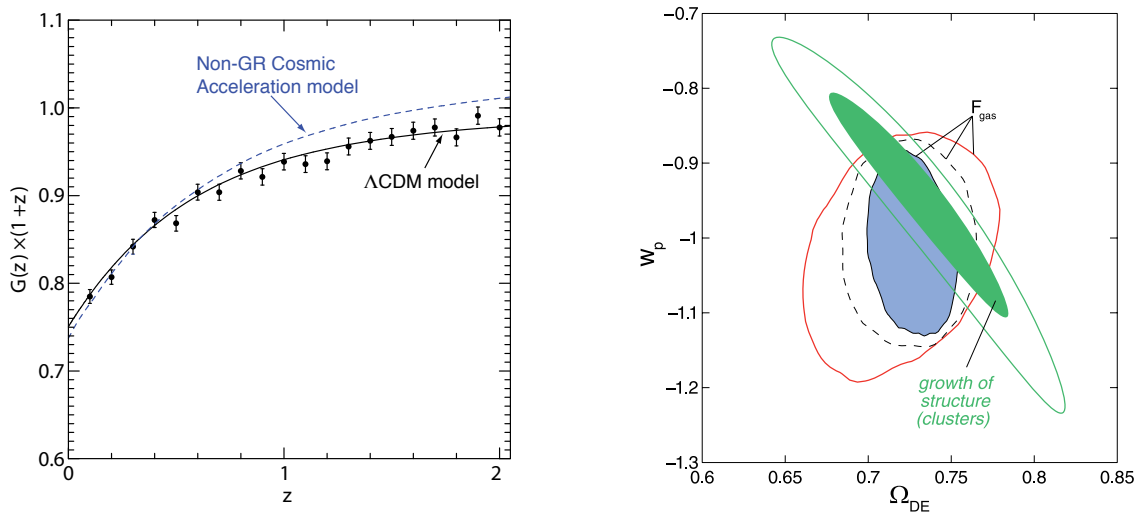

**Figure 2.15.** *Left: The growth of structure as a function of redshift for simulated observations with IXO of 2000 clusters at z=0-2. The dash line shows the growth expected in a non-GR model of cosmic acceleration. Right: The two green uncertainty ellipses are the 68% and 95% confidence contours for the parameter determination based on the structure growth experiment shown on the left panel. The three other ellipses show the constraints based on the study of $f_{gas}$ in 500 galaxy clusters in the redshift range z=0 to z=2 assuming 2% (optimistic), 5% (default), and 10% (pessimistic) systematic uncertainties in predicting $f_{gas}$ (true) as a function of redshift. The constraints shown include the use of PLANCK CMB priors.*

## Galaxy clusters as cosmological yardsticks

Massive galaxy clusters are large enough to capture fair samples of cosmic matter reflecting the cosmic ratio of dark to baryonic matter, with most baryons in X-ray luminous gas. Treating the gas mass fraction in rich clusters as invariant with redshift (with some corrections derived from simulations) and using the d(*z*)³/² dependence of its measurement one can obtain further constraints on the geometry of the Universe which are competitive to those of SN Ia observations (Allen et al. 2008; Rapetti et al. 2008). In a similar way, the combination of X-ray and Sunyaev-Zel'dovich Effect measurements provide cosmic geometry tests (Bonamente et al. 2006). Such studies rest on the ability to obtain precise X-ray measurements for hundreds of massive clusters in the redshift range 0.5<*z*<2, which is only possible with IXO (Rapetti et al. 2008; see also Figure 2.14 of Arnaud et al. 2009). Current measurements in this redshift range are limited to small numbers of clusters and require very long observations.

## 2.2.5    Chemical evolution through cosmic time

The most abundant heavy elements in the Universe are the so-called alpha elements (O, Ne, Mg, Si, S, Ar, Ca) and Fe-group elements. These elements are mostly released by supernovae, and some are even created during the explosions. The production of alpha-elements is dominated by core collapse SNe, the explosions of massive stars, whereas Fe-group production is dominated by Type Ia (thermonuclear) SNe, which are thought to be exploding CO white dwarfs. The progenitor systems of Type Ia SNe are still poorly understood. A simplified characterisation of element production assumes that there are two channels (Mannuci et al. 2006), one prompt channel in which the delay time between stellar birth and explosion is of the order $5 \times 10^8$ yr and a late channel with delay times of up to $4 \times 10^9$ yr. The relative importance of these channels is not well known, nor is it clear whether the designation "channels" is justified, or whether there is a continuum distribution in the delay times  (Greggio 2010). Note that there are still major problems in explaining both the Type Ia SN rate and the relative ratio of Fe-group over alpha-elements (de Plaa et al. 2007; Maoz 2008). According to stellar population and binary evolution calculations, the Type Ia rate and contribution to the Intra-cluster Medium should be much lower than observed. Two other elements of interest, besides the alpha and Fe-group elements, are carbon and nitrogen. The enrichment of carbon and nitrogen are less well understood, as both low mass stars (AGB winds) and massive stars (stellar winds and SNe) contribute.

X-rays provide one of the best means of studying these elements, as they have their K-shell transitions in the energy range 0.1-10 keV, in addition Fe has its L-shell emission in the 0.7-1.3 keV range. The line formation properties of these lines are well known and are similar for all these elements. This makes X-ray spectroscopy a powerful tool to measure abundances in tenuous plasmas. X-ray spectroscopy will enter a new era with IXO, which will bring the study of elemental abundances to a new level, both for those objects where the elements have just been released, supernova remnants (see Section 2.4.1), and the largest reservoirs of hot gas in the Universe, clusters of galaxies (see Section 2.2.3). This approach to trace the production of chemical elements along cosmic time will be complementary to studies of star formation and SN rates as function of redshift by future facilities as JWST, ALMA and the E-ELT.

Clusters of galaxies are the largest gravitationally bound structures in the Universe, with most baryonic matter in the form of hot, X-ray emitting gas (Section 2.2.3). Surprisingly, the metallicity of the gas is about 50% of that of the Galactic interstellar medium, whereas the gas mass to stellar mass ratio is one to two orders of magnitude larger than that in galaxies. This suggests either that the star formation rate in cluster galaxies was much higher than in field galaxies like our own, or galaxies do not retain most of the elements they produce due to galactic winds and/or AGN outflows. Galactic winds may be in particular important for less massive galaxies, which have shallower gravitational potentials. Indeed, the metallicity of galaxies scales with galactic mass. To first order, the abundance ratios are Solar-like. Since no gas escapes from a given volume that collapses into a massive cluster, clusters contain all elements ever produced by their member galaxies, and most of these elements are in the Intra-cluster Medium instead of locked up in stars.

IXO will address the following key questions concerning chemical evolution, as traced by the intracluster medium: **(1)** when were the alpha and Fe-group elements and carbon, nitrogen formed, **(2)** how and at which redshifts were these mainly dispersed into the Intra-cluster Medium, **(3)** what is the main production mechanism? More specifically, we can attempt to separate the contribution of core collapse and Type Ia SNe, identify the contribution from different Type Ia channels, identify the origin of nitrogen, AGB-winds or massive stars, and determine if there is a contribution from ongoing low-level star formation inside clusters.

The way to address these questions is by making abundance maps of unprecedented quality in present-day clusters, due to the IXO-XMS spatially resolved high spectral resolution combined with its high throughput. This will make possible, for instance, to determine temperatures from line ratios alone (e.g. K-alpha over K-beta line ratios), thereby improving the accuracy of the abundance determinations. A new window on contributions from different SN types will come from routinely measuring abundances of trace elements, such as chromium, manganese, and nickel (see Section 2.4.1). The high quality abundance maps together with temperature maps will help to identify locations of active enrichment, such as AGN driven outflows (Figure 2.16), or locations of recent galaxy stripping. IXO will also further extend abundance studies to fainter, less massive clusters and groups. This may reveal trends in abundances as a function of cluster mass, and determine whether early star formation was dependent on the initial overdensities prior to cluster/group formation (and help test whether groups and less massive clusters retained all metals produced in their galaxies). Another important contribution that IXO will make is the determination of abundances in the low surface brightness outskirts of clusters. The abundances there will be closer to the pristine abundances of the intergalactic gas as it was prior to cluster formation.

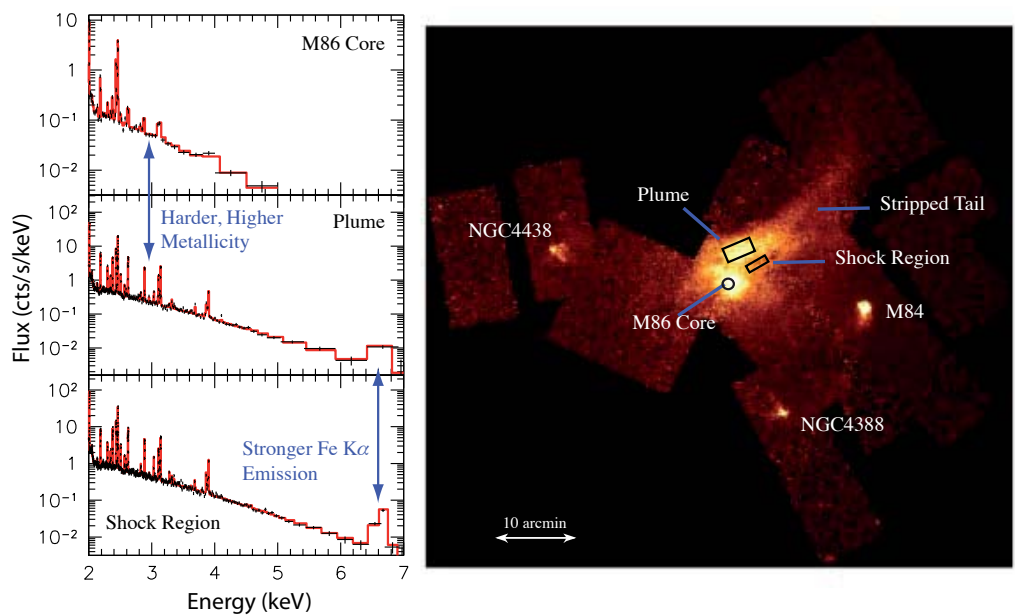

**Figure 2.16.** *IXO will measure position-dependent metal abundances of the intra-cluster medium, linking them to the possible chemical enrichment scenarios.* **Left panels:** *Extracted spectra from IXO-XMS simulations of three regions in M86, according to the model by Finoguenov et al. (2004). The exposure time is 50ks.* **Top panel:** *0.5' x 0.5' region around the core, with a two-temperature plasma ($kT_1 = 0.68$ keV, $kT_2 = 1.19$ keV), metallicity 0.76 solar:* **Middle panel:** *2' x 4' region from the "plume", with again a two-temperature plasma ($kT_1 = 0.79$ keV, $kT_2 = 1.57$ keV), metallicity 1.18 solar.* **Bottom panel:** *1' x 3' region just to the south of the plume, with $kT_1 = 0.83$ keV, $kT_2 = 1.37$ keV, metallicity 0.98 solar. The main differences between these spectra are a noticeable lack of emission lines between the core and plume regions. This is a result of the harder spectrum and higher metallicity in the plume region. Additionally, there is stronger Fe Kα emission in the region south of the plume than in the other regions.* **Right:** *X-ray emission from the Virgo elliptical galaxy M86 and its 380 kpc ram pressure stripped tail dominates this mosaic of Chandra images. Sensitive, high-resolution X-ray spectroscopy will measure the metallicity along the tail and in the surrounding cluster gas.*

Elemental abundances in clusters as a function of redshift, up to redshifts as high as $z\sim2$, when clusters began assembling much of their current mass, are also a key goal of IXO studies of chemical enrichment. Tentative results up to $z\sim1$ suggest that the Fe abundance is steadily decreasing with redshift (Balestra et al. 2007; Maughan et al. 2008), which could be the result of the ongoing contribution of the long delay channel Type Ia SNe. However, it could also mean that Fe and other elements were already inside the clusters, but confined to the galaxies, and were slowly released into the Intra-cluster Medium by ram pressure stripping (Kapferer et al. 2007), or by AGN driven outflows from member galaxies. In particular, giant central elliptical galaxies may have retained the metals produced by their SNe for a long time, due to their deep gravitational potentials. It will, therefore, also be important to measure the abundances of alpha-elements as function of redshift, as these will show the enrichment contribution of core collapse SNe. Since most core collapse SNe exploded before cluster formation, alpha-elements will make the distinction between continuing enrichment by Type Ia SNe, or release of previous enriched gas from cluster galaxies.

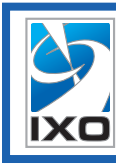 IXO high-resolution spectroscopy of supernova remnants, where chemical elements are produced, and of galaxy clusters, the largest gas reservoirs in the Universe, will reveal when and where chemical elements were created and eventually dispersed through cosmic time.

## 2.3    Matter under extreme conditions

### 2.3.1    Strong gravity and accretion physics

Black Holes of stellar mass, of intermediate mass (see Section 2.3.2), or supermassive ones as existing in galaxy centres (Section 2.1) create the strongest gravitational fields in the Universe. Matter trapped in the gravitational potential of such objects is spun and heated, emitting copious X-rays which betray the presence of the huge gravitational field. X-ray observations are uniquely suited to witness and reveal this hot component of the Universe, thereby revealing how the laws of General Relativity apply in the strong field limit and also how matter behaves under such intense gravity fields.

General relativity has been successfully tested in the weak field regime, but probes of General Relativity in the strong-field limit are extremely rare. X-ray observations of active galactic nuclei and stellar mass compact objects (Galactic Black Holes, GBHs) offer a unique opportunity for this, since they probe accretion and ejection flows in the regions close to the event horizon, where strong gravity effects such as gravitational redshift, light bending and frame dragging, shape the observed phenomena. Irradiation of the inner accretion disk by hard X-rays produces emission lines and a characteristic disk "reflection" spectrum. The most prominent line is typically Fe-K$\alpha$ at 6-7 keV, but Fe-L and lines from lower-Z elements are also seen below a few keV, albeit less frequently. The lines are accompanied by a "reflection" spectrum with a flux excess peaking at 20-30 keV, which is actually due to Compton backscattering. Relativistic Doppler shifts and gravitational redshifts endemic to the inner disk around black holes act to skew the shape of disk lines and the "reflection" spectrum. The shifts grow more extreme with increasing black hole spin, as the innermost stable circular orbit gets closer to the black hole (Figure 2.17), and hence modeling the line and reflection components can be used to measure black hole spin.

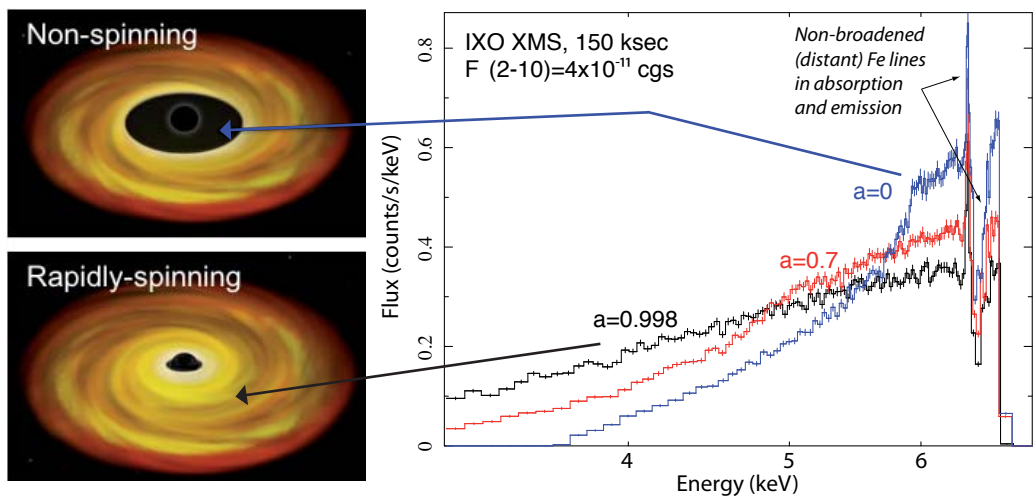

**Figure 2.17.** *Left: X-ray illumination of the inner accretion disk provides a unique probe of the strong gravity environment near black holes. If the black hole is spinning the disk extends closer to the event horizon, resulting in a much broader, redshifted line profile. Right: Simulated IXO-XMS spectrum of a bright AGN, showing various broad Fe emission line profiles for different SMBH spins (with an equivalent width of 330 eV), superposed on the narrow emission and absorption features (typically of < 10 eV) resulting from more distant material. This shows that the XMS will be able to separate the narrow features from those produced by strong gravity.*

IXO's focal plane instrumentation will in fact enable the measurement of black hole spins in many bright GBHs and AGN using at least five independent techniques (which are described in Section 2.3.1.2). Such methods entail a certain amount of generic knowledge of the physics and geometry of the inner accretion flow around black holes, but they do not need a fully detailed understanding of those. Nonetheless, high spectral resolution X-ray observations will give us a fully detailed understanding. The five independent methods will be used to "calibrate" black hole spin measurements with Fe-Kα lines for application to much larger samples of faint AGN, to constrain their merger and accretion histories, providing an important link with the co-evolution of black holes and their host galaxies (see Section 2.1.1). The high-throughput of IXO will provide sufficient effective area to track the orbits of test particles near the event horizon, via narrow Fe-Kα fluorescent emission modulated on the orbital timescale. In addition, the outflows and radiation thought to drive AGN feedback in evolving galaxies ultimately have their origin in physical processes close to the central black hole. It is only by probing this region directly with IXO that can we hope to understand the remarkable link between these processes and galaxy evolution on scales more than 10 orders of magnitude larger.

### 2.3.1.1   Probing General Relativity in the strong field regime

In luminous black hole systems, the accretion flow is most probably in the form of a thin disk of gas orbiting around the black hole. To a very good approximation, each parcel of gas within the disk follows a circular test-particle orbit. This geometrical and dynamical simplicity makes accretion disks useful for probing the black hole potential and, hence, the predictions of General Relativity. In a small number of AGN, current observations already hint at the power of orbit-by-orbit traces using emission lines (Figure 2.18). With its superior throughput, IXO will enable the detection of iron line variability on sub-orbital timescales (from 100s to 1000s seconds) in approximately 20-30 AGN. Any non-axisymmetry in the iron line profile (e.g. associated with the expected turbulence in the disk) will lead to a characteristic variability of the iron line, with "arcs" being traced out on the time-energy plane[*] (Figure 2.19).

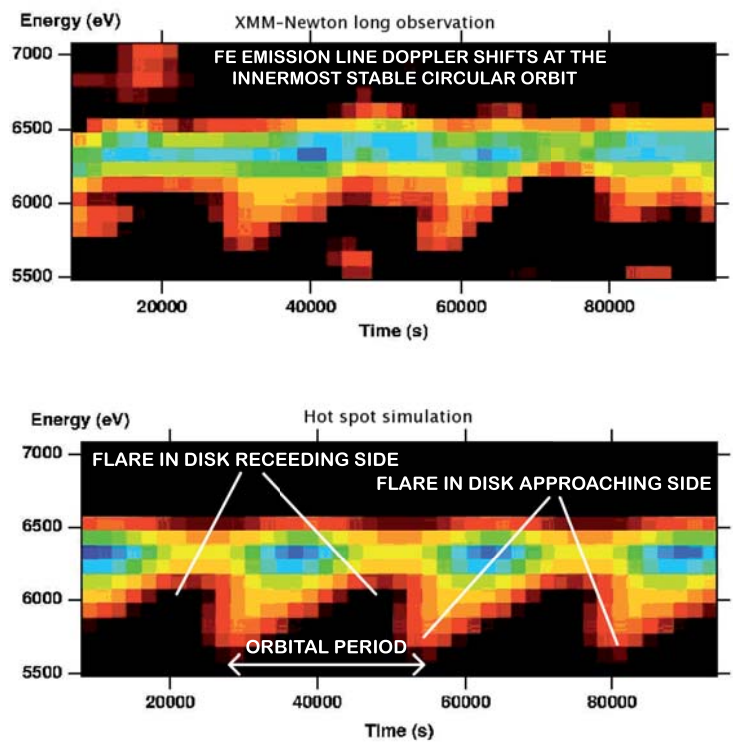

**Figure 2.18.** *Current observations have barely revealed emission line variations on the orbital timescale at the innermost stable circular orbit (DeMarco et al. 2009).* **Top panel**: *The figure depicts XMM-Newton measurements of Doppler shifts in an iron emission line consistent with Keplerian orbits at the innermost stable circular orbit in the Seyfert AGN NGC 3516 (Iwasawa, Miniutti, & Fabian 2004). The saw-tooth pattern is exactly that expected* (**bottom panel**) *for orbital motion in the innermost relativistic regime around black holes. IXO will be the first observatory with sufficient effective area to measure these motions in dozens of AGN covering more than 2 decades in black hole mass.*

---

[*]   *A video of an MHD simulation of a turbulent disk, and the time resolved iron lines that it would produce, can be seen at:*
       *http://ixo.gsfc.nasa.gov/documents/resources/posters/aas2009/brenneman_plunge_30i.avi*

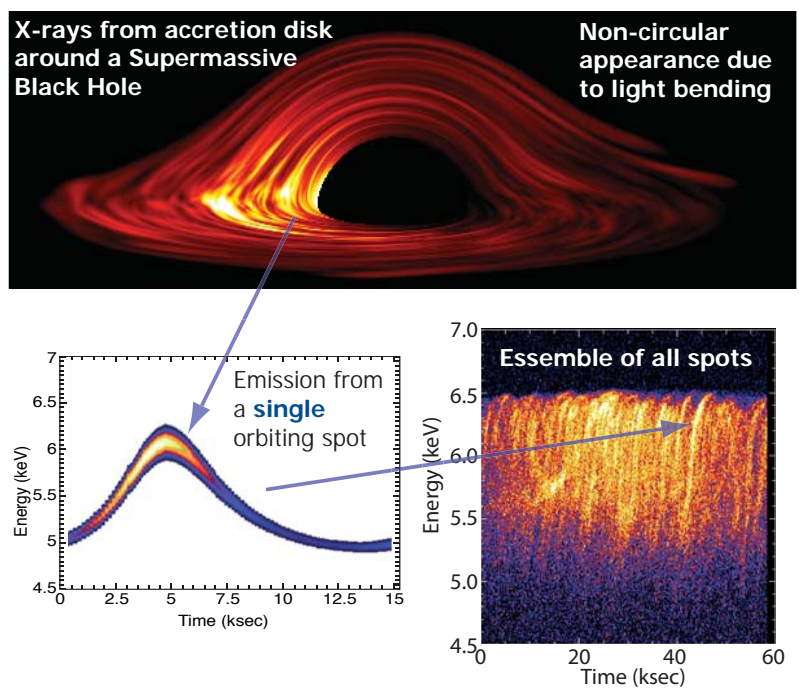

**Figure 2.19.** ***Top panel***: *IXO will resolve multiple hot spots in energy and time as they orbit the SMBH, each of which traces the Kerr metric at a particular radius.* ***Bottom left***: *In the time-energy plane, the emission from these hot spots appears as "arcs", each corresponding to an orbit of a given bright region. These arcs can be directly mapped to test particle like orbital motion of hot spots in the disk.* ***Bottom right:*** *Simulated IXO observation, assuming a $3 \times 10^7 M_{\odot}$ black hole and a flux characteristic of a bright AGN (adapted from Armitage & Reynolds 2003).*

General Relativity makes specific predictions for the form of these arcs, and the ensemble of arcs can be fitted for the mass and spin of the black hole, and the inclination at which the accretion disk is being viewed. A second kind of emission line variability will occur due to the reverberation (or "light echo") of X-ray flares across the accretion disk (see Section 2.3.1.2). Reverberation observations offer unambiguous proof of the origin of the X-ray lines as reflection features, allowing us to map the geometry of the X-ray source and inner accretion flow, and determine the mass of supermassive black holes. The path and travel time of photons close to the black hole is also strongly affected by space-time curvature and frame-dragging. In systems with very rapidly rotating black holes, the region of the accretion disk capable of producing line emission can extend down almost to the event horizon, so we can probe time-delays along photon paths that pass close to the horizon. These photon paths create a low-energy, time-delayed "tail" in the General Relativity reverberation transfer function. The nature of this tail is insensitive to the location of the X-ray source but is highly sensitive to the spacetime metric. Their characterisation will thus provide another probe of General Relativity, this time based on photon orbits rather than matter orbits (Cunningham & Bardeen 1973; Young & Reynolds 2000). The reverberation of individual flares will be accessible to IXO in the brightest few AGN. However, reverberation will be detected in many more AGN and GBHs/NSs via the use of Fourier techniques aimed at detecting the lag between the driving continuum emission and the strongest fluorescent emission lines.

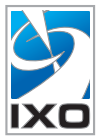

IXO will probe a number of predictions of General Relativity in the strong field limit, in particular those introduced by gravitational redshift, space-time curvature and frame-dragging. Time-resolved spectroscopy, enabled by the IXO throughput, will probe both General Relativity in the strong-field regime and the geometry of the immediate environment around the black hole.

## 2.3.1.2   Measuring Black Hole Spin

Despite their extreme nature, only two parameters describe black holes in an astrophysical context: mass and spin. Ever since the seminal work of Penrose (1969) and Blandford & Znajek (1977), it has been realized that black hole spin may be an important energy source in astrophysics, potentially allowing the extraction of energy and angular momentum from the black holes themselves. The spin may also impact on the apparent radio-loud radio-quiet 'dichotomy' in AGN, in the powering of jets and/or winds in AGN and GBHs (Rees et al. 1982), in the formation of gamma-ray burst engines, in producing gravitational waves from merging BHs, and may tell us about the final stage of stellar evolution (supernovae) and on larger scales on galaxy merger and accretion histories (see Section 2.1.1). Despite its importance, however, we are only now gaining our first tantalizing glimpses of black hole spin in a few objects. Unlike measurements of black hole mass, spin measurements require observables that originate within a few gravitational radii of the black hole. The observational signatures needed to determine spin are found in the X-ray band, and IXO will be able to measure spins in stellar mass and supermassive black holes with at least five independent techniques.

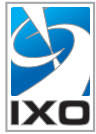 *IXO will measure the spin of stellar-mass and supermassive black holes in 5 independent ways: Fe line spectroscopic profiles, reverberation, accretion disk spectral continuum fitting, quasi-periodic oscillations and polarimetry. Although these methods entail some amount of modeling, they can be cross-calibrated and used to strengthen the reliability of spin measurements.*

### Time-averaged fitting of Iron disk lines

Relativistic disk lines are common in Seyfert AGN, in stellar-mass black holes, and even in the spectra of accreting neutron stars (e.g. Miller 2007) and, as discussed earlier, the profiles of lines can be used to measure the spin of the black hole (Figure 2.20, left). Because the lineshifts scale with gravitational radii ($GM/c^2$), the mass of a given black hole and its distance are not needed to measure its spin using this technique. The effects of spin on the disk reflection spectrum are not subtle, but the disk spectrum must be decomposed from other complexity such as continuum curvature or the effects of photoionized absorbers. For this reason, current studies (with *XMM-Newton* and *Suzaku*) have been limited to a handful of objects. The unprecedented sensitivity and energy resolution of IXO will overcome these uncertainties and allow us to measure the spin in hundreds of AGN, GBHs and neutron stars, going beyond the nearest Seyfert 1 galaxies and the brightest GBHs.

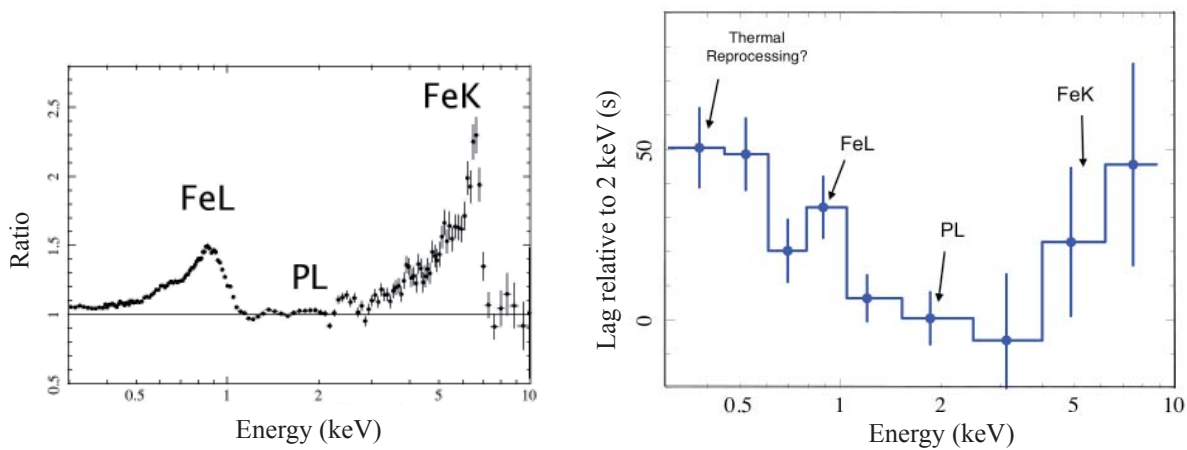

**Figure 2.20.** *Left: Reflection spectrum with broad and ionized iron L and K lines of the NLSy1 1H0707-495, seen in a very long (4 orbits) observation of XMM-Newton. Right: Time lag relative to the power-law continuum as a function of energy calculated for the same data shown left. This plot shows suggestive features corresponding to the reverberation of the reflection components, on a timescale corresponding to a few times the light-crossing time of a gravitational radius, which is expected to be 5s for a $10^6 M_\odot$ black hole.*

## Reverberation

As mentioned in Section 2.3.1.1, measuring the light travel time between flux variations in the hard X-ray continuum and the lines that it excites in the accretion disk provides a model-independent way to measure black hole spin. The time delay simply translates into distance for a given geometry. If a black hole has a low spin parameter, iron emission lines in the 6.4-6.97 keV range should have a characteristic lag of approximately 6 GM/c³; if the black hole is rapidly spinning, lags can be as short as 1 GM/c³. Clear evidence for such a reverberation lag of 30s between the direct X-ray continuum and the FeL emission accompanying the relativistic reflection, has only recently been found (Fabian et al. 2009) in the NLSy1 1H0707-495, thanks to a week-long observation with *XMM-Newton* (Figure 2.20, right). This is exciting also because it demonstrates the possibility to measure spin and reverberation using the bright soft X-ray band where the effective area of IXO is at its maximum. Close to spinning black holes, the path light takes will be strongly impacted by spacetime curvature. When very close to the black hole, an otherwise isotropic source of hard X-ray emission will have its flux bent downward onto the disk. An observable consequence of these light-bending effects is a particular non-linear relationship between hard X-ray emission and iron emission lines (Miniutti & Fabian 2004). At present, there is tantalizing evidence for this effect in some Seyfert AGN (e.g. Ponti et al. 2006) and stellar-mass black holes (e.g. Rossi et al. 2005). The IXO-HTRS, (for galactic black holes and neutron stars, see Figure 2.27, left) and the XMS/WFI (for AGN) have the effective area, time resolution, broad energy range, energy resolution, and flux tolerance needed to make careful studies of lags that can lead to spin measurements and clear detections of gravitational light bending.

## Accretion disk continuum fitting

Thermal continuum emission from the accretion disk may be used to measure the spin of stellar-mass black holes. An accretion disk around a spinning black hole is expected to be hotter and more luminous than a disk around a black hole with low spin, because the innermost stable circular orbit is deeper within the gravitational well. New spectral models have recently been developed that exploit the corresponding changes in the shape of the continuum to measure spin. If the mass and distance to a black hole are known, these models may be applied to spectra in order to measure the spin of a stellar-mass black hole (see, e.g., McClintock et al. 2006). By virtue of its flux tolerance, the HTRS is the best suited instrument for measuring black hole spins using the disk continuum. As shown in Figure 2.21, the simulated statistical uncertainty on the black hole spin measured through continuum fitting is about an order of magnitude smaller than the one derived by fitting the relativistic iron line profile, but both measurements will be available simultaneously with IXO. Stellar mass black holes are bright enough in outburst that both methods can be used, which has recently been done and yields consistent results (Steiner et al. 2010).

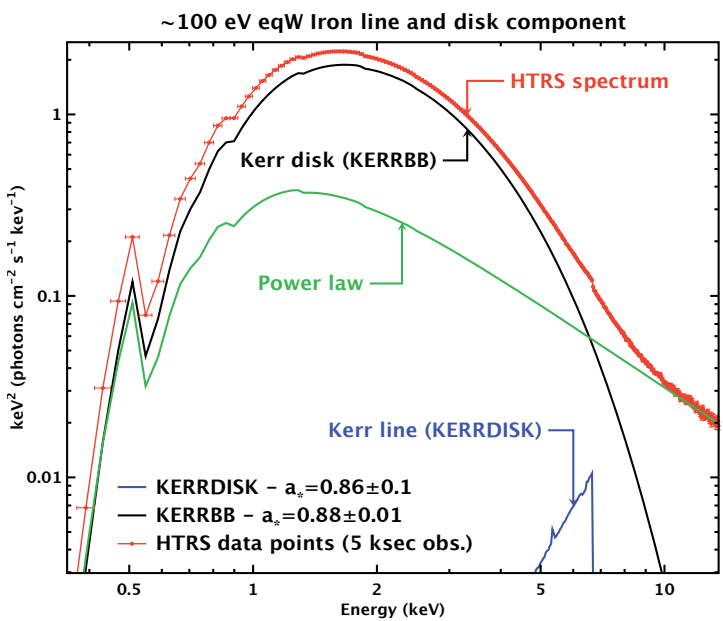

**Figure 2.21.** *A typical high state spectrum of a GBH (e.g. GX339-4) as observed by the IXO-HTRS. Both a disk component and a weak iron line are present in the spectrum, together with a power law (all spectra are absorbed by photoelectric absorption from the interstellar medium along the line of sight). Simultaneous fitting of the disk component and line should yield similar values for the black hole spins. The statistical accuracy on the spin measurement from the disk component is about ten times smaller than for the relativistic iron line. Only the statistical error is given.*

## Quasi-periodic oscillations

The X-ray flux from accreting black holes and neutron stars is sometimes modulated at frequencies commensurate with Keplerian frequencies close to the compact object. The oscillations are not pure but quasi-periodic, due to small variations in frequency and phase as expected for gas orbiting in a real fluid disk with internal viscosity. X-ray quasi-periodic oscillations probe the strongly curved spacetime around compact objects and constrain their mass, spin and radius. They therefore hold the potential to test some clear predictions of General Relativity, such as the existence of an innermost stable circular orbit. Most current models associate the quasi-periodic oscillations with the general relativistic regime. When applied to the existing observations such models appear to point to maximally spinning black holes (Figure 2.22).

IXO will make a large leap in sensitivity, opening the way to detection of strong quasi-periodic oscillations on timescales closer to the coherence time of the underlying oscillator, and to detect the weakest features predicted in models (harmonics, sidebands), allowing us to identify the origin of the peaks in the power spectrum. Moreover IXO will reveal quasi-periodic oscillations in a much wider variety of objects, such as ultra-luminous X-ray sources (possibly harbouring intermediate mass black holes), and AGN, for which a claim of a quasi-periodic oscillation detection was recently reported (Gierliński et al. 2008). In the IXO timeframe, theoretical understanding of accretion disk physics as well as global disk simulations will advance (e.g. quasi-periodic oscillations are now beginning to appear in 3-dimensional magnetohydrodynamic simulations, e.g. Romanova & Kulkarni, 2009). This will provide the necessary framework for exploiting the potential of quasi-periodic oscillations for probing General Relativity, compact object parameters and accretion disk physics.

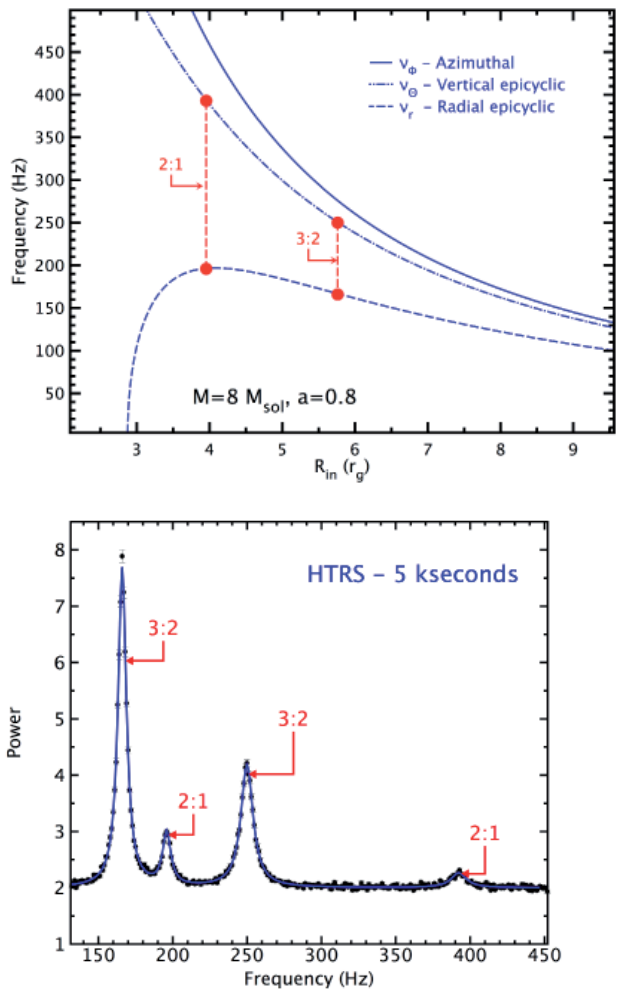

**Figure 2.22. *Top left:*** *The three General Relativity frequencies around an 8 $M_\odot$ black hole with a dimensionless spin parameter of 0.8. There exist two radii in the disk where the ratio of the vertical and radial epicyclic frequencies are equal to 3:2 and 2:1. In a particular class of models, resonance should be excited at these particular radii.* ***Bottom left:*** *The power density spectrum simulated for an HTRS observation of 5 ks (0.8 Crab source), in which the 4 quasi-periodic oscillations are simultaneously detected at a significance level larger than 6σ. As of today, only the quasi-periodic oscillations corresponding to the 3:2 ratio have been reported from RXTE. If the mass of the black hole is known (from the dynamical mass function estimate of the binary system), solving the General Relativity frequency equations for the black hole spin should yield an estimate with an error of about 10%.* ***Bottom right:*** *Azimuthal, radial and vertical motion does not occur at the same frequency in general relativity, leading to epicyclic orbits.*

## Polarimetry

Polarization properties are significantly altered when photons travel on null geodesics in a strong gravity field, as in the vicinity of a black hole or a neutron star (Connors & Stark 1977). In practice, this results in a rotation of the polarization angle as seen by a distant observer. The amount of rotation depends on the geodesic parameters, and increases with decreasing impact parameter with respect to the compact object. This effect can be used to measure the spin of the black hole in both GBHs and AGN (Li et al. 2008; Schnittman & Krolik 2009). For example, in GBH the emission from the accretion disk is hotter and closer to the black hole. As a result, a variation of the polarization angle, as well as of the polarization degree with energy is expected (Figure 2.23). This rotation depends on the spin via the spin-dependence of the innermost stable circular orbit. In AGN, the disc thermal emission is much softer, outside the X-ray band. However, strong gravity effects may manifest themselves through time-dependent, rather than energy-dependent, rotation of the polarization angle of the Compton reflection component (Dovciak et al. 2004), which again is spin-dependent. In the light-bending model (Miniutti & Fabian 2004), a relation with the flux of the source is also expected (Dovciak et al. 2004).

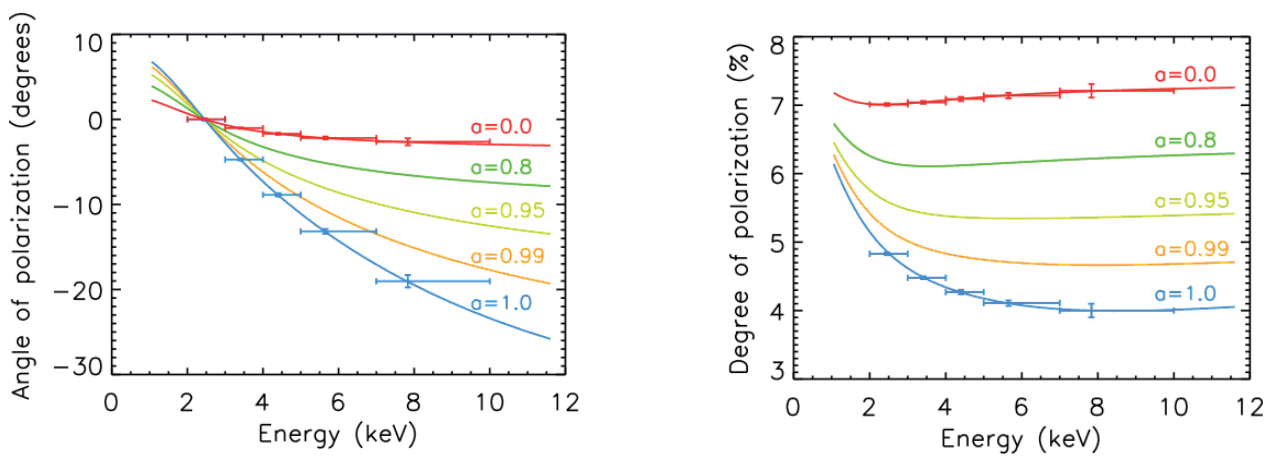

**Figure 2.23.** *100 ks IXO simulations showing the energy-resolved polarization angle (**left**) and degree (**right**) expected for GRS 1915+105, for a variety of values of the spin parameter a . These demonstrate that energy-resolved polarimetry allows to estimate the spin of galactic black holes and provides an independent tool to be used in complement to spectroscopy and timing.*

### 2.3.1.3   Gas dynamics under strong gravity effects

Accretion disks and massive, high velocity outflows are known to be forming in the innermost regions of AGN and GBHs, but we do not yet understand the physics behind the flow patterns and energy generation around black holes. These issues are intertwined with those related to General Relativity and only a significant step forward in observational capability will enable the important physical effects to be disentangled. To perform physical tests of accretion and ejection mechanisms requires the ability to obtain X-ray spectra at or below the relevant orbital timescales. This, in turn, requires a combination of large collecting area, high spectral and timing resolution, and drives several of the IXO mission requirements.

### X-ray "tomography" of disk inner regions

Time-resolved spectroscopy of several bright AGN has revealed some clear cases of eclipses of the X-ray source lasting a few hours. They are thought to be due to obscuring clouds with column densities of $10^{23}$-$10^{24}$ cm$^{-2}$ crossing the line of sight with velocities in excess of $10^3$ km/s. This opens up the possibility for a new, unique experiment of "tomography" of the X-ray source: while passing, the cloud covers different parts of the X-ray source, revealing its structure (geometry, emissivity, and associated relativistic effects). If an occultation by a Compton-thick cloud is observed in a source with a broad relativistic line, the line profile is expected to change during the eclipse, as we would probe the approaching and receding parts of the line emitting region separately. Such an experiment might just be within current reach for one object (NGC 1365; Risaliti et al. 2009, see Figure 2.24). Conservative estimates predict that these measurements will be relatively straightforward with IXO for several sources with great accuracy, providing a unique opportunity to probe the relativistic effects due to strong gravity and fast orbital motion in the innermost regions of AGN.

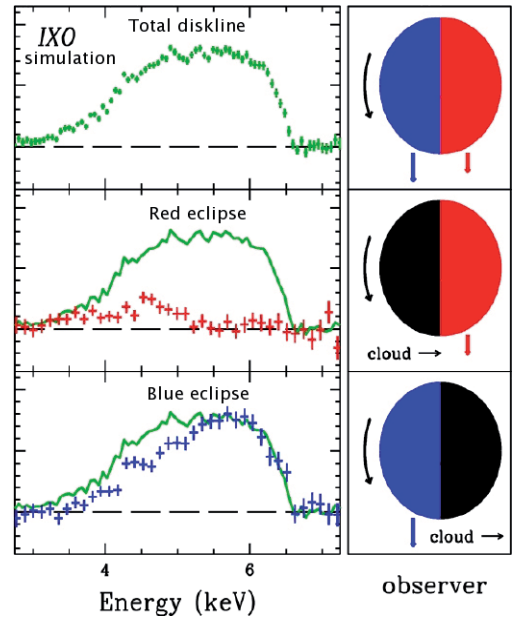

**Figure 2.24.** *X-ray tomography of the accretion disk around the black hole in NGC1365. IXO simulation of the iron line profiles are shown at different phases of the disc occultation due to an orbiting cloud. It is assumed that NGC 1365 is observed with IXO for ~200 ks, and that during this time a total eclipse occurs for 60 ks, in agreement with what was actually observed with XMM and Suzaku. From top to bottom, green is the disk line profile obtained from the non-eclipsed part of the observation, while the red and blue profiles correspond to the "left" and "right" halves of the X-ray source. IXO will allow a detailed time-resolved analysis of such eclipses, providing a direct proof of the GR effects in the vicinity of the central black hole.*

## Outflows and the Physics of Feedback

AGN outflows can release a huge amount of mechanical energy from the innermost region nearest the black hole into the surrounding host galaxy environment. Indeed such outflows may be the key to understanding the physical origin of the feedback process discussed earlier (Section 2.1.3), which can regulate the relative growth of black holes (see Section 2.1.1) and their host galaxies and heat and shape the intracluster medium (see Section 2.2.3). *XMM-Newton* and *Chandra* spectroscopy of nearby AGN and quasars have shown extraordinary details below 2 keV of the so-called "warm absorbers" (Crenshaw, Kraemer & George 2003). These have been interpreted in terms of multi-temperature, stratified, photo-ionized outflowing gas. In addition, unexpected absorption lines in the ~7-10 keV energy range have also been detected in about a third of all radio-quiet AGN where good enough spectra have been obtained (see, e.g., Chartas et al. 2002). These lines have been commonly interpreted as due to resonant absorption from Fe XXV and/or Fe XXVI associated with a zone of circumnuclear gas photo-ionized by the central X-ray source, with ionization parameter log $\xi$~2–5 (with $\xi$ in erg s$^{-1}$ cm) and column density $N_H$~$10^{22}$–$10^{24}$ cm$^{-2}$. These extreme values and their typical velocities of ~0.1c (with values up to ~0.3c), indicate the existence of previously unknown ultra-fast outflow in AGN (Pounds et al. 2003; Tombesi et al. 2010).

The mass outflow rates appear to be comparable to the accretion rate and their kinetic energy represents a significant fraction of the bolometric luminosity (Reeves et al. 2004). Therefore, they are in principle able to profoundly influence their host galaxies (as discussed in Section 2.1.3), providing the feedback linking the evolution of AGN and galaxies. At present it is possible to detect the ultra-fast outflows in relatively low *z* (i.e. *z*<0.1) bright AGN (with the exception of a few gravitationally lensed broad absorption line quasi-stellar objects), but with the superior throughput and resolution of IXO, it will be possible to extend the study of these outflows out to high redshift, where the bulk of the galaxy evolution occurs (Section 2.1.1). Higher energy resolution will resolve the line profiles and follow the line variations in time, thereby constraining definitively the outflow properties (ionization, density, velocity, turbulence, abundance) and geometry (Figure 2.25).

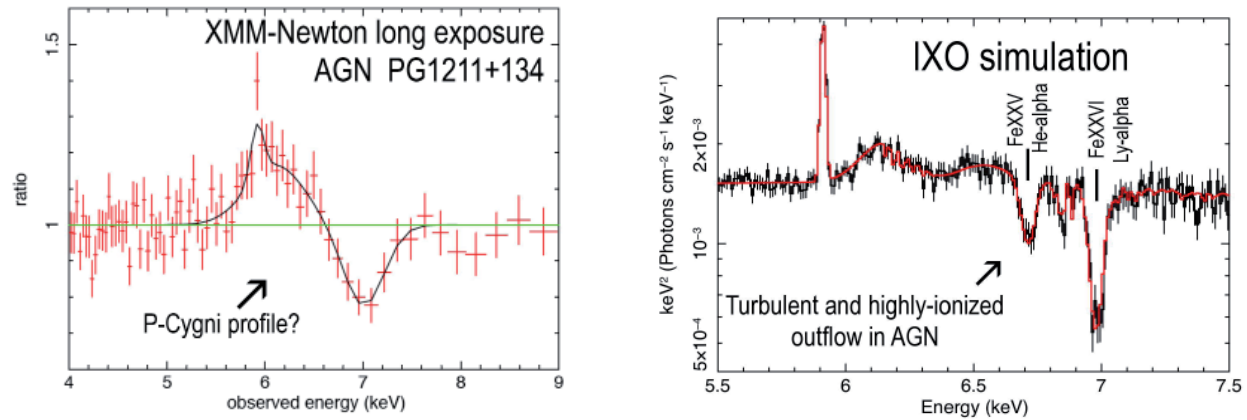

**Figure 2.25.** *Left*: *XMM-Newton spectrum of quasar PG1211+143 (z=0.08) indicating the presence of a P-Cygni profile in the Fe K line, a signature of a fast AGN outflow. **Right**: IXO 100 ks simulation expected from the model measured with XMM-Newton (left panel), i.e. $N_h=10^{23}cm^{-2}$, $\log \xi=4$ and $v_{out}=0$. A turbulence of 3000 km/s is assumed. The absorption lines are clearly resolved into their multiple species, and their energy width is also resolved, unlike with XMM. A narrow 6.4 keV (rest-frame) emission line is included in the model.*

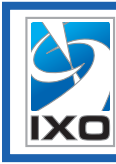

Thanks to its large throughput and spectral resolution, IXO will measure the energy released by AGN through massive outflows out to high redshift where the bulk of galaxy formation occurs.

### 2.3.2    Neutron star equation of state

#### 2.3.2.1    Probing Quantum Chromodynamics through the neutron star equation of state

Seven decades after the first speculation on the existence of gravitationally bound neutron stars, we still know very little about their fundamental properties. Initial modelling attempts were based on the assumption that matter in these objects can be adequately described as a degenerate gas of free neutrons, but it has become progressively clear that the cores of neutron stars must in fact be the stage for intricate and complex collective behaviour of the constituent particles. Over most of the range of the density/temperature phase plane, Quantum Chromodynamics is believed to correctly describe the fundamental behaviour of matter, from the subnuclear scale up. The ultimate constituents of matter are quarks, which are ordinarily bound in various combinations by an interaction mediated by gluons to form composite particles. At very high energies, a phase transition to a plasma of free quarks and gluons should occur, and various experiments are currently probing this low-density, high temperature limit of Quantum Chromodynamics. Likewise, the Quantum Chromodynamics of bound states is beginning to be quantitatively understood; recently, the first correct calculation of the mass of the proton was announced.

The opposite limit of high densities and low (near zero, compared to the neutron Fermi energy) temperature Quantum Chromodynamics has been predicted to exhibit very rich behavior. At densities exceeding a few times the density in atomic nuclei ($\rho \sim 3\times10^{14}$ g cm$^{-3}$ ), exotic excitations such as hyperons, or Bose condensates of pions or kaons may appear. It has also been suggested that at very high densities a phase transition to strange quark matter may occur. When and how such transitions occur is of course determined by the correlations between the particles, and the ultra-high-density behavior of matter is governed by many-body effects. This makes the direct calculation of the properties of matter under these conditions extremely difficult. The only possible way to probe the high density, low temperature limit of Quantum Chromodynamics is by observations and measurements of the densest material objects in nature, neutron stars. As in the case of the measurement of black hole spins, IXO will provide multiple, complementary and/or redundant measurements of the neutron-star mass/radius (*M-R*) relation. Crucially, this relation will be explored over the wide range of masses required to distinguish between the various competing models, and in a wide range of environments and conditions (e.g. isolated stars, accreting binaries, quiescent cooling systems, Eddington-limited X-ray bursters).

The relation between pressure and density, the equation of state, is the simplest way to parameterize the bulk behaviour of matter, and governs the mechanical equilibrium structure of bound stars. Conversely, measurements of quantities like the mass and the radius, or the mass and the moment of inertia, probe the equation of state. Figure 2.26 shows the *M-R* plane for neutron stars, with a number of illustrative relations based on various assumptions concerning the equation of state (Lattimer and Prakash 2007). Neutron stars have been the subject of intensive radio observations for forty years, and precise radio pulse arrival time measurements on double neutron star binaries have produced a series of exquisite mass determinations, with a weighted average stellar mass of $M_{NS} = 1.413 \pm 0.028\ M_\odot$. Such a mass can be accommodated by virtually all equations of state and it is obvious that definitive constraints can only be derived from simultaneous measurement of masses and radii of individual neutron stars. Looking at Figure 2.26, it is also clear that one needs to probe higher neutron star masses where the differences in the equation of state predictions are more striking. Neutron stars in mass-transferring binaries will give us access to a wider range of neutron star masses (of order a solar mass of material can be transferred over the lifetime of a low-mass X-ray binary), therefore addressing this fundamental issue. As a reference, effective discrimination between different families of hadronic equations of state will require a relative precision of order 10% in mass and radius, and similar requirements apply to the strange matter equation of state (Figure 2.26).

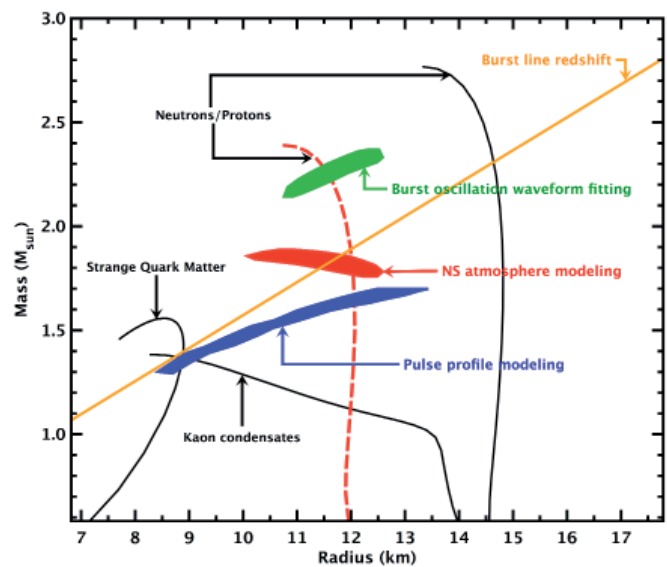

**Figure 2.26.** *The mass-radius relationship for neutron stars reflects the equation of state for cold superdensed matter. Trajectories for typical equations of state are shown as black curves, for standard nucleonic matter, self- bound quark stars and Kaon condensates. For illustrative purposes, the constraints derived from IXO are from i) pulse profile fitting of X-ray burst oscillations (green filled ellipse, IXO-HTRS, adapted from Strohmayer 2004), ii) waveform fitting of pulsations obtained from a 2 hr observation of an accreting millisecond pulsar (blue filled ellipse, IXO-HTRS, Courtesy of Juri Poutanen), iii) hydrogen atmosphere model fitting of the X-ray spectrum of a quiescent NS in the globular cluster Omega Cen (red filled elipse, 99%, IXO-XMS & IXO-WFI, 100 kseconds) and the detection of a gravitational redshift (z=0.35) from burst spectroscopy (orange line, IXO- XMS & IXO-HTRS). In this plot, only one M-R relation would match the above constraints, leading to the conclusion that matter at supra-nuclear densities is made of standard nucleonic matter.*

## 2.3.2.2   Measuring Mass and Radius of neutron stars

The X-rays generated around neutron stars will be used to constrain their masses and radii, using multiple and complementary diagnostics for a wide range of objects. These diagnostics, which require both high and medium spectral resolution and high count rate capabilities, drive in part the instrumentation required IXO. Here we describe the *M-R* diagnostics, bearing in mind that in many cases several will be provided for the same object:

***Waveform fitting of X-ray burst oscillations.*** Observations with *RXTE* show brightness oscillations at around 300 to 600 Hz during thermonuclear X-ray bursts (see Strohmayer et al. 1998 for a review). These oscillations are likely caused by rotational modulation of a hot spot on the stellar surface and can therefore be used to infer the neutron star spin frequency. The emission from the hot spot is affected by Doppler boosting, relativistic aberration and gravitational light bending. By fitting the waveform, it will be possible to investigate the spacetime around the neutron star, and simultaneously constrain its mass and radius. Thanks to its throughput and high count-rate capability, the IXO-HTRS will detect 10 to 20 times more photons than the *RXTE*-PCA for X-ray bursts up to 10 times the Crab intensity, allowing the study of individual oscillation cycles. Simulations indicate that stacking the folded light curves of a few X-ray bursts will lead to the determination of the neutron-star mass and radius with an accuracy of about 10-20% (e.g. Strohmayer 2003). Similar X-ray pulse profile modeling can also be performed for accreting millisecond pulsars (e.g. Poutanen 2008), taking advantage of the pulsations in the persistent emission, and the fact that these objects are found in binary systems, where the mass of the neutron star can be constrained from the dynamics of the binary. Using X-ray bursts, additional constraints may be obtained through continuum spectroscopy of those bursts showing photospheric radius expansion. Assuming that the touch down flux is equal to the Eddington flux and that the emission comes from the whole surface of the star, a constraint on the M-R plane can also be obtained, provided that both the atmospheric opacity and the color correction factor are known (Özel et al. 2010).

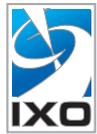

IXO will be the first observatory with enough throughput and time resolution to enable waveform fitting of X-ray burst oscillations with the statistical accuracy required to obtain meaningful constraints on the mass and radius of tens of neutron stars.

***Emission line modeling and variability:*** Just as AGN and galactic black hole systems, neutron-star binaries also exhibit emission lines (e.g. Fe Kα) broadened by Doppler and General Relativity effects, (e.g. Cackett et al. 2010). The IXO-HTRS will observe bright X-ray binaries without suffering pile-up up to several times the intensity of the Crab, providing energy resolution comparable to *XMM-Newton* and *Chandra*. Modeling the line profile yields the inner disk radius, which in turn sets an upper limit on the neutron star radius. Simultaneously, kilohertz quasi-periodic oscillations will be detectable down to amplitudes 3 to 5 times lower than currently achieved by the *RXTE*-PCA. Most models attribute the quasi-periodic oscillations to the orbital frequency at the inner disk radius (van der Klis 2006 for a review). A combination of the kilohertz quasi-periodic oscillation frequencies with radii provided by simultaneous measurements of the line yields the mass of the neutron star (Cackett et al. 2010). In addition, quasi-periodic oscillations are effective probes of the strongly curved spacetime around neutron stars, as they take place in the last tens of kilometers of the accretion flow. The evolution of the quasi-periodic oscillation parameters (coherence and amplitude) appear to be related to the existence of the innermost stable circular orbit, which is a key prediction of strong-field general relativity (Barret et al. 2008). The IXO-HTRS will detect quasi-periodic oscillations on their coherence timescales, which will provide important information about the properties of spacetime near the central object. The cycle waveform, which can be reconstructed statistically, depends on the Doppler shifts associated with the local linear velocity of the radiating matter in the emitting region, convolved with curved-spacetime propagation effects. If the frequency of the orbit is known, modeling of the waveform yields the mass of the compact object and the radius of the orbit (hence an upper limit on the neutron star radius), even if the quasi-periodic oscillation is not produced at a special radius such as the innermost stable circular orbit.

***Quasi-periodic oscillations and continuum time lags:*** The IXO-HTRS will provide frequency resolved spectra on the time scales of the fastest variability. High-time resolution spectroscopy of the associated quasi-periodic oscillations will yield absolute sizes of the disk, if light-travel time effects can be resolved via continuum reverberation. A time-delay spectrum within the frequency range of the variability (profiting from the fact that the HTRS will observe below 2 keV where the disk reprocessed emission dominates), immediately provides the physical size of the inner rim of the accretion disk, and hence an upper limit to the neutron-star radius (see Figure 2.27, left).

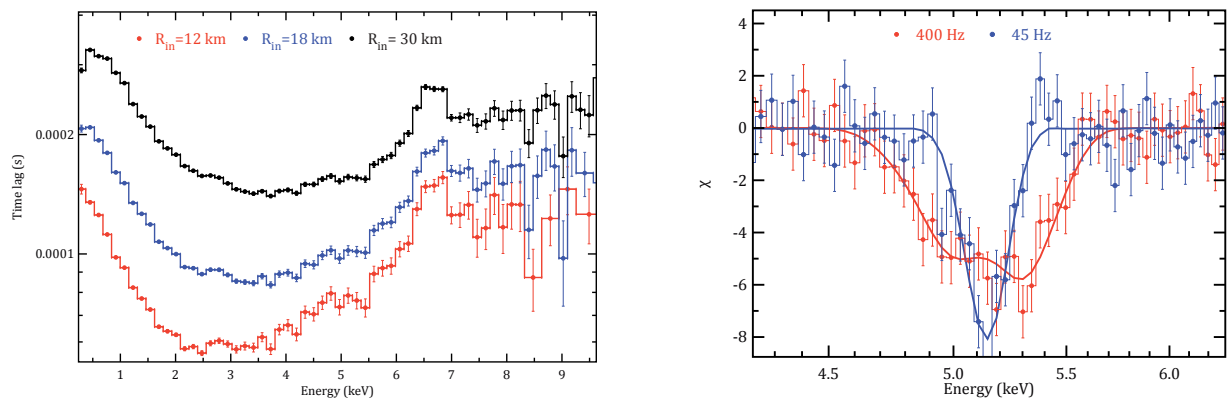

**Figure 2.27.** *Left: The lag vs energy for different inner disc radii for a NS KHz QPO observed with HTRS (the closer the disk inner rim the shorter the time delay). The frequency band chosen is 850−950 Hz, typical for the kHz QPO in NS. These bands were selected to optimise the signal detection, while ensuring that the source is observed at high-enough frequencies that light-travel time effects likely dominate the lags. (Figure courtesy of Phil Uttley). Right: Absorption-line spectrum (Fe XXVI Ly α) of a 1.4 $M_\odot$ and 11.5 km neutron star observed with the HTRS during a moderately bright (1 Crab) X-ray burst. The line profile includes the effects of light bending, rotational Doppler splitting, Doppler boosting and gravitational redshift. The figure shows the fit residuals (in units of sigma) with respect to the underlying continuum spectrum (excluding the line). The red curve is for a 60 s exposure and NS spin of 45 Hz, while the blue curve is for a 120 s exposure and a NS spin of 400 Hz. For each case, the black curves show the theoretical line profiles plotted at the resolution of the HTRS. The HTRS will be able to observe bursts up to rates of ~10 Crab. The XMS can be used for weaker bursts (up to ~0.3 Crab).*

***Gravitational shifts during X-ray bursts:*** Weak (~10 eV equivalent width) absorption lines (e.g. Fe Lyα and Heα) have been predicted in type I X-ray bursts (Chang et al. 2005). The profile of the line is distorted by magnetic (Zeeman or Paschen-Back) splitting by the star's magnetic field, longitudinal and transverse Doppler shifts, special relativistic beaming, gravitational redshifts, light bending, and frame dragging. Spectroscopic observations of X-ray bursts with the high spectral resolution of IXO-XMS of slowly rotating neutron stars and with the moderate spectral resolution of HTRS for fast rotating neutron stars (e.g. Barret et al. 2007) will enable us to search for and detect weak absorption features, if present (see Figure 2.27, right). As discussed in Bhattacharyya (2010), despite the surface lines being broad and asymmetric, a redshift (M/R) can still be measured down to a few % accuracy. In addition to providing the redshift, if the broadening is mostly rotational, measurements of the line can also constrain the stellar radius through the measured surface velocity (proportional to $\Omega R$). On the other hand, if the neutron star is slowly rotating, then Stark pressure broadening, proportional to $M/R^2$ is likely to dominate. This implies that from a single detection of a gravitational redshift, in either the rotational or pressure broadening limits, the two unknowns M and R can be determined uniquely.

***Continuum modeling:*** Another promising way of inferring neutron-star radii is through observations of quiescent X-ray emission from neutron stars for which the distance can be estimated reliably (e.g. those in globular clusters). Whether the energy reservoir is the heat deposited deep in the neutron star crust during the outburst phase of the transient, or emission is sustained by a low-level of radial accretion, the X-rays originate from the atmosphere of the neutron star. The resulting X-ray spectrum will therefore depend on the chemical composition of the neutron-star atmosphere, as well as on the strength and structure of the magnetic field. Early nonmagnetic hydrogen atmosphere models provide adequate fits to the quiescent X-ray spectra of several neutron star systems, yielding plausible values for the neutron-star radius (typically around 10 km). Models accounting for self consistent surface gravity and effective temperature, incorporating thermal electron conduction and self-irradiation by photons from the compact object, have been recently applied to the *XMM-Newton* data of the three best known quiescent neutron stars in globular clusters (e.g. Webb and Barret 2007). The gain in sensitivity offered by IXO will reduce the allowed *M-R* region dramatically compared to *XMM-Newton* (see Figure 2.26). IXO's angular resolution and sensitivity will extend such observations to a much larger sample of objects; population studies indicate that galactic globular clusters may contain up to one hundred quiescent neutron stars.

In summary, the proposed techniques will yield mass and radius estimates for a few tens of neutron stars in accreting X-ray binaries, and similar numbers of quiescent low-mass X-ray binaries and isolated neutron stars. The X-rays generated at the surface or in the immediate vicinity of the neutron star encode information about the neutron star mass and radius. For the first time, IXO will bring a collection of instruments with unique capabilities to enable a leap in spectral and timing sensitivity which is required to finally open up the window into Quantum Chromodynamics by determining the neutron star equation of state.

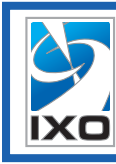 Thanks to its unique timing and spectral capabilities, IXO will enable multiple independent constraints to be obtained on the mass and radius of each neutron star observed, thus giving insights on the densest form of observable matter in the Universe.

### 2.3.2.3　Probing Quantum Electrodynamics in magnetized neutron stars

The objects previously described are weakly magnetized, but the extreme physics of strongly magnetized neutron stars will also be revealed via time resolved spectroscopy (HTRS) and polarimetry (XPOL). This is particularly the case for Soft Gamma-ray repeaters. These are X-ray sources pulsing at periods in the 5-8 sec range, showing rare active episodes (typically a few months) during which they emit many very powerful short bursts of hard X-rays. They are believed to be young neutron stars with extreme magnetic fields ($10^{14}$ G) and have been dubbed "magnetars". There are numerous predicted processes such as photon splitting that are only important in such strong fields, and which cannot be tested in terrestrial experiments.

Magnetic fields will be measured directly with X-ray spectroscopy through the detection and identification of cyclotron resonance scattering features, as well as through unique polarimetric signatures. In addition, the vacuum polarization in high magnetic-field neutron stars will be detectable, as it is expected to alter significantly the surface emission and induce unique polarization signatures in X-rays (Lai et al. 2010; van Adelsberg & Perna 2009). Finally, quasi-periodic oscillations in the X-ray emission following flares in soft gamma-ray repeaters, presumably due to torsional vibrations of the star crust, will constrain the mass and radius of those neutron stars.

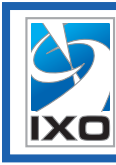 In highly magnetized neutron stars, IXO X-ray spectroscopy will measure the surface magnetic field via cyclotron resonance scattering. Vacuum polarization produced by such strong magnetic fields will also lead to polarimetric features that IXO can measure.

## 2.4　Life cycles of matter and energy in the Universe

Besides the overarching IXO science goals described in the previous sections and that drive the mission parameters, IXO will enable much relevant studies of additional questions in contemporary Astrophysics. A sample of such important issues, to which IXO will be key to make progress on, is presented in this section. Such questions and the foreseen IXO contributions are by no means minor, but none of them drives any of the IXO parameters. A summary of these questions was presented in the bottom section of Table 2.1.

### 2.4.1 Supernova remnants: formation of the elements, shock heating and particle acceleration

The most abundant massive elements in the Universe are released by, and in many cases formed in, supernova explosions. These explosions are also the most important source of energy for the interstellar medium, both in the form of kinetic/thermal energy and in the form of cosmic rays. IXO will allow us to study these explosions, their products and their immediate surroundings either by directly observing X-ray emission from extra-galactic SNe, or by studying supernova remnants (SNRs) in the Galaxy and the Local Group. Understanding SNe/SNRs is important for the rest of the IXO science programme, as SNe are of key importance for the chemical evolution of the Universe (see Section 2.2.5), and often mark the creation of a neutron star or a black hole. Moreover, the physics of collisionless shock waves and cosmic ray acceleration, which can be studied by IXO in nearby SNRs, are directly relevant to the large-scale shocks that formed the Warm-Hot Intergalactic Medium (Section 2.2.1) and heated clusters of galaxies (Section 2.2.3). Key questions that IXO will address regarding SNe & SNRs are:

- What is the chemical production of SNe of different types, and what is its intrinsic variation?
- What is the origin of thermonuclear SNe (Type Ia SNe)?
- How do core collapse SNe (Type Ib/Ic/II/IIb) explode?
- What is the physics of collisionless shocks, and how do they accelerate particles?

#### 2.4.1.1 Supernovae, explosion mechanisms, and nucleosynthesis products

For Type Ia SNe, IXO will observe both the shocked circumstellar medium and SN ejecta, and thus will probe both the SN explosion material and the surroundings as shaped by the progenitor system. A crucial element in Type Ia SNRs is iron. Unlike current instruments, IXO will be able to resolve the important Fe-L emission line complex around 1 keV. This will provide powerful plasma diagnostics, and will lead to accurate measurements of the temperature, velocity (through Doppler shifts/broadening) and abundance structure. The modelling of the SNR spectra has now advanced far enough to distinguish between energetic, normal and sub-energetic Type Ia explosions, based on the iron content and energy content of the SNRs (Badenes et al. 2008a). However, current and future theoretical models stress the importance of the 3D structure of Type Ia SNe (Kasen et al. 2009), which only IXO-XMS can obtain using Doppler mapping.

For core collapse SNe the explosion process itself is not understood. Most of the energy released by the collapse of the stellar core will be in the form of neutrinos. Part of the neutrino flux may drive the explosion, but rotation, magnetic fields, and acoustic instabilities may also play a role in the explosion process. This results in significant deviations from spherical symmetry. Young core collapse SNRs have X-ray spectra dominated by O, Ne, and Mg. The masses of these elements relate directly to the main sequence mass of the progenitor. These O/Ne/Mg rich SNRs show pure metal ejecta of more massive elements (Ar, Ca, Fe) that are distributed in an irregular pattern all over the SNR, and sometimes very close to the shock front. This suggests high velocities for complete and incomplete silicon burning products, which are synthesized deep inside the star, and whose velocities are not reproduced by current SN simulations (e.g. Kifonidis et al. 2006). High spectral resolution imaging spectroscopy with IXO-XMS will make it possible to reconstruct the 3D explosion properties by measuring Doppler shifts and broadening, even for velocities as low as 300 km/s. Current X-ray detectors restrict Doppler measurements to velocities in excess of 2000 km/s (e.g. Delaney et al. 2010).

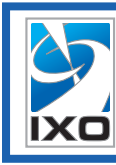 Spatially resolved high-resolution X-ray spectroscopy of supernova remnants with IXO will reveal velocity shifts and broadening of various ion species, which will then be used to reconstruct the supernova explosion properties in 3D.

Recently, the study of nucleosynthesis has been extended to include trace elements such as chromium and manganese (Tamagawa et al. 2009), extending our means to study the explosion physics. In particular, the Cr/Mn ratio, but also the Fe/Ni, can be used to learn about the electron fraction during nucleosynthesis. For Type Ia SNRs this is linked to the progenitor's original metallicity (Badenes et al. 2008b), whereas for core collapse SNe it tells us something about the exposure of the interior SN ejecta to the extreme neutrino flux from the collapsing core. In addition, weak line emission can be expected from radioactive $^{44}$Ti. This element is predominantly produced in core collapse SNe, when, due to high expansion velocities deep inside the SN, the $^4$He fraction freezes out. This provides a powerful diagnostic of the inner dynamics of the explosion, and the mass cut between SN ejecta and neutron star. $^{44}$Ti, as detected in Cas A in hard X-rays/ Gamma-rays should also be clearly visible in X-rays due to K-shell emission from its daughter product $^{44}$Sc at 4.1 keV. The predicted flux is sufficient to map the spatial distribution and kinematics of $^{44}$Ti inside Cas A, even for the unshocked, cold ejecta component. For SN 1987A, in the Large Magellanic Cloud, a detectable flux is predicted of 0.04 ct/s, for the expected initial $^{44}$Ti mass of $10^{-4}$ M$_\odot$.

Currently, our knowledge of SNRs is based on the very biased sample of Galactic SNRs, and on the population of Magellanic Cloud SNRs. IXO will be able to make a census of SNRs in the Local Group. Although detailed imaging may not be possible, IXO's spectral resolution will make it possible to accurately measure abundances, temperatures and kinematic ages (from Doppler broadening, Figure 2.28). This will even allow for sub-typing of the SNRs. This can, for example, be used to investigate the relation between Type Ia properties and local star-forming history (Badenes et al. 2009).

**Figure 2.28.** *A simulated IXO-XMS spectrum of a typical young Type Ia SNR in the Local Group galaxy M33 with an exposure of 100 ks. The type Ia nature is clear from the bright Fe-L complex. The line width is influenced by the 1800 km/s broadening due to the bulk motion of the expanding shell. For comparison the blue line also gives a model without any line broadening, showing off the IXO-XMS spectral resolution.*

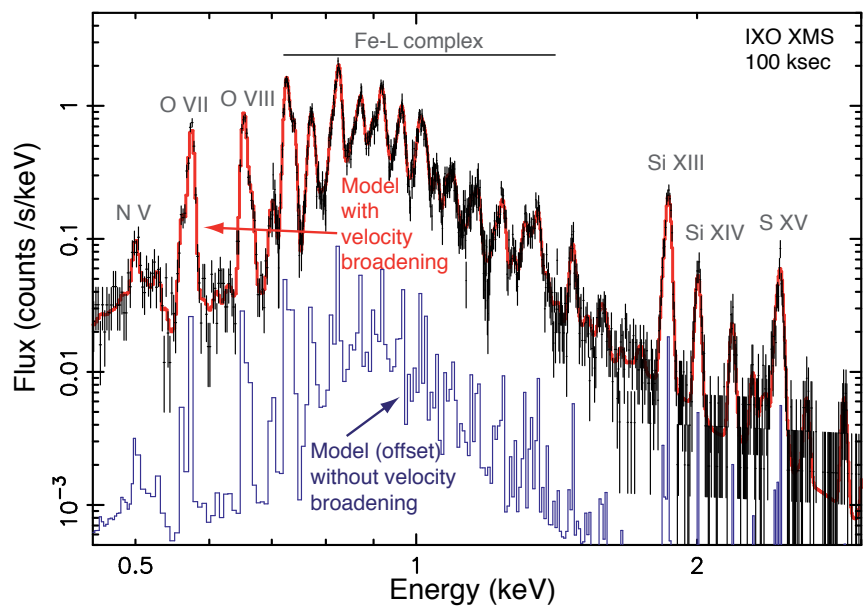

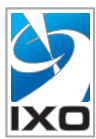 IXO will make an unbiased census of supernova remnants in the Local Group, and will measure elemental abundances and broadening. Weak lines from radioactive elements such as $^{44}$Ti and its daughter product $^{44}$Sc will provide unique insight into the detailed physics of the explosion.

### 2.4.1.2  Shock heating and particle acceleration

SNRs emit X-rays due to heating by high Mach number shocks that are driven into the surrounding medium and back into the explosion debris. As in most astrophysical shocks, heating occurs through plasma waves. Apart from being a major heating source of the Inter-stellar Medium, SNR shocks are probably responsible for most of the cosmic rays observed on Earth, at least for energies up to $10^{15}$ eV. The theoretical expectation that efficient cosmic ray acceleration modifies the heating and hydrodynamics of SNRs has been confirmed (Decourchelle et al. 2000), mainly as a result of X-ray observations (Helder et al. 2009). Cosmic ray acceleration takes away energy, resulting in colder plasma temperatures for a given shock velocity. Cosmic rays may also escape the SNR, thereby extracting energy from the SNR shell. This results in a different hydrodynamic structure and evolution of the SNR.

The physics of the shocks can be investigated by measuring simultaneously the temperature of the electrons and the ions. Collisionless shocks may heat ions to higher temperatures than electrons. There is evidence based on optical (e.g. Ghavamian et al. 2007) and X-ray spectroscopy (Vink et al. 2003) that for shock velocities above 500 km/s electrons are indeed cooler than protons/ions by factors of ten. Because IXO-XMS will offer imaging spectroscopy with high spectral resolution, thermal Doppler broadening can be measured close to the edges of SNRs, thereby providing direct measurements of ion temperatures that can then be compared to the more easily determined electron temperatures. Moreover, combining these measurements with measured shock velocities will reveal whether shock energy has been transferred to cosmic rays, resulting in lower plasma temperatures for a given shock velocity. This will be in particular interesting for those SNRs that show X-ray synchrotron emission, a signature of fast cosmic rays acceleration. In fact, the X-ray emission from some SNRs (RX J1713.7-3946 and "Vela Jr") seems to be entirely caused by synchrotron emission. Is this a result of electron temperatures below $10^6$ K, for which no X-ray thermal emission is expected (Drury et al. 2009) or is it just the density in those SNRs that is very low, giving rise to very weak line emission? The latter hypothesis can be tested by IXO, whose high throughput and high spectral resolution will make it possible to detect weak line emission from regions dominated by synchrotron emission. If detected, these lines can then be used to diagnose plasma temperatures and densities.

Finally, IXO will improve our understanding of cosmic ray acceleration by identifying synchrotron emission in hard X-rays, through the combination of the WFI and HXI. The hard X-ray synchrotron emission is expected to be very sensitive to magnetic field fluctuations due to the steepness of the electron spectrum near the spectral cut off (Bykov et al. 2009). Moreover, the emission from these flickering spots can be highly polarized. Hard X-ray mapping and polarization measurements with the HXI and XPOL can therefore provide direct measurements of the long wavelength modes magnetic field turbulence spectrum, which is an essential ingredient of acceleration physics. IXO studies of the thermal and non-thermal emission will complement TeV studies of SNRs with future ground based Gamma-ray telescopes (e.g. CTA), allowing us to constrain the magnetic field energy densities and maximum particle energies in SNRs.

## 2.4.2  Characterising the Inter-stellar Medium in the Galaxy

The elemental abundances fix the relative number of the different types of supernovae and the amount of star formation in a galaxy, and this has been a primary tool for unravelling star formation histories in the Milky Way and other galaxies. If we are to comprehend the evolution of the elements in the Milky Way, we must be able to determine elemental abundances in both the stars and the gas, including the dust phase. The heavy elements needed for making Earth-like planets derive from the Inter-stellar Medium, while its physics and chemistry of molecular clouds such as the catalysis of organic compound production by dust is vital for life-forming organic compounds. Improving our understanding of a vast range of processes from nucleosynthesis to planet formation will be enabled by accurate measurements of the gas and dust phase abundances of the Inter-stellar Medium.

X-ray spectral observations provide a unique tool for the determination of elemental abundances. Abundance determinations in the optical-UV rely on absorption or emission lines from low opacity sightlines and particular ionic states, with the subsequent large corrections for ionization state and depletion onto dust grains. In contrast, X-ray observations probe all ions from neutral to H-like, covering a window from the cold to $10^8$K Universe; can measure absorption across the ionization edge of an element through inner-shell absorption, which is only weakly dependent on the ionization state of the gas. In addition, the amount of dust up to equivalent hydrogen columns of $10^{24}$ cm$^{-2}$ is optically thin to X-rays, so metals in grains are also revealed by their weak absorption line properties. Because of this combination of factors, abundances measured from X-rays are more accurate and less subject to modelling uncertainties than abundances from optical-UV measurements. These absorption techniques have been used for several sightlines, but a much richer future awaits the sensitive observations possible with IXO. Spectral resolution must be sufficient to assess the X-ray absorption fine structure signature of dust composition but the sensitivity is also important to ensure different sightlines, which will enable the Inter-stellar Medium chemical uniformity and mixing or enrichment can be measured.

High-quality spectra will not only determine elemental abundances, but ionization state distributions and dust properties as well, in a way that complements infrared spectroscopy. For absorption at the ionization edge of a species, there is a shift in the energy of the absorption edge with ionization state. The ionization "edge" really consists of a series of edges, which if resolved, reveals the abundance of the individual ionic species along the line of sight. A resolution of ~3000 (such as that provided by IXO-XGS) allows to distinguish different iron compounds (see Figure 2.29), and to measure their physical state (bond lengths and charge states). Such resolution is needed to probe the composition of grains in various environments, by observing hundreds of lines of sight and providing critical constraints on the grain formation. To obtain abundance measurements with high precision (few %) over a range of densities on ~1000 objects it is important to sample sufficient lines of sight, but these observations do not need to be dedicated – IXO-XGS pointings to other targets will be useful for this purpose.

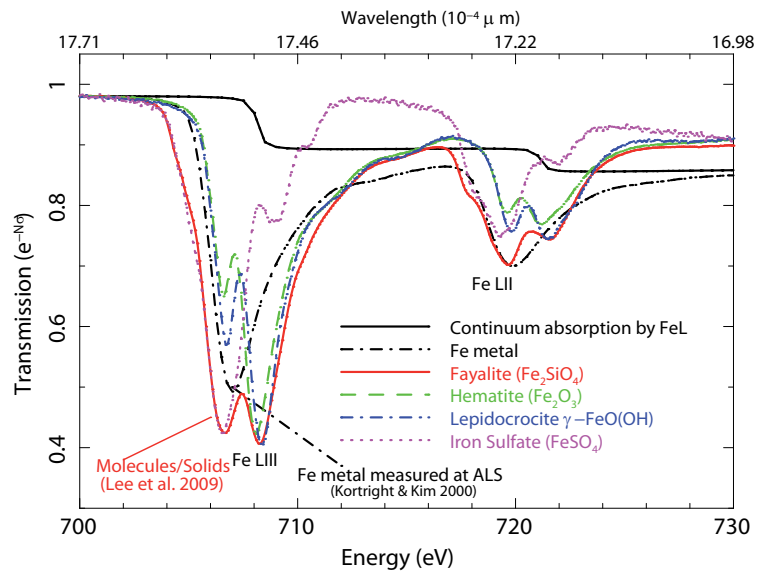

**Figure 2.29.** *Absorption curves at IXO-XGS resolution (R=3000), determined from laboratory measurements of iron ions and crystals. The spectra can distinguish between Fe in molecules (red solid), various types of metals (dot-dashed) and continuum absorption (black solid, from Lee et al. 2009).*

Finally, X-ray observations will yield information on the sizes of dust grains, especially for the larger grains, which account for most of the dust mass. This information is contained in X-ray scattered light, which produces an X-ray halo around a bright point source. The halo can be measured as a function of energy, leading not only to mass constraints on the large grains, but to their composition. X-ray observations are vital because most of the Inter-stellar Medium dust resides in dense molecular clouds that are not amenable to UV and optical measurements.

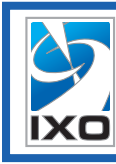

IXO's grating spectroscopy of bright X-ray sources will reveal the physical aggregation state (e.g. molecular, atomic or metallic) of ion species in the interstellar medium of the Galaxy, through the shape of their absorption edges.

## 2.4.3 The Galactic Center and its surroundings

The closest massive black hole to Earth, Sagittarius A* (Sgr A*) with 4 million $M_\odot$, lies at the dynamical center of our Galaxy. Its accretion luminosity is many orders of magnitude lower than for a typical AGN, and is thereby important since supermassive black holes spend most of their time in such a quiescent state. The X-ray flares observed from Sgr A* are likely to originate from within a few Schwarzschild radii of the event horizon, and can be well studied with the IXO-WFI, HXI and polarimeter to reveal the energetics and geometry of matter in this strong gravity region.

The large field-of-view of the IXO-WFI also allows us to observe simultaneously the neighborhood of Sgr A*, which is one of the richest regions of the sky with numerous types of extended (e.g., diffuse emission, supernovae remnants) and compact (e.g., X-ray binaries) astrophysical objects. It can also reveal past X-ray activity of Sgr A* through reflection on neighbouring molecular clouds. In particular, fluorescent X-ray emission from the cloud Sgr B2 is explained if Sgr A* was a low luminosity AGN ($L_x \sim 10^{39}$ erg/s) just 300 years ago (Koyama et al. 1996), when it would have been the brightest object in the X-ray Sky. This can be tested with IXO-XPOL observations of Sgr B2 and similar clouds since reflection is polarization sensitive. IXO observations of Sgr A* and its environment therefore provide a unique window on the current and past activity of a low luminosity active galaxy.

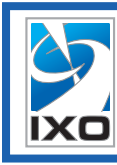

The IXO polarimeter will study for the first time the X-rays from Sgr A* as reflected in some of the surrounding structures in the Galactic centre.

## 2.4.4 Stars and planets

Being an X-ray facility open to the broad community, and because almost all astrophysical objects emit X-rays through very different physical processes, IXO holds great potential to study a broad range of systems, from young stellar objects to extra-solar planets with wide ranging implications. Here we highlight a few examples in the field of stars and planets, for which IXO can provide major breakthroughs, by extending measurements beyond the brightest objects and into time resolved processes.

**X-ray emission from Young Stellar Objects.** Understanding the X-ray emission of Young Stellar Objects is essential for studying the dynamics, structure and chemistry of their circumstellar environment. For example, the investigation of powerful flaring phenomena will show how ionisation affects magneto-rotational instability and how the resulting distribution of turbulence influences the evolution of the protoplanetary disk and, eventually, planetary system formation. A major sensitivity improvement, as provided by IXO, is necessary to constrain X-ray irradiation effects in hundreds of Young Stellar Object circumstellar disks that will be explored in all nearby star forming regions. Such an effect is the Fe K$\alpha$ emission that has now been observed from about a dozen of Young Stellar Objects, (e.g. Tsujimoto et al. 2005). Its prevalence in accretion disks and the inferred absorption geometry show that the fluorescence is produced in the circumstellar disk. Higher quality time-resolved X-ray spectroscopy is needed to understand the connection between the central continuum X-ray source and the emission line on time scales of the Young Stellar Object rotational period. Time delays measured with reverberation mapping can give access to the geometry of the system. The Fe K$\alpha$ line can be simultaneously detected from a large sample of Young Stellar Objects, while the intensity of deeply penetrating hard (10-30 keV) X-rays must be exploited to evaluate the effects on turbulence and disk "dead" zone.

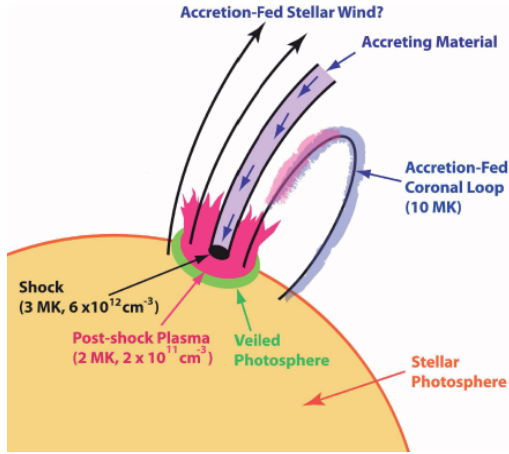

**Figure 2.30.** *Left: Illustration of accretion shock for TTauri star (Brickhouse 2010). Bottom: IXO-XGS simulated spectrum (red line), in the OVII triplet region, of TW Hya ($L_x = 7 \times 10^{29}$ erg s$^{-1}$ at 50 pc), with an exposure of 5 ks, compared to a 500 ks exposure of the same source with Chandra-HETGS (black line). Thanks to its large throughput, IXO will enable, for the first time, dynamical studies of the emitting plasma to be conducted through sequences of short exposures.*

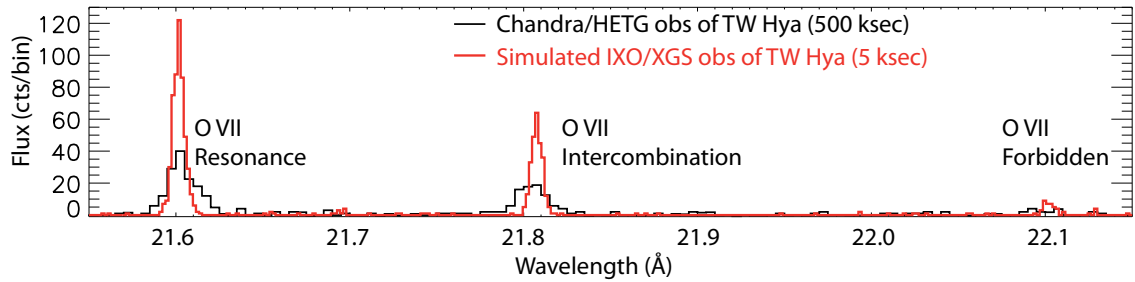

Only recently it has been found that soft X-ray emission (E ≤ 1 keV) in some Young Stellar Objects, instead of being of coronal nature, is likely due to accretion shocks from material falling onto the stellar surface (Figure 2.30, left). High-resolution spectroscopy of forbidden line features indicates a high density environment. The vastly improved resolution and sensitivity of the IXO-XGS spectra (Figure 2.30, bottom) will allow the studies to be extended to many more star forming regions, with time resolved spectra to observe geometric modulation in the accretion process of the brightest objects, as well as the direct Doppler shift of accreting material.

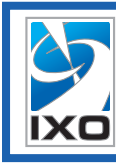
IXO's time-resolved high-resolution X-ray spectroscopy of Young Stellar Objects will indicate the geometry of the circumstellar disk via reverberation mapping of the various spectral components.

**The complex X-ray emission of solar system bodies.** X-rays emitted by planets like Jupiter, Saturn, Venus and Mars have revealed a variety of emission processes (charge exchange between energetic ions and hydrogen molecules in the planet upper atmosphere, scattering and fluorescence of solar X-rays, electron bremsstrahlung) and emission regions (aurorae near the poles, disk, exosphere). IXO's superior capabilities in terms of spectral resolution, sensitivity, broad band coverage will enable breakthrough observations to explore particle populations (Figure 2.31), acceleration mechanisms and their response to solar activity using 'next door' examples of extreme and widespread scenarios in astrophysics (Branduardi-Raymont et al. 2010).

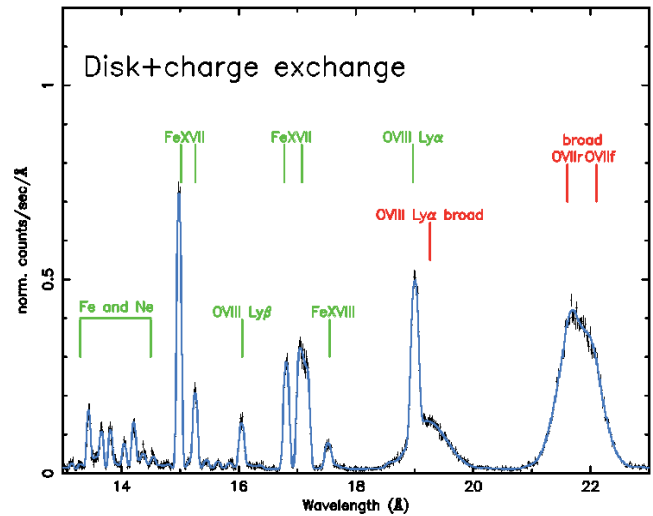

**Figure 2.31.** *The IXO-XMS spectrum of Jupiter will reveal hundreds of lines with sufficient signal to map the upper atmosphere through the planet's 36 ks rotational period. Repeated visits, particularly at different periods in the solar activity cycle and several days after powerful solar flares, should show varying ratios of the different emission components elucidating the complex physics of solar-planetary planetary interactions.*

**Atmospheric evaporation in extra-solar planets.** The atmospheric conditions, and eventually the habitability of planets is likely to be regulated partly by the evolution of the ultraviolet and X-ray emission of their host star, and may also be affected by solar wind and flare particles (Sanz-Forcada et al. 2010). Thousands of exo-planets will be known by 2020. IXO can measure both the quiescent and flare activity of specific stars with known planets in the habitable zones. Combined with stellar activity evolutionary trends and planetary atmospheric modelling, IXO findings will give unique insights into the atmospheric history of these potentially habitable planets.

**Mass loss, rotation and magnetic field of massive stars.** The global feedback mechanism described earlier (see, e.g. Section 2.1.3) can also be traced via X-ray emission from massive stars resulting from their energetic winds. To determine the wind distribution of hot plasma, the mass-loss rates, the degree of wind inhomogeneity, and even the geometrical shape of wind clumps requires high S/N lines profiles over a large representative sample of OB stars; then a major increase in sensitivity is necessary. With these diagnostics the factor of 10 discordance of deduced mass-loss rates found by traditional observational diagnostics (e.g. Hα and UV lines, radio free-free emission e.g., Fullerton et al. 2006) may be resolved. This is vital to better understand stellar evolution, the physics of the massive stars and to quantify massive star feedback that energizes and enriches the Inter-stellar Medium.

The importance of magnetic fields in massive stars has been highlighted in the last few years by direct measurements of surface fields up to kG strength (e.g. Bouret et al. 2008), in a handful of the closest objects. With IXO-XMS it will be routinely possible to study the variability of the X-ray emission lines over the relevant time scales and to probe its connection to stellar rotation and magnetism. This will be important for further advances in the theory of dynamos in massive stars (Spruit 2002), and the inclusion of magnetic fields in stellar evolution models (e.g. Maeder et al. 2009).

The head-on collision of winds in massive binary stars (e.g. Stevens et al. 1992) produces much harder X-ray emission than in single massive stars. The high sensitivity and spectral resolution of IXO is needed to observe the orbital changes of the resulting Fe K emission line for mapping the wind interaction region, similar to Doppler tomography studies performed in the optical domain (e.g. Thaller et al. 2001). This will provide unique information on the conditions at the shock between the stellar winds and on the efficiency of the cooling mechanisms. Understanding colliding wind binaries is pivotal in inferring the properties of the massive, mostly in binary systems, rapidly evolving stars in the Universe. IXO will allow us to extend these X-ray measurements to extragalactic massive binary stars featuring different metallicities and hence different stellar wind properties.

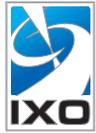

IXO's combination of high throughput and spectral resolution will be crucial for measuring mass loss rates in massive stars, stellar rotation and magnetic effects, through exquisite time-resolved emission line shapes.

**Probing non-standard dynamos and emission mechanisms in Brown Dwarfs.** In late-type stars, X-rays are a unique probe of the strength of the magnetic activity and the underlying dynamo, which are known to depend on effective temperature and on the evolutionary state of the stars. Critical transitions are expected as the stars become fully convective (0.3 $M_\odot$) and in the substellar regime (brown dwarfs) where atmospheres are mostly neutral with limited capacity for coupling with the magnetic field. The improved sensitivity of IXO-WFI will enable to study statistical samples of brown dwarfs including the youngest ones hidden behind strong extinction in star forming clouds that have remained inaccessible so far. In addition, the IXO-WFI will allow to extend brown dwarf studies to cooler temperature to spectral type L (Audard et al. 2007), to assess brown dwarf X-ray temperatures and search for evidence of additional emission mechanisms such as accretion shocks (Stelzer et al. 2010).

# 3 Science Requirements

In this section we derive the scientific performance requirements that will enable the science objectives described in Section 2 to be fulfilled. In turn these requirements are used to define the instrument requirements for the IXO payload. Naturally for an observatory class mission that addresses a wide range of science investigations, no single instrument can provide all the necessary features; the different requirements for imagers, spectrometers, polarimeters and high time resolution detectors have to be treated separately. The scientific investigations which drive key performance requirements in each science case are presented. The requirements identified per science area, together with an assessment of the observing programme, are provided in Appendix 1.

## 3.1 Performance Requirements

### 3.1.1 Point Source Detection Sensitivity

The requirement on the limiting sensitivity is determined by the search for the first supermassive black holes (Section 2.1.1). The driving science goal is to detect significant numbers of low luminosity AGN at redshifts in the range $z$=6-10. Discrimination between different models of the evolution – and hence the growth mode of black holes in the early universe – requires the detection of sufficient numbers of AGN with luminosity down to a luminosity of $10^{43}$ erg s$^{-1}$ (2-10 keV, rest frame) in the redshift range $z$~7-8. The required ultimate point source detection sensitivity is therefore ~$10^{-17}$ erg cm$^{-2}$ s$^{-1}$ (0.5-2 keV, observed frame).

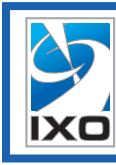 Characterising supermassive black hole growth in the early Universe drives IXO sensitivity at 1 keV (and thence effective area, angular resolution and low background at that resolution), as well as the wide imaging field of view.

Simulations (Aird et al. 2010) show that several 10's to 100's of AGN are required in each redshift bin to characterise the complete luminosity function. This is best satisfied with a multi-tiered survey, which gives full coverage of the L-$z$ plane. A typical strategy given the angular resolution and field of view (see Section 3.1.2 and Section 3.1.5) is to perform the following WFI pointings: 24x100 ks, 12x300 ks and 2x1 Ms, for a total of 8 Ms. This will yield 10's to 100's of moderate luminosity AGN at $z$>6 from dedicated surveys, and even more via serendipitously observed deep fields. Such objects are extremely difficult if not impossible to identify by means other than X-ray observations.

Formally, *Chandra* can reach close to the required limiting sensitivity in its deepest observations, but only over a small area close to the central aim point. IXO will be able to reach the required sensitivity in much shorter observing times, greatly improving the survey capability. For example, IXO will reach a CDF 2 Ms flux sensitivity of ~$2\times10^{-17}$ erg cm$^{-2}$ s$^{-1}$ (0.5-2 keV) in only 200 ks. For the deepest fields, the IXO sensitivity will substantially exceed that of *Chandra*, particularly when angle-averaged. A comparison between the deep survey capabilities of *Chandra* and IXO is shown in Figure 3.1.

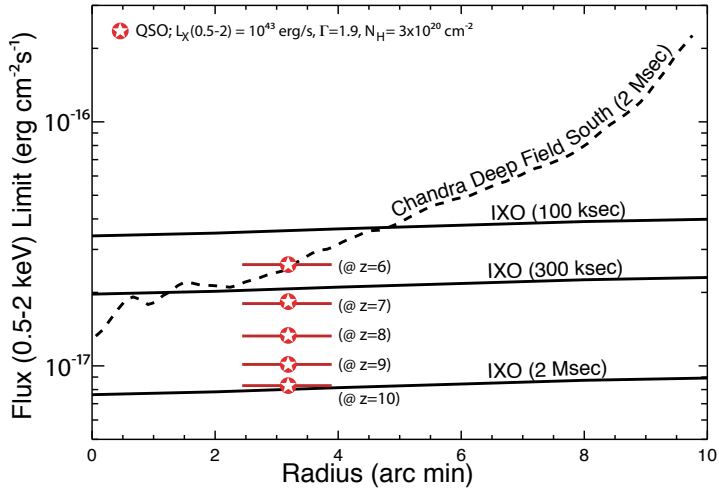

**Figure 3.1.** *The flux limit of IXO-WFI for the 3 exposure times that will be used for the multi-tiered survey approach to meet the SMBH growth science goal, showing very modest variation with radius within the WFI field of view. The Chandra ACIS 2Ms sensitivity (Luo et al. 2008), which is severely affected by vignetting and off-axis PSF degradation, is also shown for comparison. Fluxes of low-luminosity AGN at redshifts z=6-10 are displayed for reference.*

## 3.1.2    Angular Resolution

The angular resolution requirement is also driven by the detection of early supermassive black holes (see Section 2.1.1). The deep field sensitivity is affected by two different components: the confusion limits and the solid angle for collecting background in the source detection cell. Position centroiding for multi-wavelength follow-up, together with source feature extraction in extended objects, provide comparable resolution requirements:

**1)** An increased background that occurs with larger detection cells can be compensated partly by increasing exposure time. This trade-off eventually is limited by source confusion. The ultimate detectable source density depends upon the relative strengths of luminosity and density evolution with look-back time. This has been investigated with simulations of many scenarios. A simple estimate for the increasing confusion between sources with decreasing flux can take as an example the Georgakakis et al. (2008) extrapolation of CDF source density. For a confusion limit of one source per 40 beams, a 5 arcsec beam is confused at a flux limit of ~7x10$^{-18}$ erg cm$^{-2}$ s$^{-1}$ (0.5-2keV). This is compatible with the deep fields sensitivity requirement noted above. However the flux limit of sensitivity degrades very rapidly due to the combined effects of increased background and confusion. Detailed simulations show a steep dependence of limiting flux with PSF (approximately like HEW$^{2.5-3}$) and therefore the angular resolution requirement cannot be relaxed significantly (Figure 3.2).

**Figure 3.2.** *The effect of PSF degradation on total Deep Survey observing time for a survey of $L_X > 10^{43}$ erg s$^{-1}$ SMBH at redshifts ~6, 7 and 8. The dashed line indicates the nominal duration for a multi-tiered survey, such as 24x150 ks = 3.6 Ms, 15x300 ks = 4.5 Ms , 2x1 Ms = 2 Ms that has been posited to provide a good trade-off between field area and depth of observation and thus constrain the evolution of AGN at the highest redshifts.*

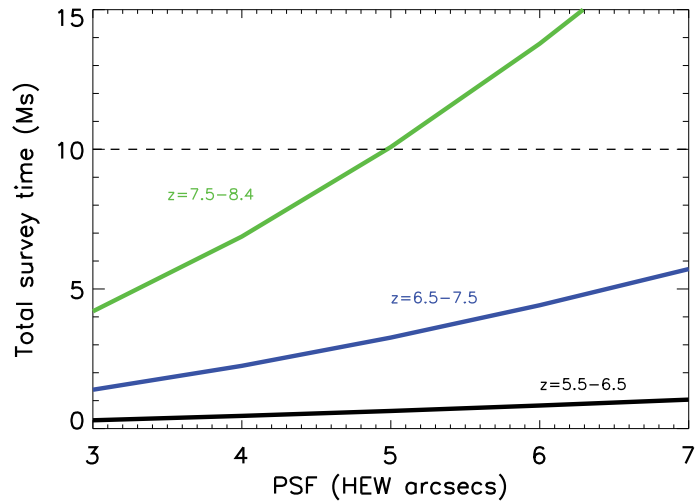

**2)** For the characterisation of the IXO high-*z* sources at longer wavelengths (e.g. E-ELT spectroscopy), the 5 arcsec HEW of IXO is sufficient for identification of the X-ray source with the corresponding IR source. At $J_{AB}$=26, the mean source separation is ~1-1.5 arcsec (Maihara et al. 2001). For an IXO X-ray source detection of 40 counts, the statistical error on the X-ray centroid is ~0.3 arcsec, small compared with the IR source separation.

**3)** High redshift galaxy clusters will be observed for spatially resolved spectral data (Section 2.2.3). The scale of AGN feedback (Section 2.1.3) over 10's kpc is consistent with a typical cool core radius of ~40-70 kpc. In the standard cosmological model, a HEW of 5 arcseconds is required to resolve this scale at ***all*** redshifts.

**4)** The grating spectrometer relies in part on the telescope angular resolution to define the spectral resolving power. The spectral resolution requirement of the baseline design has assumed a modest sub-aperturing and a HEW of 5 arcseconds to achieve E/ΔE of ~3000. To some extent the spectrograph design can be modified to account for the eventual telescope resolution, but as the grating implementation becomes fixed later in the IXO development programme, the resolution will scale with the eventually achieved telescope angular resolution.

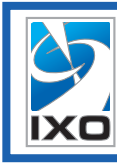

> The ability to probe to the faintest sensitivity limits requires a sustained effort to achieve the highest feasible angular resolution.

### 3.1.3   Spectral Resolution

The required spectral resolution varies strongly with science case, and hence with complementary science drivers such as time resolution and field of view, that vary by orders of magnitude between target classes. Therefore the different instrument classes can have different spectral resolution requirements.

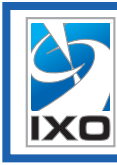

> Spectral resolution in spatially resolved high-resolution spectroscopy is required by the need to trace the dynamics of hot gas in groups and clusters out to high redshift and measure elemental abundances to sufficient accuracy.

**1) Narrow Field Imaging:** The requirement is driven by spectroscopy of clusters of galaxies in order to study how the hot baryons evolve in their potential wells, when the excess energy in these systems was injected (Section 2.2.3) and also how feedback from AGN affects groups and clusters (Section 2.1.3). The measurement of turbulence and velocity-shifted features greater than 100km/s is needed; see Figure 3.3. High spectral resolution also maximises the accuracy of abundance measurements and allows discrimination against Galactic line emission. These foreground lines dominate the low energy 0.2-0.35 keV band where we expect the O line emission for a cluster at 1<*z*<2. To separate the many (>10) background lines brighter than the cluster lines then less than 20% of the energy band should be confused by these lines, and this gives a δE<3 eV requirement. Finally, the cluster line detection over the continuum at 5 σ (for reasonable accuracy on abundance) requires a resolution of δE<3 eV.

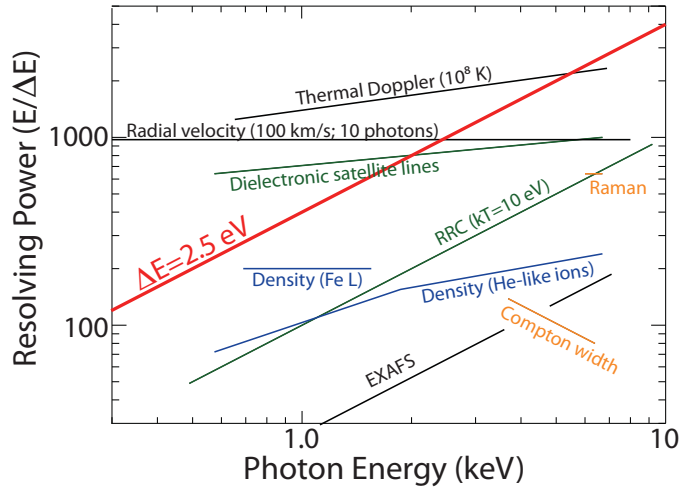

**Figure 3.3.** *Resolving power needed to measure a number of physical phenomena out of spectral features in the 0.3-10 keV band-pass. The spectral resolution of the central array of IXO-XMS is shown for reference. In particular, it will be able to diagnose hot plasmas using the O He-like triplet, and measuring velocity broadening and motions down to 100 km/s using the Fe K line at 6 keV.*

**2) Wide Field Imaging:** The specific case of the growth of SMBH (Section 2.1.1) requires modest spectral resolution to identify detect emission lines (e.g. of iron) and determine absorption properties. Typical wide field science of modest brightness sources requires CCD-class resolution to determine temperatures and centroid Fe-line for redshift determination; the required resolution is 150 eV FWHM at 6 keV to measure redshifts to 5% accuracy.

**3) Point Source Dispersive Spectroscopy:** A number of bright emission line spectral investigations can utilise higher resolution spectroscopy than the Narrow Field Imaging case. However the main driving requirement is for observing lines in absorption against a continuum. The driving science is for measuring the cosmic web of baryons (Section 2.2.2). OVII and OVIII resonance absorption lines redshifted to a range of energies (0.3-0.5 keV) will have equivalent widths of a few mÅ, but with velocity structures imposed by galactic superwinds. The detection sensitivity is a function of both the width of resolution bin and the effective area, but detecting a Doppler shift of 100 km/s imposes a resolving power requirement of 3000.

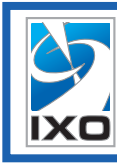

Finding the missing baryons in the Warm-Hot Intergalactic Medium and determining its velocity structure drives the spectral resolution in high throughput dispersive spectroscopy.

**4) Hard X-ray Imaging:** The most driving requirement is the use of hard X-ray imaging capability simultaneously with wide field imager measurements of SMBH spin (see Sections 2.3.1.2 and 2.1.1). This requires the accurate determination of the underlying power-law continuum, most notably the position of the Compton reflection 'hump', at much higher energies than the Fe line. A spectral resolution of 1 keV at the upper energies 30 keV is sufficient to provide this lever arm.

**5) High Time Resolution:** Time resolved spectroscopy of bright persistent or transient X-ray binaries will enable a measurement of the spins of a few tens of black hole systems, through several complementary techniques (relativistically broadened iron line, disk continuum spectral fitting, reverberation mapping, see Section 2.3.1.2). It will also enable IXO to constrain the mass and radius in neutron star systems using also different probes (high-time resolution spectroscopy of the kHz quasi-periodic oscillations, detection of rotationally broadened absorption line features, iron line variability, etc, see Section 2.3.2.2). The spectral features to be used are relatively broad (e.g. ~ keV for the iron line, a few hundreds of eV for the predicted absorption lines). For reverberation mapping studies, the disk reprocessed emission peaks below 2 keV. The main HTRS requirements are therefore a spectral resolution better than 200 eV at 6 keV (ideally 150 eV), a band pass from 0.3 to 15 keV, provided together with the ability to cope with extremely large count rates, up to at least 1 million events per second or 10 Crabs[*], with negligible pile-up and deadtime (less than 2% at 1 Crab). Most X-ray binaries exceed a few hundreds of millicrabs in their persistent emission, while transient sources can reach several tens of Crab. Similarly type I X-ray bursts of interest exceed typically 1 Crab.

## 3.1.4   Effective area

Effective area is related to the sensitivity and angular resolution parameters, especially at the peak of fluxes around 1-2 keV. There are other drivers due to the photon-limited science such as iron line reverberation mapping. Finally at high energies the effective area requirement is set by the need to determine the high energy continuum in AGN.

**1) 1-2 keV:** As discussed above, the sensitivity couples the area requirement to angular resolution and survey exposure duration. The need to meet the sensitivity limit with the baseline angular resolution in a time that is commensurate with an observatory programme containing 8 Ms of surveys (see Section 3.1.1) requires an instrument effective area of 2.5 m². This can be achieved with a mirror area of <3 m² assuming typical detector efficiencies and grating obscurations. In the limit the mirror effective area could be reduced to 2.5 m² by assuming a combination of modestly increased field of view and/or observing time. The latter of course impacts on the overall balance of the general observing programme.

**2) 6 keV:** Given the typical target fluxes of 1-10x10$^{-11}$ erg cm$^{-2}$ s$^{-1}$ (2-10 keV), orbital timescales of 1 to 300 ks, and Fe line equivalent widths of 200 eV, an effective area of 0.65 m² at 6 keV will ensure > 100 photons in the Fe line per  SMBH orbital bin (assuming at least 10 phase bins throughout the orbit).The centroid of narrow and varying iron lines can then be measured for a range of expected SMBHs fluxes (see Sections 2.1.1 and 2.3.1.2).

> 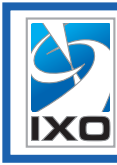 Measuring supermassive black hole spins as well as the dynamics of gas under their strong gravity fields, drives the effective area at 6 keV and the spectral resolution at that energy. It also drives the effective area at 30 keV.

**3) 30 keV:** Multi-tiered measurements of time-averaged Fe line profiles will be used to determine the spin of hundreds of SMBH in AGN (see Sections 2.1.1 and 2.3.1.2). This requires an accurate determination of the continuum under the Fe line, which can be obtained when the effective area of ~150 cm² at 30 keV is achieved. The effective area goal of 350 cm² at 30 keV combined with a goal 5″ IXO angular resolution which is substantially better than that of *Suzaku* and *NuSTAR* will allow a substantial advance in science areas such as cosmic ray acceleration in SNR shocks, the resolution of the hard X-ray background and inverse Compton scattering off cluster relativistic particles. The measurement of reflection components typical of Compton-thick sources evolving with *z*, is needed to be able to resolve 80-90% of the currently undetermined hard X-ray background (Section 2.1.2).

## 3.1.5   Field of View

**1) Wide Field:** From the previous analysis, the driving requirement is the determination of the distribution of growing supermassive black holes in the early Universe. The multi-layered survey requires a sensitivity in 200 ks of  ~3x10$^{-17}$ erg cm$^{-2}$ s$^{-1}$ (0.5-2 keV) over an area >200 arcmin $^2$ (assuming moderate vignetting) Individual objects such as galactic SNRs and low redshift clusters have a typical extent ~10 arcmin diameter, and with additional field of view for contemporaneous background extraction require a comparable sized field. (e.g. SNR diameters Cas A 3´, Crab 4´, Kepler 4´, Tycho 7´, SN1006 25´). For typical r$_{200}$ of ~1 Mpc any cluster more distant than *z*~0.1 is fully encompassed within a 10 arcmin field.

---

[*]  *1 Crab, the flux of the Crab supernova remnant, is equivalent to 2x10$^{-8}$ erg cm$^{-2}$ s$^{-1}$ in the 2-10 keV band.*

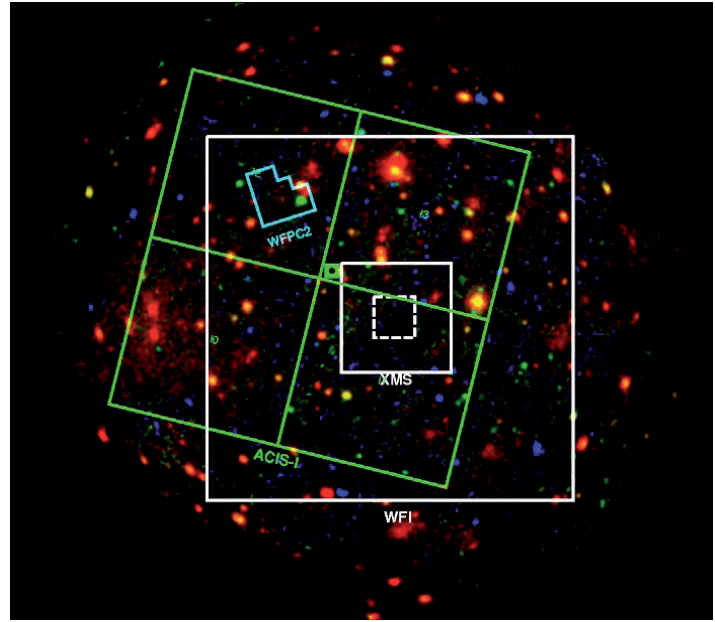

**Figure 3.4.** *XMM-Newton image of the CDF-S field. Superposed are the original WFPC2 (cyan) & Chandra ACIS-I (green) instantaneous fields of view. The baseline fields of view of the IXO-WFI and XMS (core dashed) are shown in white.*

**2) Narrow Field High Resolution:** The cluster science requires calorimeter quality spectra be obtained for clusters at *z* of 1-2 out to a cluster diameter corresponding to a density contrast of 500 times the critical density. According to Section 2.2.3 the driving requirement is to encompass a poor cluster at *z*~2 or cover a typical *z*~1 cluster to the virial radius. For the highest quality measurements this requires a field of view of 2 arcmin. To provide a contemporaneous field to subtract the sky background components a total field of view of 5x5 arcmin is necessary, otherwise the observatory must mosaic observations with additional observing time and potential systematic errors due to response and background variations. The larger field is also important in order to study nearby and larger clusters.

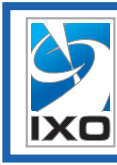

A huge survey grasp advantage over *Chandra* for deep field science is enabled by nearly constant angular resolution across the wide field of view.

## 3.1.6   Time Resolution and Count Rate Capability

While many of the above science requirements have focused on the requirements for faint targets, the very high effective area also enables a statistically high accuracy on traditionally bright targets for time resolved phenomena (see Section 2.3). The X-ray emission of bright X-ray binaries shows aperiodic variability on the dynamical timescales of the innermost region of the accretion flow, whose frequency in the case of neutron star systems exceeds 1 kHz and in the case of black hole systems, several hundreds of Hz. Similarly during type I X-ray bursts, or in the persistent emission of accreting millisecond pulsars, the X-rays can be modulated at the neutron star spin frequency (typically 300-400 Hz) or at twice the spin frequency. The study of the aperiodic and periodic signals can be used to place constraints on black hole spins (e.g. with high frequency QPOs), and the neutron star compactness (through waveform fitting the X-ray pulsations). The main HTRS requirements are therefore to time-tag X-ray photons with sub-millisecond time resolution (10 microseconds), while observing X-ray sources producing extremely large count rates, up to more than 1 million events per seconds (about 10 Crab), with negligible pile-up and deadtime (less than 2% at 1 Crab). Sources or type I X-ray bursts of interest can easily exceed 1 Crab.

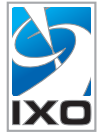

High time resolution is needed to obtain the power spectrum of matter orbiting black holes up to orbital frequencies and to constrain mass and radius of neutron stars.

Understanding the emission process of milisecond pulsars requires the capability to compare the folded X-ray light curve with the one observed at other wavelengths, such as in radio or in gamma-rays. For this purpose, an absolute time accuracy of 100 microsecond is required.

At fluxes above a few mCrab, the cryogenic narrow field spectrometer reaches a limit in the per pixel counting rate and the spectral resolution begins to degrade. This can be overcome to some extent by defocusing (requires an intrafocal diffusing optic) which would extend the range up to ~hundred mCrab and total rates of $10^4$ ct s$^{-1}$ , but above this the HTRS will take over.

For WFI, bright point sources are not expected to be drivers for count rate capability (as HTRS will provide comparable spectral performance). we consider the moderately extended objects (bright Galactic SNRs) to be the critical driver. A count rate capability of ~100 ct s$^{-1}$ per PSF is required.

### 3.1.7   Polarisation Capability

The polarization map of SNRs should reveal highly polarized structures with typical scales of about 10 arcseconds (Section 2.4.1.2). Since each clump should have a different polarization, the requirement on the angular resolution is to be better than 10 arcseconds to avoid the dilution of the average polarization. The field of view will be minimum 2.6x2.6 arcminutes square to image significant regions of SNRs such as Cas A or RX J1713.7-3946. The polarimetry requirement for black hole spin (Section 2.3.1.2) is the capability to detect a rotation of the angle of polarization larger than 1 degree/keV. This allows to detect the rotation of the angle of polarization due to relativistic effects for a number of systems with different observer's orientations (Dovciak et al. 2008). The emission from highly magnetized neutron stars (Section 2.3.2.3) is expected to be strongly polarized at the level of 5% in 18 bins of the light curve allowing us to study the strength and geometry of the magnetic field. It is required that the accuracy on the angle of polarization of SgrB2  (Section 2.4.3) is below 2 degrees to constrain tightly the position of the external illuminating source. The energy resolution will be higher than 20% FWHM at 6 keV to detect a possible variation of the polarization (in degree or direction) with energy.

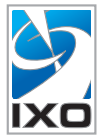

Polarimetric sensitivity is driven by the need to measure galactic black hole spins and vacuum polarisation effects around highly magnetised neutron stars.

### 3.1.8   Charged Particle Background

Again the faint target science (Sections 2.1 and 2.2) drives the requirement on background. With an orbit location of L2, the limiting flux of Galactic cosmic rays is relatively well known, while some physical (GEANT) modelling of typical detector and spacecraft structures is able to predict the quiescent level of background components that can be expected. While this is irreducible, the detector system design must take care to ensure no excess generation of secondaries, as well as refining rejection techniques to discriminate between the X-ray signal and prompt effects of charged particle background. The preceding analyses have all assumed such performance, where for the wide field imaging, a flat spectrum of unrejected background ~5x10$^{-5}$ cts/s/keV/mm$^2$ is required. The equivalent requirement for the narrow field spectrometer has been scaled from low earth orbit measurements of *Suzaku*, and is required to be <~2.4x10$^{-4}$ cts/s/keV/mm$^2$.

### 3.1.9   Straylight requirement

Off axis sources can be falsely imaged by single bounce reflection off the mirrors. In order for the in-field background contribution from this effect to be <10% of the unresolved AGN background at an energy range of 0.3-2 keV, then the contribution of all sky background in the annulus 0.5-1.5 degrees radius must be rejected by a factor 5x10⁻⁴.

## 3.2   Summary of Requirements

The matrix of performance requirements that flow down from the driving science cases has been distilled into a single table summarising all the science instrument performance requirements (see Table 3.1). A draft Observing Plan has been constructed, using a sample of the typical objects representative of the science case and with a distribution suggested by the observing durations presented in the tables above.

For a single year of operations the target distribution around the sky was used to define durations for slewing and focal plane instrument changes, as well as contingencies, solar flare losses, etc. The resulting observational efficiency was predicted to be ~85%. For a 5 year science mission duration carried out with this observational efficiency, the core science cases (~110 Ms) leaves ample margin for routine internal and celestial calibrations, targets of opportunity (10 Ms allocated), and a large additional time for "general observatory" science (~15%).

| PARAMETER | REQUIREMENTS | SCIENCE |
|---|---|---|
| **Effective Area** | **2.5 m² @ 1.25 keV** <br> **0.65 m² @ 6 keV** with a goal of 1m² <br> **150 cm² @ 30 keV** with a goal of 350 cm² | **Black hole evolution, large scale structure, cosmic feedback** <br> **Strong gravity, Equation of State** <br> **Cosmic acceleration, strong gravity** |
| **Spectral Resolution (FWHM)** | **ΔE = 2.5 eV (@ 6 keV)** within 2 x 2 arcmin <br> **ΔE = 10 eV (@ 6 keV)** within 5 x 5 arcmin (bandpass **0.3-12 keV**) <br> **ΔE = 150 eV @ 6 keV** within 18 arcmin diameter **(0.1-15 keV)** <br> **E/ΔE = 3000 (0.3-1 keV)** with an area of 1,000 cm² and a goal of 3,000 cm² for point sources <br> **ΔE = 1 keV** within 8 x 8 arcmin **(10-40 keV)** | **Black hole evolution & cosmic feedback** <br><br> **Large scale structure** <br><br> **Missing baryons using tens of background AGN** |
| **Angular Resolution** | **5 arcsec** HPD **(0.1-7 keV)** <br> **30 arcsec** HPD **(7-40 keV)**;  goal of 5 arcsec | **Large scale structure, cosmic feedback, black hole evolution, missing baryons** |
| **Count Rate** | **1 Crab with > 90% throughput** <br> **ΔE < 200 eV @ 6 keV (0.3-15 keV)** | **Strong gravity** <br> **Equation of State** |
| **Polarimetry** | **1% MDP on 1mCrab, 100 ksec, 3σ** <br> **(2-6 keV)** | **AGN geometry** <br> **Strong gravity** |
| **Astrometry** | **1.5 arcsec at 3σ confidence** | **Black hole evolution** |
| **Absolute Timing** | **100 μsec** | **Neutron star studies** |

**Table 3.1.**   *Summary of requirements.*

# 4 Payload

The science aims of IXO are highly demanding and require state of the art instrumentation. First of all a X-ray mirror of at least 2.5 m$^2$ effective area at 1 keV with a spatial resolution < 5 arcsec, and a focal length of 20 m is required. The focal plane plate scale for this optic equals 0.1 mm/arcsec. Secondly, the requirements on spatial resolution, field of view (FoV), energy resolution, energy range, quantum efficiency, count rate capability, and polarisation sensitivity cannot be met by a single focal plane instrument. Therefore a movable platform located at the telescope focus allows the illumination of different focal plane instruments. The instruments comprise:

- *An X-ray imaging Microcalorimeter Spectrometer (XMS)* that covers the 0.3-12 keV energy range with unprecedented energy resolution, a 5x5 arcmin FoV, and relatively low count rate capability.

- *A Wide Field Imager (WFI)* covering the 0.1-15 keV energy range with a large, 18 arcmin diameter FoV, excellent spatial resolution and efficiency, good energy resolution, and adequate count rate capability.

- *A confocal Hard X-ray Imager (HXI)* that covers an 8 arcmin FoV with excellent spatial resolution and efficiency in the 10-40 keV energy range, in combination with good energy resolution and count rate performance.

- *A non-imaging High Time Resolution Spectrometer (HTRS)* that covers the 0.3-15 keV energy range with good energy resolution, and ultra-high count rate capability.

- *An imaging X-ray Polarimeter (XPOL)* with a modest FoV, modest energy resolution, and excellent sensitivity to polarization in the 2-10 keV energy range.

- *An X-ray Grating Spectrometer (XGS)* for the 0.3-1 keV range with 1000 cm$^2$ effective area and a resolving power of >3000.

In this section the key elements of the X-ray telescope and the focal plane instruments are described. Here features and performance of both the baseline and back-up mirror technologies are noted. Next the essential features for each science instrument, their driving requirements, operating principles and outline of implementation and resource requirements are described.

## 4.1 IXO X-ray optics

The IXO mission requires a large effective area (>2.5 m$^2$ at 1.25 keV) and high angular resolution (<5 arcsec HEW, at <7 keV). To allow for such a large telescope within the mission mass constraints two innovative X-ray optics technologies are under advanced development. The baseline technology for IXO is the silicon pore optics, SPO (developed by ESA) and the back-up technology is the segmented glass optics, SGO (developed primarily by NASA and with further alternate designs by ESA). Both technologies are compatible with the design of the mirror assembly module and can be accommodated in the spacecraft design. The selection of the SPO as baseline allows the following definition phase to proceed with a detailed development programme. The identification of a viable back-up reduces the programmatic risk on the programme. A confirmation of the IXO optics technology is planned by 2012, based on actual technical achievements.

### 4.1.1   IXO optics baseline (SPO)

The ESA system level study assumed the SPO baseline for the IXO mirror. SPO is a highly modular concept; it has demonstrated an excellent effective area-to-mass ratio and is based on a set of compact individual mirror modules (MMs). About 2000 MM are required to populate the IXO mirror, and the total mass allocated to them is less than 700 kg, including the isostatic interfaces and the sub-system maturity margins. The full production process for SPO has been demonstrated employing a consortium of industrial partners.

The SPO solves the IXO mass-area-resolution challenge by introducing a new approach to mounting the X-ray mirrors into a matrix-like structure. The resulting MMs are intrinsically very stiff and robust, keeping the figure of the mounted X-ray mirrors stable to the arc second level. Unlike the conventional and established X-ray optics technologies, which use a limited number of interface points to attach the mirror optics elements to the support structure, the SPO technology relies on a much stronger inter-linkage of the X-ray mirror elements via integrated ribs.

The individual mirror plates are assembled into a matrix-like structure, whereby the ribs of a mirror plate bond to the surface of the preceding mirror plate (see Figure 4.1). The individual mirror plates form a monolithic structure, of mono-crystalline silicon. These mirror plate stacks are therefore very rigid. Each mirror plate in the stack is slightly inclined to the previous one, as required for the Wolter geometry of the telescope design.

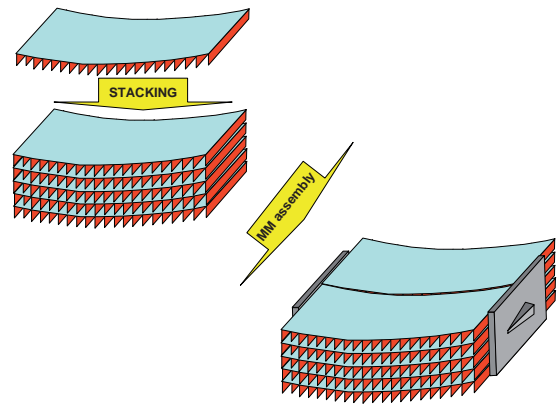

**Figure 4.1.** *Mirror plates are stacked and bonded together to form monolithic blocks of mounted mirror plates. Two such stacks are then assembled into the mirror module, employing two brackets. The MM provide attachment points for the isostatic mounting to the telescope structure.*

Two such mirror plate stacks are then assembled to form a MM, using two brackets. Very lightweight coefficient of thermal expansion (CTE) matched brackets are used to provide a stiff and permanent alignment of the mirror plate stacks. These brackets also include the interface elements for the isostatic mounts used to decouple the loads between the MM and the optical bench. Once assembled, a MM effectively forms the equivalent of a "lenslet" and the tolerance requirement for its alignment within the optical bench is much lower and achievable via conventional engineering techniques.

Considering the large number of required optics modules for IXO, it was very important to take into account mass production aspects right from the beginning of the technology development. Due to the small size of the SPO modules the production equipment can be kept compact, ensuring the cost effective implementation of a production line, including the associated cleanroom infrastructure.

The mirror plate production begins with 300 mm silicon wafers that are superpolished on both sides. Such wafers are readily available and mass-produced for the semiconductor industry. Furthermore the subsequent processing steps are also tailored from existing industrial processes, including dicing the pores, angular wedging, masking and coating, cleaning and packaging. The silicon plate surfaces can be bonded, normally using hydrophilic bonding that occurs between two oxide layers, providing a high bonding strength.

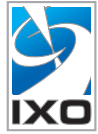 The baseline silicon pore optics concept relies on standard semiconductor industry techniques and equipment. The super-polished silicon plates with X-ray quality reflecting surfaces are commercially available off-the-shelf.

The fully automated assembly robot (Figure 4.2) is specifically developed to stack SPO and is a combination of standard semiconductor systems and newly developed tools. The complete system has a footprint of a few m² only and is installed in a class 100 clean area. The robot selects a plate for stacking and inspects it for particles. The plate is then handed over to the actual stacking tool, which will elastically bend it into a cylindrical or conical shape, then align and bond it to the underlying mirror plate. The stacking is done from outer radii inwards, thus always exposing the last integrated mirror surface to the metrology tools.

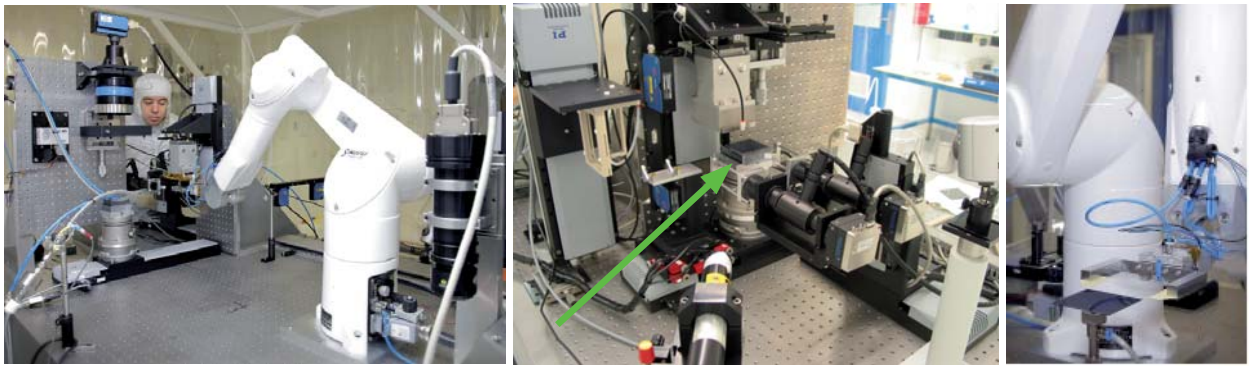

**Figure 4.2.** *Stacking robot inside the class 100 clean area at cosine Research BV. The system is installed on a vibration isolated table, consists of more than 16 axes, is fully automated and is designed to build stacks up to 100 plates high. The plates can be positioned with μm accuracy and automatically be bent into the required shape.*

The integration of two mirror stacks into a MM in flight configuration requires the alignment of the two stacks with an angular accuracy below one arc second. This is achieved using a dedicated integration setup and X-ray beam metrology. Figure 4.3 shows a set of mirror plate stacks and an assembled SPO mirror module. The integration of MMs into a full size petal was already demonstrated within the former XEUS activities.

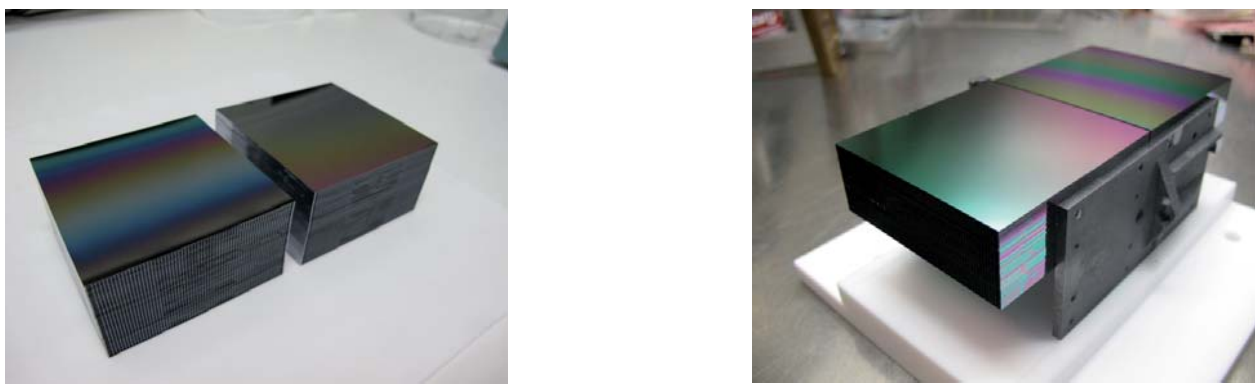

**Figure 4.3.** **Left**: *a pair of SPO stacks, consisting of 45 bonded mirror plates each, as required for the IXO telescope baseline design.* **Right**: *A complete SPO mirror module, consisting of a pair of mirror plate stacks permanently joined by two CTE matched silicon carbide brackets (cosine Research BV).*

Since the development of SPO started in 2002, it followed a holistic approach, tackling the important aspects of the entire production chain, including consideration of the eventual mass production in a flight programme. Consequently the processes and equipment relating to production and characterisation of the SPO from plate up to petal level were iterated, with the resulting MMs constantly improving. In parallel, elements like the integrated baffles, grounding issues and the application and performance of reflective coatings (e.g. $B_4C$ on Pt, Ir), have been tackled. In particular the process has been verified with patterned reflecting surfaces (Pt and Ir) between the ribs. Furthermore, coating of multilayers on Si-substrates has been demonstrated and measured for the hard X-ray extension of the instrument.

X-ray testing of the MMs, mounted in a flight representative configuration as they will be mounted into the petal, was performed at the XPBF (X-ray Pencil Beam Facility), a dedicated beamline in the PTB laboratory at the synchrotron radiation facility BESSY II and at PANTER. The impressive track record is summarised in Figure 4.4, showing the improvement in measured Half Energy Width (in 3 keV X-rays) for 4 representatively mounted mirror plate pairs, as a function of time. The latest measured MM shows a HEW of 7.5 arcseconds for the first 4 mirror pairs, and 9 arcseconds for the first 10 mirror pairs. Note that in Figure 4.4 we arbitrarily display the data for 4 plate pairs (4pc) in each case to track the corresponding improvement with time.

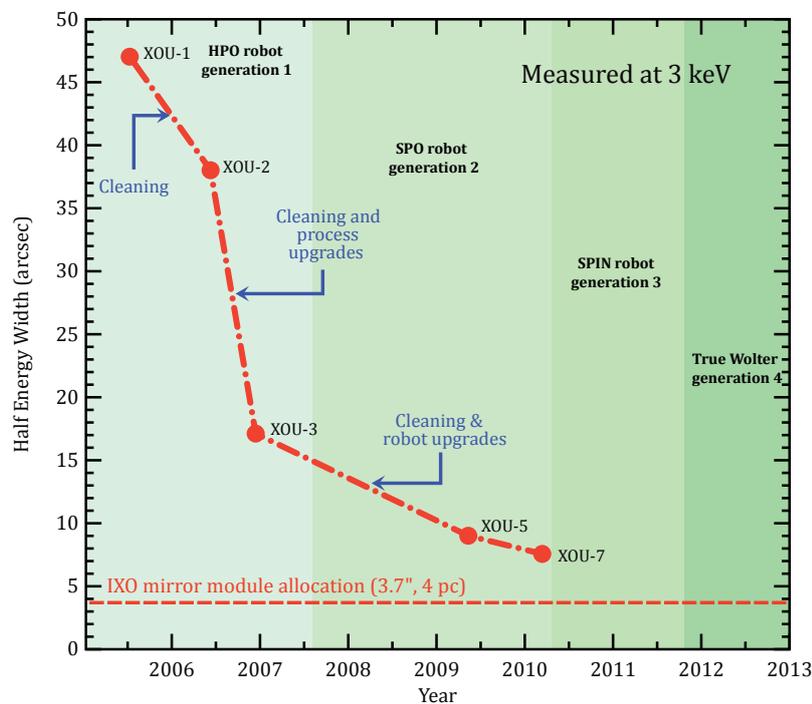

**Figure 4.4.** *The measured performance of the SPO mirror modules as function of time. The HEW in arcseconds, measured with 3keV X-rays in double reflection, for the full area of 4 representatively mounted mirror plate pairs (4pc) is shown. With the first generation equipment significant progress was made on the mirror plate production (High Performing Optics contract - HPO), leading to a steep improvement between mirror modules XOU-1/XOU-2 and XOU-3. After that the 2nd generation stacking robot was developed and brought on-line (Silicon Pore Optics contract - SPO), which further improved the performance of the mirror modules produced. For the last module produced, XOU-7, a HEW of 7.5 arcseconds was measured for the first 4 plate pairs.*

The future plans for upgrading the production process have already been identified and are in the process of being implemented. Three major technical improvements foreseen for the next two years are: a new cleaning system based on semiconductor industry standard, better wedge profile productions and a true Wolter profile that will significantly improve the HEW (Silicon Pore Industrialisation contract - SPIN).

The effective area of the optics, measured on uncoated optics from 0.3 to 3 keV, matched within 5% the theoretical expectation, confirming the good surface roughness and alignment achieved in the constructed MMs, satisfying the IXO design requirements.

Dedicated optical metrology equipment is employed throughout the plate stacking procedure, and this is verified and complemented with X-ray measurements. The combination of these metrologies identifies the particular parts of the stack process that might fail to meet the 5 arcsecond performance. Having a proper understanding of the current limitations, good progress continues to be made towards the IXO requirement of 5 arcseconds. A detailed technology development plan was established for this purpose.

As for the previous development contracts, the next phase of SPO performance improvements will simultaneously address a number of modifications, in parallel, to the production process. A non-exhaustive list that has been defined in the Technology Development Plan recently submitted, or already implemented but not yet completed in final MM stacks, includes:

| METHOD | EFFECT |
|---|---|
| **Marangoni cleaning station** | Remove the effect of incomplete bonding areas at the plate edges. |
| **New mandrel coating** | Remove effects of Si particles sticking to the mandrel plates. |
| **Introduce Wolter 1 shapes** | Remove conical approximation effect, by introducing minor secondary curvature on parabola/hyperbola mandrels. |
| **Implement reduced inner radii** | Already demonstrated 300 mm radius of curvature stacks without excessive strain. Develop new mandrels and trial optimised membrane thickness to show innermost modules performance. |
| **Implement increased outer radii** | Shorter MMs require modified handling jigs, etc. Demonstrate the outermost MMs have desired performance. |
| **Modify stack plate numbers** | Reduced stack heights may have lower accumulated errors, but also can be mounted in optimised bracket for increased packing efficiency. Trial yield performance. |
| **Modify reference surfaces** | Improve stacking metrology and enhance QA aspects for traceability. |
| **Improved wafers** | Lower total-thickness-variations now available will result in improved wedge angles and reduce stacking errors. |

**Table 4.1.** *The future activities that have been included in the Technology Development Plan, in order to ensure the SPO reaches the required TRL by 2012, and to ensure sustained improvement in angular resolution performance.*

Dedicated activities concentrate on improving the ruggedised mounting system that fulfils the mechanical and thermal requirements for the IXO spacecraft. A first design was developed and verified by model calculations. First environmental tests to verify the models are being prepared for 2010, in a dedicated technology development activity, and it is planned to achieve TRL>5 by the end of the definition phase activities (A/B1).

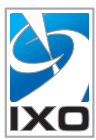 The silicon pore optics have always been measured in X-rays with a flight-like configuration. The sustained improvement has been achieved with step-wise developments of cleaning and stacking tools. Further identified development activities will allow the required 5 arcsecond resolution to be met with the new generation assembly robots under construction.

### 4.1.2   IXO optics back-up (SGO)

The NASA system level studies have developed a segmented glass optics (SGO) approach for the IXO mirror. The SGO assembly comprises highly nested grazing incidence glass mirror segments, assembled into 61 MM that are aligned and mounted onto a wheel-like primary structure. The majority of the optical system is dedicated to the low energy (0.1 to 15 keV), high-resolution (4.63 arcsec HEW below 7 keV) response. The high-energy response (10 to 40 keV) is provided by a single MM located at the centre of the mirror assembly. The mirror utilizes a Wolter I optical design consisting of 349 nested shells. The modules are arranged in three rings. The mirror segments are coated with iridium to enhance X-ray reflectivity. The segments of the central MM are coated with Pt/C multilayers to optimise their reflectivity at high energies.

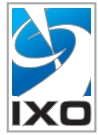

The segmented glass optics is well founded on concepts already demonstrated for *NuSTAR*, with attention now focusing on improved mounting techniques.

Segments are produced by thermally slumping 0.4 mm thick borosilicate glass (the same as that manufactured for flat panel displays) onto precisely figured fused quartz mandrels. This process is shown schematically in Figure 4.5. In order to meet the angular resolution requirement, these mandrels must be figured to a half power diameter (HPD) of 1.5 arcsec or better, a range easily achievable by commercial optics manufacturers. Each slumped segment must have 2.3 arcsec HPD or better; this is a focus of an ongoing technology effort. The complete mirror requires ~16,000 such segments plus spares. Segment mass production is being demonstrated by *NuSTAR* for which ~8,000 segments are required and the current weekly production rate is ~200.

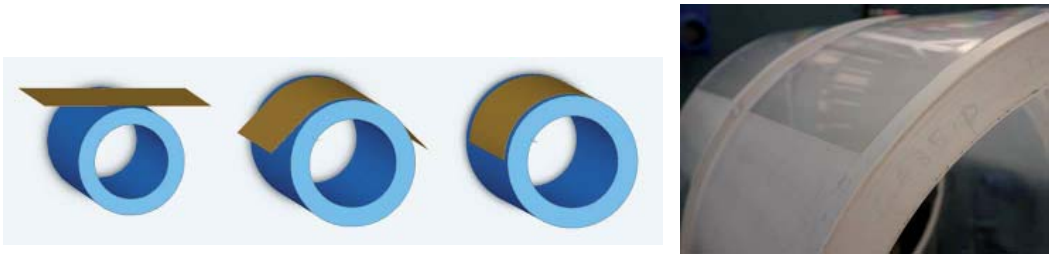

**Figure 4.5. *Left:*** *Graphic illustration of the glass slumping process.* ***Right****: this panel shows two mandrels with substrates on them that have come out of the oven.*

The methodology for precisely aligning and mounting the mirror segments into the MM housings is also the focus of an on-going technology development effort. Segments must be mounted as free of stress as possible in order to minimize the introduction of figure errors. Each segment is aligned and bonded to the MM structure independently, so no stack-up errors associated with the mirror segment bonding should develop. Forward of the segments, each MM has a stray light baffle, designed to substantially reduce the incidence of single-bounce photons from sources outside the field of view into the focal plane.

The MM housings transfer the loads to the MA primary structure, while isolating the mirror segments from the deformations of the primary structure through the use of a kinematic mounting interface. Segments are mounted to the MM structure at either 6 or 8 bonding points, depending on the segment size. At each mounting/bonding point location, a rail is fastened to the MM structure. During the segment mounting process, tabs positioned along these rails are bonded to the mirror segment. The MM housings will be constructed of material with a close CTE match to the glass comprising the segments. Materials under consideration include Titanium (Ti-6Al-4V), Kovar, and CTE-matched Carbon Fiber Reinforced Plastic (CFRP). Figure 4.6 shows schematics of modules.

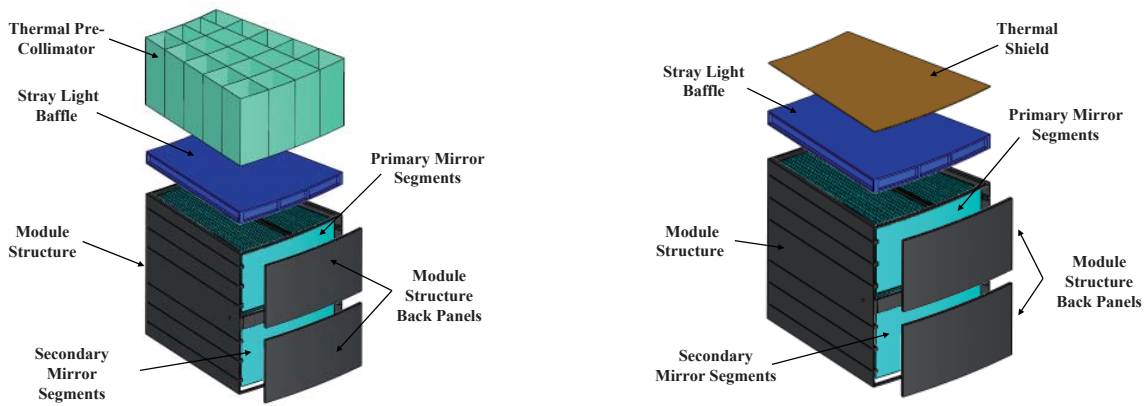

**Figure 4.6.** *Views of an outer or middle module (**left**) and an inner module (**right**). The outer and middle modules have open thermal precollimators to maximize throughput of low energy X-rays (<0.5 keV); the inner modules have thin aluminized thermal shields.*

Technology development efforts are divided between improvement of the formed segments and demonstrating a repeatable alignment/mounting approach. Forming mandrels with the required 1.5 arcsec HPD figure have been fabricated; the quality of the formed substrates is limited by roughness imparted on size scales of 0.2-2.0 cm by the release layer coating the mandrels. Steady progress has been made in improving the release layer quality: recent segment pairs have 4.5 arcsec HPD, approaching the error budget allocation of 3.3 arcsec for a mirror pair (this represents the HPD of a perfectly aligned pair). The improvement of the segments over time is depicted in Figure 4.7.

**Figure 4.7.** *Measurement by optical metrology of improvement in slumping glass shells. The error allocation for this component is ~3.3 arc-seconds. Effects of aligning and mounting shells in flight-like configuration is not accounted for.*

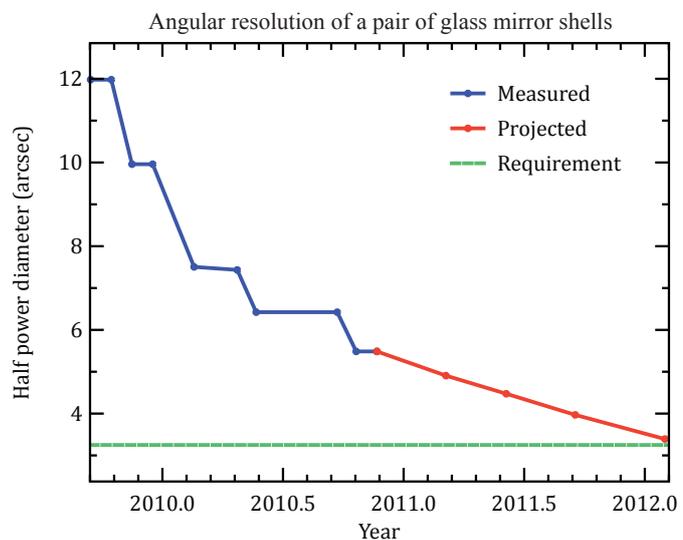

Two alignment/mounting approaches are being pursued: a "passive" approach wherein segments are rigidized in a stress-free mount, aligned, and permanently mounted using flight-like mounting surfaces so that the as-formed figure is preserved; and an "active" approach in which minor low order figure errors (slope error, roundness) are removed prior to bonding (Figure 4.8). Both approaches have demonstrated alignment of segment pairs to ~1.1 arcsec, which surpasses the allocation in the error budget. Efforts currently focus on distortion-free permanent mounting. The best permanently mounted pair to date, using the "passive" approach, has a HPD measured in the X-ray of 9.7 arcsec. This HPD is consistent with the prediction from optical metrology, indicating that the contribution to the blur of the alignment and mounting is negligible at the ~10 arcsec level. Mirror segment pairs with the best HPD to date, 5.5 arcsec, will be aligned and X-ray tested by the end of the calendar year 2010.

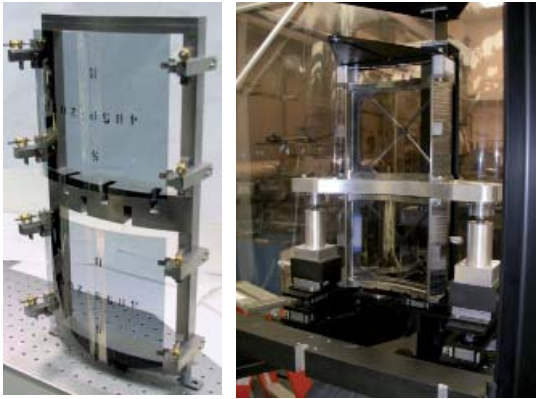

**Figure 4.8.** **Left**: *A pair of permanently bonded mirror segments in the mirror housing simulator used for demonstrating passive mounting.* **Right**: *An aligned and bonded mirror pair in the housing used for demonstrating active mounting.*

The technology development plan leading to TRL 6 entails a systematic build up of components with performance demonstrations and environmental testing until a flight-like prototype is demonstrated. Near term work concentrates on demonstrating alignment and mounting of single mirror segment pairs using flight-like fixtures. Once alignment and mounting has been demonstrated, multiple, closely spaced mirror pairs will be mounted in a medium fidelity housing to demonstrate coalignment and the absence of cross talk between sets of mirror pairs as additional pairs are aligned. The final step (TRL 6) is a high fidelity demonstration module, fully populated with a combination of aligned mirror segment pairs and dummy segments. At each step, X-ray performance measurements will be carried out, before and after a battery of environmental tests.

## 4.2   The X-ray Microcalorimeter Spectrometer (XMS)

### 4.2.1   Instrument requirements

The XMS requirements are listed in Table 4.2. A top level instrument design is shown in Figure 4.10.

| Parameter | Requirement | |
|---|---|---|
| | **Inner array** | **Outer array** |
| Energy range | 0.3 – 12 keV | 0.3 – 12 keV |
| Energy resolution:     E < 7 keV<br>E > 7 keV | $\Delta E \leq 2.5$ eV<br>$E/\Delta E \geq 2800$ | $\Delta E \leq 10$ eV<br>$E/\Delta E \geq 700$ |
| FoV | 2x2 arcmin | 5x5 arcmin |
| Good grade events | > 80% @ 50 counts/sec/pixel<br>($\Delta E$ <2.5 eV) | > 80% @ 2 counts/sec/pixel<br>($\Delta E$ < 10 eV) |
| | **Full array** | |
| Quantum efficiency     @ 1 keV<br>@ 7 keV | > 60 %<br>> 80 % | |
| Energy scale stability | 1 eV / h (pk-pk) | |
| Non X-ray background | $2 \times 10^{-2}$ counts/cm²/keV/s | |
| Total count rate all events | 10000 counts/sec/array | |
| | (assumes defocusing optics applied) 1eV / h (pk-pk) | |
| Continuous observing time | > 31 hour | |

**Table 4.2.** *XMS Instrument Requirements.*

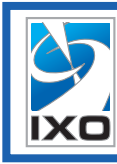

The XMS microcalorimeter instrument is a powerful X-ray "integral-field spectrometer". It provides a revolutionary new capability for plasma diagnostics.

### 4.2.2  Detectors

Two consortia are currently developing detectors for the XMS. A group including SRON (NL), INAF (I), ISDC (Switzerland), GSFC (USA), NIST (USA), ISAS (Japan) and MTU (Japan) are developing arrays using Transition Edge Sensor (TES) thermometers to sense the energy deposit in the absorbers. Starting from *Suzaku*/XRS MIS design, a second group, led by CEA in France, is developing metal-insulator based sensor (MIS) arrays exploiting the heritage of the PACS bolometers on board the Herschel satellite.

**Principle**

The operating principle of a calorimeter is shown in Figure 4.9. The detector works by sensing the heat pulses generated by X-ray photons when they are absorbed and thermalised. The temperature increase is directly proportional to photon energy and is measured by the change in electrical resistance of the sensor.

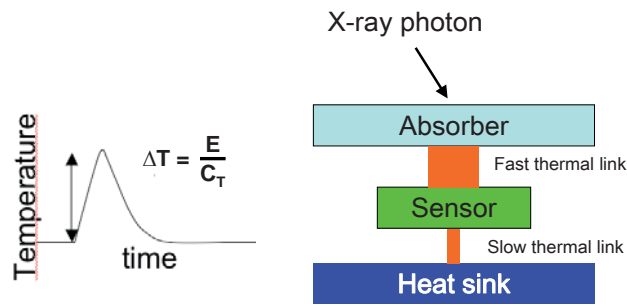

**Figure 4.9.**  *Principle of a microcalorimeter. The absorption of an X-ray photon heats both the absorber and the sensor. The resulting signal represents the total energy deposited. The system goes back slowly to its original state through a weak thermal link with a heat sink.*

**Figure 4.10.**  *Conceptual design of the focal plane assembly of a microcalorimeter, showing the various steps in the cooling system. The FPA is laid out as a cube. The inner, outer, and anti-coincidence detectors are mounted at the upper horizontal plane of it. Along the sides of the cube the cold electronics are mounted and connected to the electrical harness. The central cube is surrounded by thermal and magnetic shields at the various temperature levels. Kevlar suspension is used for thermal isolation of these temperature stages.*

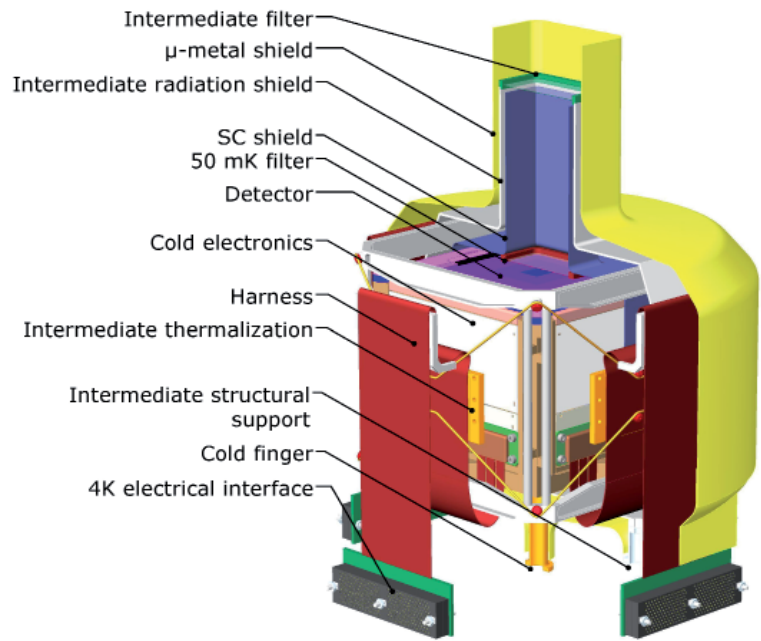

### 4.2.3  TES based calorimeters

By operating a Mo/Au bilayer TES in the range of its transition between super conducting and normal resistance, an energy resolution of <2 eV has been demonstrated. To meet the 5x5 arcmin FoV for a realistic amount of pixels and electronics, two different types of pixels are used for the inner and outer array. In the inner (2x2 arcmin) array the pixels are 300 μm squared and each is sensed by a separate TES. In the outer (5x5 arcmin) array the pixels are 600 μm and a set of four is sensed by a single TES and risetime discrimination is implemented by using a different thermal coupling between the pixels and the sensor.

For an array with 2176 TESs it is mandatory to limit the thermal harness load by using multiplexed read-out of the pixels. From three different read-out topologies the time domain multiplexing (TDM) scheme has currently been selected as this has the highest technological readiness and meets the read-out requirements (multiplex 32 pixels in a single channel). Pixels in the same column are sequentially read-out in TDM. Other, more performing topologies for the read-out are under development in which the pixels are continuously sensed and separation is based on either multiplexing in the frequency domain (using AC biasing of the pixels) or using code division multiplexing (using standard TDM technology with continuous read-out flipping the signal polarity).

The focal plane assembly (FPA) provides the thermal and mechanical support to the detector proper: three separate chips for the inner array, outer array, and the anti-coincidence detector just underneath the detector. In addition it accommodates the cold electronics and provides the appropriate magnetic shielding. A magnetic field attenuation of $1.6 \times 10^5$ has been achieved by two shields: a super conducting Nb shield and a cryo-perm shield at 4 K. The design, given in Figure 4.10, is based on flight qualified components and technology.

In Figure 4.11 the measured spectra of a Fe[55] source are shown for an 8x8 pixel array of which 2x8 pixels have been read-out by a 2x8 pixels TDM electronics. The average resolution is <3 eV with a small spread between the pixels. Scaling to the exact XMS instrument requirements (pixel size and number of pixels per read-out channel) is ongoing and no fundamental steps in the technology are required.

The time constants of the detectors (300 μs) together with the noise levels of the TDM allow for single pixel count rates of 50 c/s (inner array) with 80% of the events detected with high resolution. Insertion of a diffuser allows for observation of very bright sources with a total input count rate on the diffuser of $\sim 10^5$ c/s of which $\sim 10^4$ events/s will be detected with high resolution.

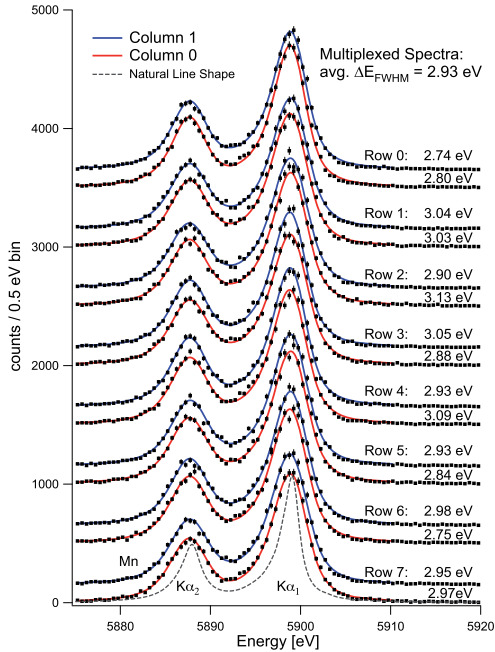

**Figure 4.11.** *Results of a 8x8 pixel array using TDM to read out 2x8 pixels. An average energy resolution of 2.93 eV for these 16 pixels has been obtained.*

| DEVELOPMENT ACTIVITY | GOAL |
|---|---|
| **32 x 32 array demonstration** | Replicate performance from smaller arrays including uniformity (requires wiring and heat sink modifications) TRL 5 by end 2010. |
| **Readout 96 channels of 32 x 32 array** | Demonstrate three units of the of time-division MUX in representative configuration by end of 2010. |
| **Demonstrate outer pixel design** | Repeat the demonstration of the readout of 4 pixels with one sensor, but on IXO representative pixel size (March 2011). |
| **Multiplex outer pixels** | Demonstrate the combination of MUX and large pixels by June 2011. |
| **High fidelity detector demonstration** | Combine all activities to meet IXO detector requirements (Sept. 2011). |
| **3-stage ADR cooler** | Deliver ASTRO-H FM (July 2012). |

**Table 4.3.** *Summary of the near term developments of the TES sensors to successively achieve TRL 5 and 6.*

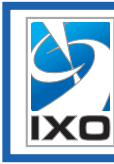 The XMS baseline of TES spectrometers already matches the required energy resolution in modest sized arrays. Developments are under way to increase the size of arrays and implement improved multiplexing techniques.

## 4.2.4  MIS based calorimeters

For the MIS based detector at a given number of pixel outputs (driven by dissipated power at 50 mK), there is a trade-off between pixel size and spectral resolution. Due to the superconducting nature of the absorber (low heat capacity); a larger pixel size can be accommodated without compromising the spectral resolution. So the MIS based camera is designed for a 5.5x5.5 arcmin² FoV with 64x64 pixels.

Building on the extensive *Herschel*/PACS heritage, the FPA is designed with modularity and testability in mind, having four exchangeable quadrants each made of a 32x32 pixel array with associated cold electronics and veto (Figure 4.12). Cosmic ray rejection is based on the detection of fast a-thermal phonons in the veto detector with similar electronics to the standard pixels. The MIS micro-calorimeters are weakly sensitive to magnetic fields and require a passive shield.

With a measured heat capacity of a MIS pixel as low as 0.03 pJ/K, an alpha (thermometer steepness) of ~5, and a noise around 2.5 nV/Hz, the target energy resolution of 2.5 eV may be met with a 32:1 (TDM) multiplexing scheme. With a 500 m pitch, the camera holds 4096 independent pixel outputs. Thanks to the moderate electrothermal feedback of the MIS sensors, the FPA needs less than 1 Watt at the 50 mK level, even taking into account the heat load from the upper stage.

A resistive ("classical") electronic concept is used for the cryogenic front-end electronics. GaAs and SiGe application-specific integrated circuits (ASICs) have been successfully developed which exhibit performances within specifications in terms of noise, frequency band pass and power consumption at low temperature.

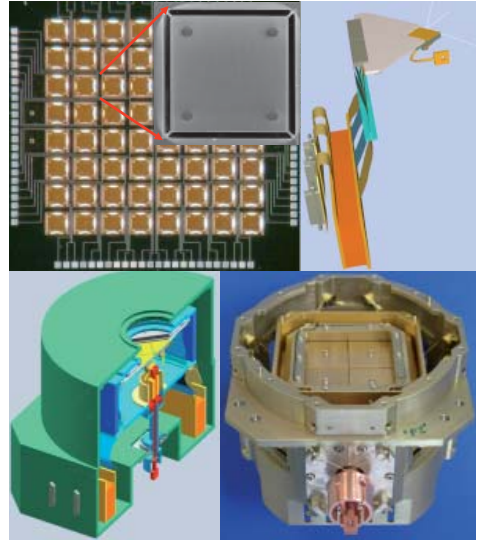

**Figure 4.12. *Top left:* *MIS array and zoomed pixel.*
*Top right:* *quadrant design with its electronics.* **Bottom
left:** *focal plane assembly design.* **Bottom right**: *SAFARI
similar design realization.*

### 4.2.5   Coolers

Operating detectors at 50 mK in space is a major challenge. Many different solutions exist at ESA, JAXA,
and NASA with varying level of flight heritage. The baseline for the XMS cooling system (Figure 4.13
and Figure 4.14) is based on the cooling system for *ASTRO-H* giving it the highest possible TRL. The
cooling chain is split into two components: the prime cooling chain for cooling the instrument from room
temperature to 4 K and the last cooling stage cooling from 4 K to 50 mK. The coolers are cryogen free,
enabling a lifetime of 10 years or more.

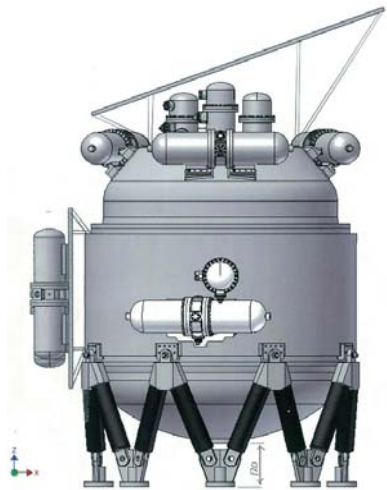

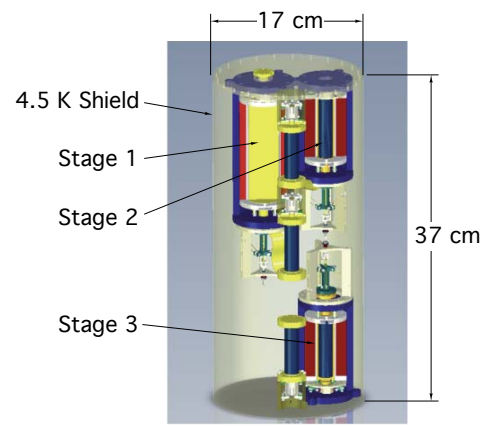

**Figure 4.13.**  *Engineering drawing of the dewar.*          **Figure 4.14.**  *Last stage cooler, based on a 3-stage ADR.*

**Prime cooling chain**
The prime cooling chain provides a continuous cooling power of 20 mW at 4 K (peak). It is a cryogen free
system and uses 2-stage Stirling coolers and $^4$He J-T coolers which all have flight heritage. In addition
to full redundancy in the drive electronics each of the mechanical coolers (compressors) are also fully
redundant. Initial estimates of the reliability are above 98%.

**Last stage cooler**
The last stage cooler provides a cooling power of 2 µW at the 50 mK level using a 3-stage adiabatic
demagnetization refrigerator (ADR), see Figure 4.14. Following the magnetization of the salt pills, cooling
power is provided by a slow relaxation of the spins in the magnetized material. An identical system (although with a different optimisation) will be flown on *ASTRO-H*. The system has a regeneration time of < 2 hours and a hold time of 31 hours meeting the instrument specifications.

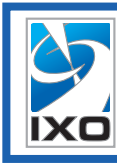

The XMS baseline cooler design is based on a modest extension of that for the cryogenic spectrometer already in development for *ASTRO-H* which is to be launched ~2014.

## 4.3  WFI/HXI

### 4.3.1  The Wide Field Imager (WFI)

The WFI is being developed by a collaboration consisting of the MPE-HLL (Germany), a consortium of German universities (ECAP, IAAT, TUD), a consortium of US universities (CfA, MIT, PSU), and partner universities in Italy (Politecnico di Milano), Great Britain (University of Leicester) and Japan (University of Osaka).

#### 4.3.1.1  Performance Requirements and Specifications

The IXO wide field imager is an imaging X-ray spectrometer with a large field of view. The purpose of the WFI is to provide images in the energy band 0.1-15 keV, simultaneously with spectrally and time resolved photon counting. Being an on-axis imaging instrument, the WFI is mounted on the Moveable Instrument Platform (MIP) and is moved into the optical path on demand. The Hard-X-ray-Imager (HXI) will be mounted onto the back of the WFI with an extra-focal distance of 2.5 cm and operates in the same MIP position as the WFI (see Figure 4.15). This allows extending the energy range of observations into the 15-80 keV range. The WFI requirements are summarized in Table 4.4.

| Parameter | WFI requirement |
|---|---|
| **Effective area** | QE @ 282 eV : 34%<br>QE @ 10 keV : 96% |
| **Readout rate** | 2.5 – 4 µs/row<br>500 – 800 frames per sec |
| **Energy range** | 0.1 – 15 keV |
| **Spectral resolution** | $\Delta E \leq 130$ eV @ 6 keV<br>3 – 5 e$^-$ ENC |
| **FoV** | 18 arcmin diameter requires 10 x 10 cm$^2$ monolithic wafer-scale detector |
| **Angular resolution** | $\leq 5$ arcsec HPD is oversampled by 100 x 100 µm$^2$ pixel size |

**Table 4.4.** *WFI Instrument Requirements.*

#### 4.3.1.2  Instrument description and configuration

The effective area of IXO in the WFI energy range is more than a factor of 20 larger than previous X-ray missions, leading to high event rates. This necessitates very high read-out rates and flexible read-out modes. Active pixel sensors (APS) based on depleted P-channel field-effect transistors (DEPFETs) are highly suited for this task, as the signal charge does not need to be transferred over macroscopic distances, but is amplified directly in each pixel of the detector. To achieve a FoV of 18 arcmin diameter and avoid WFI mounting structures in the HXI field of view, the WFI APS is integrated monolithically onto a single 6-inch wafer, covering almost the complete usable Si wafer surface (see Figure 4.16).

The device is fully depleted using the sideways depletion principle, allowing signal charges generated by radiation entering from the backside of the detector to be detected efficiently. Backside illumination allows the APS to have a geometrical fill factor of 100%. Furthermore, the use of a very thin entrance window leads to a very good quantum efficiency in the energy range below 1 keV. The large thickness of the fully depleted bulk provides for high quantum efficiency, even beyond 10 keV. A beneficial side-effect of backside-illumination is self-shielding. As the radiation absorbed in the bulk does not reach the amplifying structures on the front-side, radiation hardness is intrinsically improved.

As the large effective area X-ray mirror system does not only focus X-rays but also visible and UV light, care must be taken to avoid problematically high optical photon loads. Therefore, an optical blocking filter system consisting of a thin Al-layer on top of an UV blocking system consisting of a Silicon Oxide/Nitride multi-layer coating is foreseen.

In order to achieve a roughly 5-times oversampling of the expected PSF of IXO's X-ray mirror system, the WFI baseline design foresees a pixel size of 100x100 $\mu m^2$. Therefore, 1024x1024 pixels are needed to fully cover the 6-inch wafer. To facilitate monolithical integration, each corner of the detector has an area of the size of 128x128 pixels that is left free. As this area lies outside of the nominal FoV of the X-ray optics and is shaded by the baffle system, these free corners do not limit the usable FoV.

To achieve the high speed required, the detector is subdivided into two hemispheres, read out in parallel by 8 read-out ASICs per hemisphere. At the aimed-for read-out speed of 2.5-4 $\mu s$ per row, the raw data-rate produced by the WFI is more than 10 Gbit/s, necessitating a very efficient on-board data-reduction and compression scheme. WFI incorporates a multi-staged data-reduction pipeline based on several field-programmable gate arrays that reduces the raw data to a nominal data-rate of less than 450 kbit/s.

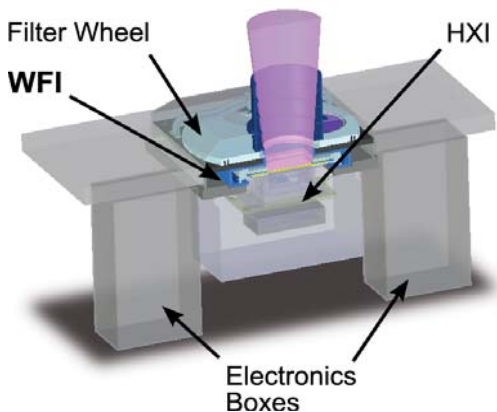

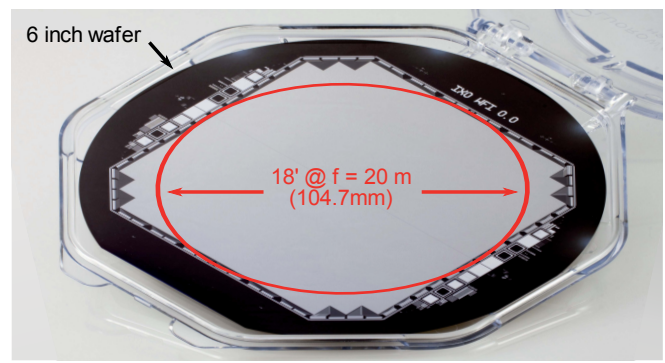

**Figure 4.15.** *Overview over the WFI / HXI mechanical layout. The X-ray radiation is entering from the top. X-rays in the 0.1-15 keV band are detected by WFI. At higher energies the WFI APS becomes transparent, and X-rays are detected by HXI.*

**Figure 4.16.** *Mechanical model of the WFI wafer-scale APS. Overplotted in red is the nominal 18' FoV. The detector is monolithical in nature and has 100 % fill factor.*

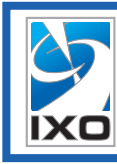 IXO-WFI is a maturing technology derived from *XMM-Newton* EPIC and developments for *BepiColombo*. It provides essential wide field survey capabilities for observatory workhorse Science and already offers the required resolution and count rate capability.

### 4.3.1.3    Technology Development Status

The current prototypes of DEPFET matrices and the associated bread-boards demonstrate high performance, already meeting all WFI science requirements addressable with this detector size (see Figure 4.17, Figure 4.18 and Figure 4.19). A demonstrator of representative size is currently under production and will show technological readiness in all aspects before the end of the definition phase.

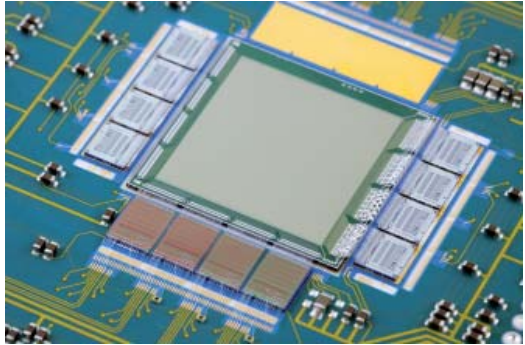

**Figure 4.17.** *DEPFET matrix of 256x256 pixels of 75x75 μm² size. The complete matrix is read out by four ASTEROID ASICs operating in parallel.*

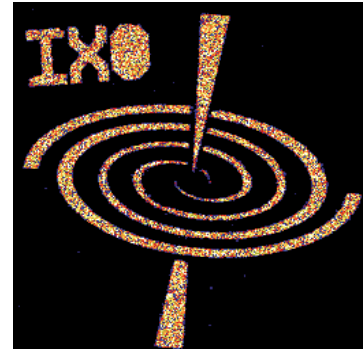

**Figure 4.18.** *X-ray image taken by the DEPFET matrix shown in Figure 4.17, demonstrating the high resolution of this prototype.*

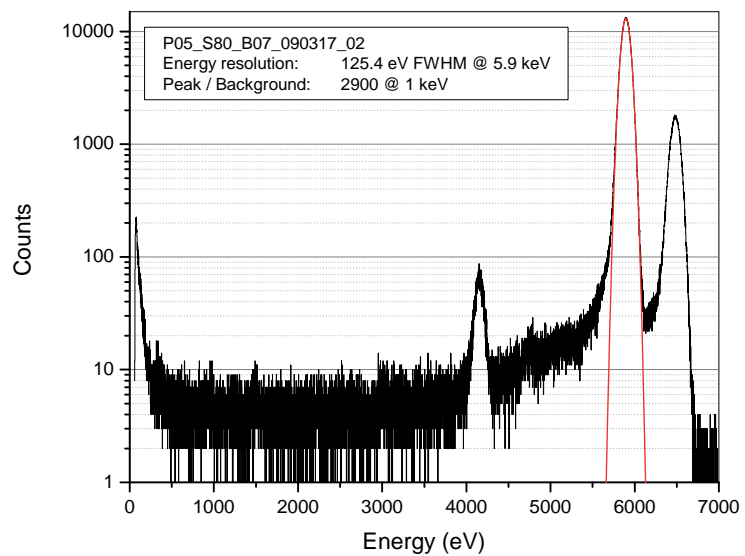

**Figure 4.19.** *Backside-illuminated ⁵⁵Fe X-ray spectrum taken with a DEPFET matrix. The resolution of 125.4 eV is nearly Fano-limited, and the measurement shows a good P/B ratio of 2900. The spectral response to the X-ray line source is very clean.*

## 4.3.2    Hard X-ray Imager (HXI)

HXI has been studied by a JAXA led team including contributions from the Universities of Tokyo, Hiroshima, Waseda, Saitama, and CEA France.

### 4.3.2.1    Performance Requirements and Specifications

The Hard X-ray Imager (HXI) is mounted on the extrafocal side of WFI, extending the energy coverage up to 40 keV (see Figure 4.20). The two instruments are mechanically coupled and electrically independent. They share the same focal plane, providing the simultaneous wide-band (0.1-40 keV) spectro-imaging capability to IXO. The HXI has an energy resolution better than 1 keV (FWHM) and a FoV of 8x8 arcmin². The instrument requirements are summarized in Table 4.5.

| Parameter | HXI requirement |
|-----------|-----------------|
| **Detector Type** | Si (0.5 mm thick) and CdTe (0.5-0.75 mm) double-sided strip |
| **Strip pitch** | 250 μm (=2.6") for both sides |
| **Energy range** | 10 – 40 keV |
| **FoV** | $50 \times 50$ mm$^2$ |
| **Energy Resolution** | $\Delta E < 1$ keV (FWHM) |
| **Non X-Ray Background** | $5 \times 10^{-4}$ counts keV$^{-1}$ cm$^{-2}$ s$^{-1}$ roughly flat |
| **Timing accuracy** | $< 10$ μs relative, $< 100$ μs absolute |
| **Typical / Max telemetry** | 11 kbs$^{-1}$ (256 kbs$^{-1}$ max.) |

**Table 4.5.** *HXI Instrument Requirements.*

### 4.3.2.2    Instrument description and configuration

The HXI system consists of two boxes, the sensor part (HXI-S) and electronics part (HXI-E). The system is designed to be compact, with a total weight of 21.5 kg, and a power of 43 W. The sensors will consist of 2 layers of double-sided Si strip detectors (DSSDs, 0.5 mm thick) and a newly developed double-sided strip CdTe detector (DS-CdTe, 0.75 mm thick) surrounded by a 2 cm thick BGO ($Bi_4Ge_3O_{12}$) active shield. The 8x8 arcmin$^2$ field of view is provided by a 2x2 array of 2.5 cm-large DSSD/DS-CdTe. Each imager device has 96+96 strips with a pitch of 250 microns. The position resolution of 250 μm corresponds to a pixel size of 2.5 arcsec. The DSSD works as a low-background medium energy detector almost free of activation working up to 30 keV, and also as a low-Z passive shield layer for WFI. In place of the DS-CdTe, a pixel CdTe array is also considered as an option. Graded-Z passive shield combined with plastic scintillators is also in consideration for the active shield. All these options are of modular design, and designed not to significantly affect the overall detector structure.

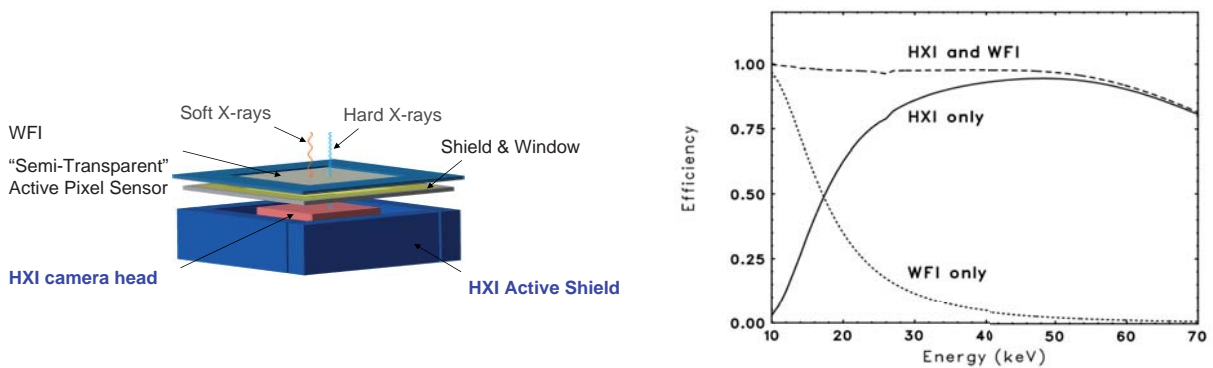

**Figure 4.20.** *Concept of HXI combined with WFI (**left**), and its efficiency (**right**).*

Major technology items of the HXI rely on the heritage of *ASTRO-H* and *Simbol-X* projects to ensure good progress. Baseline design is basically a modified *ASTRO-H*/HXI. DSSD, DS-CdTe and BGO system as well as the electronics box are of *ASTRO-H* heritage. Technology used for the *Simbol-X* development activity is also utilized and lessons learned from *NuSTAR* will be adopted.

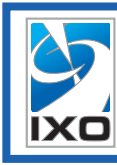

IXO-HXI provides a hard X-ray capability essential for understanding reflection phenomena and obscuration around supermassive black holes. The active pixel technology is being matured for missions such as *ASTRO-H* and *NuSTAR*.

### 4.3.2.3 Technology status of the imager part

DSSD/DS-CdTe (see Figure 4.21) are based on technologies developed for *ASTRO-H*/HXI, which is using 32x32 mm$^2$ devices with a strip pitch of 250 microns. Many DSSD models with a strip pitches of 75-400 microns are already working well in the labs. There are also several models of DS-CdTe. The signal from a photon absorbed in the detector is accumulated via strip electrodes diagonally oriented, and fed into analogue chains implemented in the ASICs. The IDeF-X technology (*Simbol-X* heritage) is considered as a base-line, while the VATA technology (of *ASTRO-H*) is a back up. Active shield technology is also in good progress. These are brand new, compact and high performance technologies with good technical readiness level.

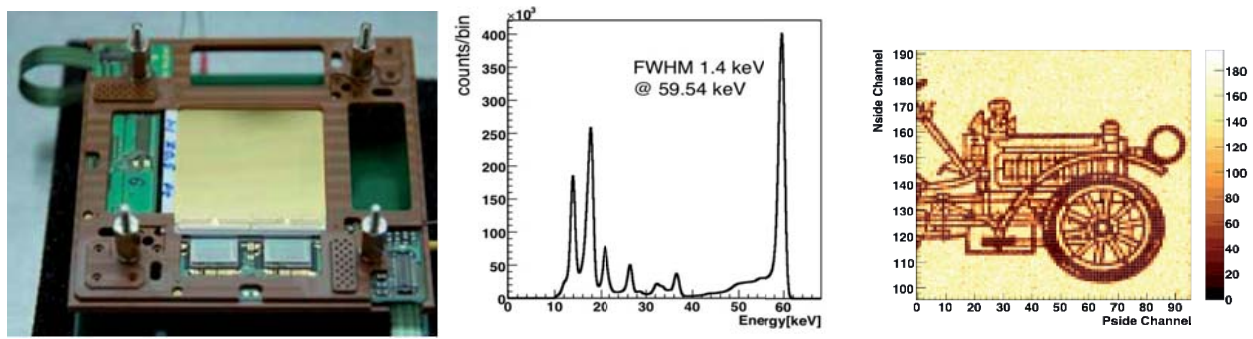

**Figure 4.21.** *Left*: Photo of a 25x25 mm$^2$ large DS-CdTe detector, with 400 um strip pitch. **Middle**: Energy spectrum obtained with a 13x13 mm$^2$ DS-CdTe. Strip pitch is 400 um, and operating temperature is -20°C. **Right**: Shadow image of 241Am source obtained with 39x39 mm$^2$ DSSD device.

## 4.4 The High-Time Resolution Spectrometer (HTRS)

### 4.4.1 Performance Requirements and Specifications

The HTRS was studied by an international consortium led by the French Space Agency and the Centre d'Etude Spatiale des Rayonnements (Toulouse). The hardware consortium involves the Observatoire de Strasbourg, the MPE-HLL (& PN-Sensor GmbH), the University of Geneva, the University of Tubingen, the University of Erlangen-Nuremberg and Politecnico di Milano.

For a number of topics of the IXO science case the capability is required to observe the strongest X-ray sources of the sky (black hole X-ray novae, type I X-ray bursters, magnetars, etc), with count rates exceeding several hundred thousands up to a few million counts per second (i.e. sources brighter than a few 100 milliCrab up to several times the Crab). This requires the HTRS capability to provide a good spectral resolution (about 150 eV at 6 keV) simultaneously with sub-millisecond timing, low deadtime and low pile-up. The top level HTRS requirements are listed in Table 4.6.

| Parameter | HTRS Requirement |
|---|---|
| **Energy range** | ~ 0.3 - 15 keV |
| **Time resolution** | 10 µsec |
| **Energy resolution at 6 keV** | ~ 150 eV |
| **Count rate capability** | > 10 Crab |
| **Deadtime & pile-up** | < 2% at 1 Crab |
| **Crab count rate** | ~170000 counts/s |

**Table 4.6.** *HTRS Instrument Requirements.*

## 4.4.2   Instrument description and configuration

In order to meet its performance requirements, the HTRS is based on a monolithic array of silicon drift detectors (SDDs) with 31 cells in a circular envelope and a sensitive volume of 4.5 cm² x 450 µm. The SDD principle uses fast signal charge collection on an integrated amplifier by a focusing internal electrical field. It combines a large sensitive area and a small capacitance, thus facilitating good energy resolution and high count rate capability. The HTRS, which will be looking at bright and unconfused point sources, is a non-imaging device. It will be operated out of focus, in such a way that the focal beam from the mirrors is spread almost uniformly over the 31 SDDs, as to increase the overall count rate capability of the instrument.

The expected performance of the HTRS in terms of the fraction of good events and the energy resolution as a function of countrate is given in Figure 4.22 and Figure 4.23, respectively. The HTRS is made out of two assemblies: the focal plane assembly, which includes the detector unit and the filter wheel, and the electronics box, which includes the pre-processing unit, the data processing unit, and the power control and distribution unit. The front-end electronics is located close to the detector array and is made of 4 high-speed analog ASICs that process the 31 detector signals. The ASICs send their output signals to the pre-processing unit. The shaping of the detector signals and the detection of the events are partially done in the readout electronics and in the pre-processing unit. For each detected photon, the pre-processing unit provides the raw data (amplitude and time) to the data processing unit. The data processing unit's main tasks are: event processing, data compression and data formatting, distributing and executing commands from the spacecraft. The filter wheel protects the detector, reduces the optical loading on the SDD array and enables instrument in-flight calibration with a radioactive source.

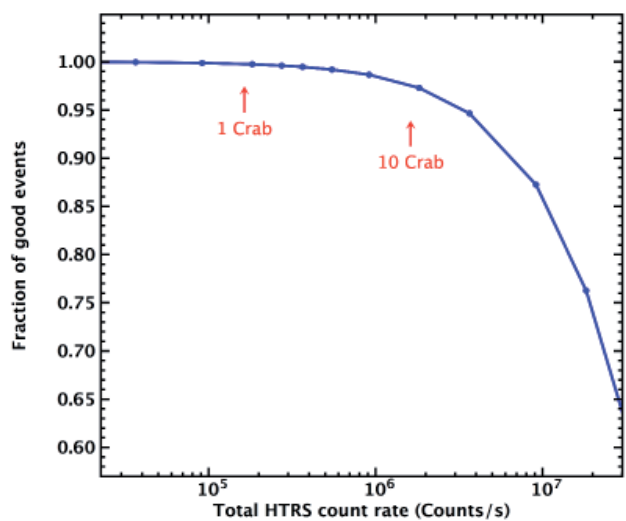

**Figure 4.22.** *Monte Carlo simulations of the HTRS throughput performance as a function of the HTRS total count rate (cumulated over the 31 Silicon drift detectors). Good events are events for which both the energy and arrival time are correctly measured by the readout electronics. This figure shows that thanks to its high speed readout electronics, the fraction of good events for a very bright source of 10 times the intensity of the Crab is still larger than 98%.*

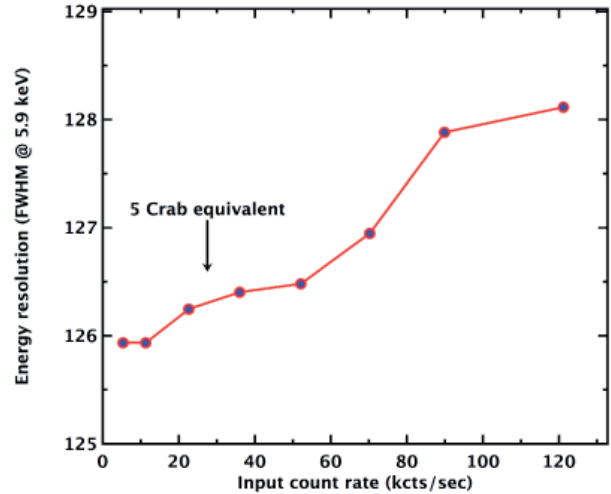

**Figure 4.23.** *Measured high rate performance of the silicon drift detector of the HTRS. The energy resolution is given in terms of FWHM at 5.9 keV. The HTRS count rate equivalent to a 5 Crab source is also indicated. In the range of count rates sampled by the HTRS, the energy resolution is stable within less than 1%.*

The CAD design of the HTRS focal plane assembly, together with the layout of the SDD array are shown in Figure 4.24.

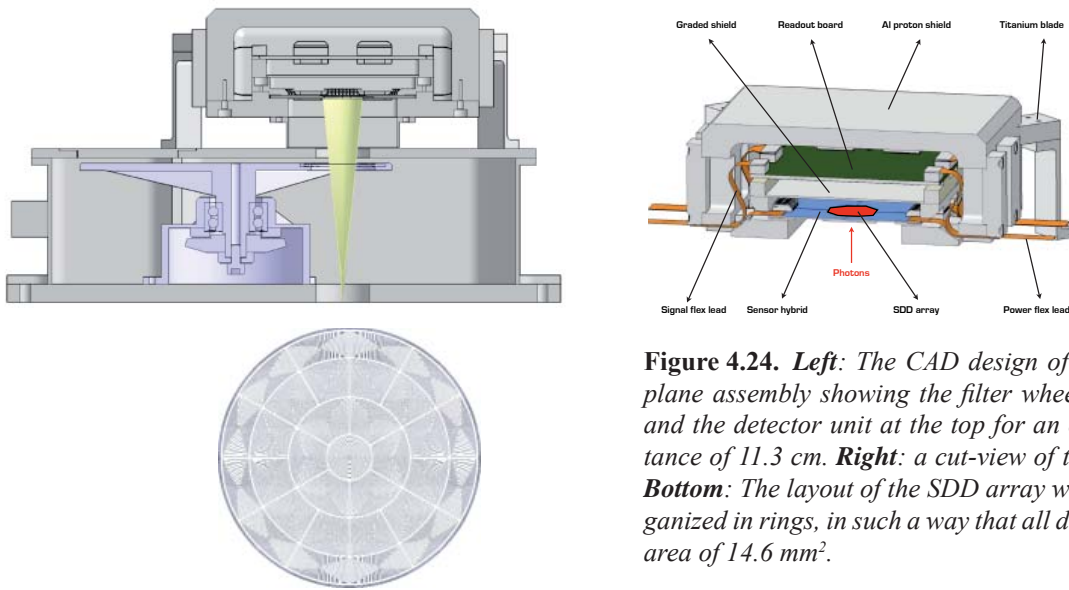

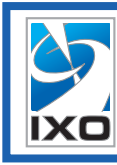

**Figure 4.24.** *Left: The CAD design of the HTRS focal plane assembly showing the filter wheel at the bottom, and the detector unit at the top for an out-of-focus distance of 11.3 cm. Right: a cut-view of the detector unit. Bottom: The layout of the SDD array with 31 diodes organized in rings, in such a way that all diodes have equal area of 14.6 mm².*

> IXO-HTRS is a mature technology from high energy physics applications and offers unprecedented count rate capability to observe dynamics around the most extreme objects in the Universe.

## 4.5 The X-ray polarimeter (XPOL)

The XPOL studies have been pursued by an Italian consortium led by INFN (Pisa) and IASF (Rome).

### 4.5.1 Performance Requirements and Specifications

XPOL provides IXO with the capability of performing polarimetry in the energy range 2-10 keV simultaneously with imaging (with resolution of about 5 arcsec), spectroscopy ($\Delta E/E \sim 20\%$ at 6 keV) and timing (at the level of µs). The basic performance specification is to enable polarimetric measurements to 1% Minimum Detectable Polarization (3σ) for a 1 mCrab source. The key characteristics of XPOL are reported in Table 4.7.

| Parameter | XPOL Requirement | Parameter | XPOL Requirement |
|---|---|---|---|
| MDP | 1% for 1 mCrab | Pixel size | 50 μm, each with an independent electronic chain |
| Energy range | 2 – 10 keV | Pixel pattern | 300 x 352, hexagonally arranged |
| Energy resolution | 20% at 6 keV | Pixel noise | 50 electrons ENC |
| Field of view | 2.6 x 2.6 arcmin² | Pixel full scale range | 30000 electrons |
| Angular resolution | 5" | Readout | Self-triggering, fetch out of the window enclosing hit pixels |
| Timing resolution | 5 μs | GEM thickness | 50 μm |
| Dead time | 10 μs per event | GEM hole pitch | 50 μm in a triangular pattern |
| Baseline mixture | Helium 20% and DME 80% at 1 bar | GPD Temperature | 5-20°C ± 2°C, maintained by a Peltier managed by XPOL |
| Detector size | 15 x 15 mm² | Telemetry | Max 0.84 Mps including HKs |

**Table 4.7.** *XPOL Instrument Requirements.*

## 4.5.2   Instrument description and configuration

XPOL is based on the gas pixel detector (Bellazzini et al. 2006) which is an advanced evolution of the micropattern gas chamber where the multi-anode read out is fully pixellated. Photons pass a thin entrance window and are absorbed in a gas cell which is used as detection/drift gap. The photoelectrons produce an ionization pattern in the gas (track), which is drifted by a parallel electric field to a gas electron multiplier (GEM) that multiplies the charge without changing the shape. The signal, amplified by a factor ~500, is collected by a finely subdivided pixel detector which eventually provides a true 2-dimensional image of the track. The small pixel pitch (50 μm) allows to resolve the track and to reconstruct the initial direction of the photoelectron, which is related through a cos² dependency to the direction of polarization, and the impact point (see Figure 4.25). The sum of the signals is proportional to the photon energy. A second spectroscopic signal comes out from the GEM upper face.

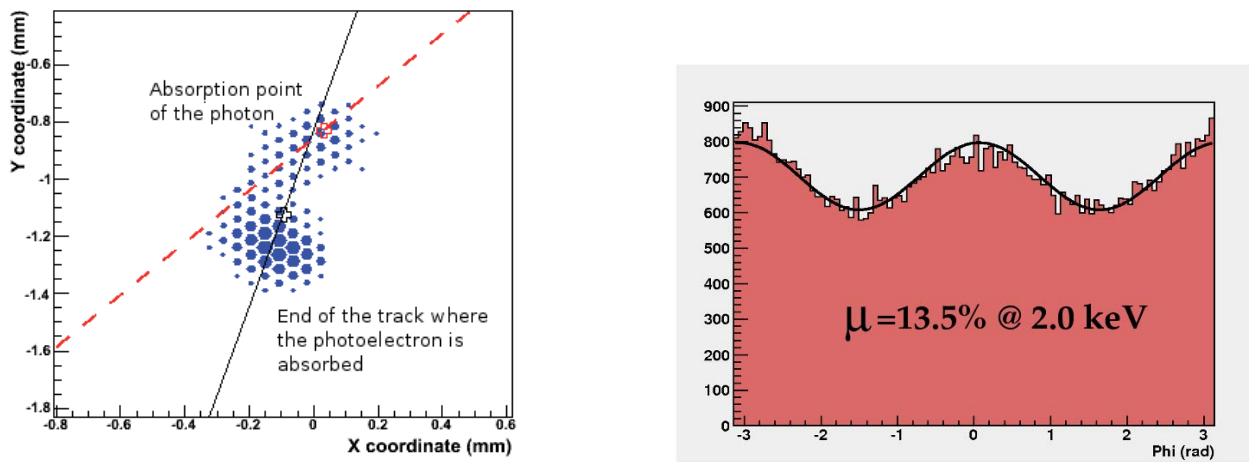

**Figure 4.25.** *Actual detection of a polarized signal.* **Left**: *Digitized charge magnitude on individual pixel elements. Black line is first-pass emission angle reconstruction, red-dashed line is second-pass reconstruction after removal of the Bragg peak at the end of the track.* **Right**: *Measured modulation factor with calibration target.*

XPOL is composed of the FPA (see Figure 4.27), located on the MIP, and of the control electronics (CE) located anywhere on MIP or FIP. The FPA is further subdivided in the gas pixel detector (GPD, Figure 4.26), the filter wheel (FW), and the back end electronics (BEE).

**FPA: GPD and FW.** The GPD in XPOL has a 50 μm thick Beryllium window and the gas cell is sealed and it is 1 cm thick. The baseline mixture is 20% Helium and 80% DME (or $(CH_3)_2O$) at 1 bar which provides the best sensitivity among tested mixtures in the 2-10 keV energy range. The GPD is in an advanced development phase. Prototypes are working in the laboratory since 2006 (see Figure 4.26) and only minor changes are needed to adapt the current ASIC to XPOL. Two unpolarized sources will be used to verify the gas gain and the absence of systematic effects throughout the whole duration of the mission. Moreover, the polarized X-ray source is designed to obtain highly polarized radiation at two energies. This will allow to check for any change in the modulation factor. Together with calibration sources, the filter wheel has an open/closed position, a Beryllium window to decrease the flux of exceptionally bright sources by a factor 30 at 2 keV, and a diaphragm to exclude bright sources in the field of view.

The BEE contains the interface electronics which is dedicated to **(1)** control and power the ASIC, **(2)** select the hit pixel and convert the analog charge content in digital signals, **(3)** tag the time of the event by reading out the signal of the GEM. The BEE contains the high voltage power supply to bias the GPD. The CE is the data handling unit which receives the raw data from the BEE and routes them to the spacecraft. In case of bright sources, when the count rate is above 950 counts/sec, the CE performs on-board the first scientific analysis on data. The direction of photoelectron emission, the impact point, the energy of the event and a quality parameter are calculated in real-time and only these data are sent to the spacecraft. This allows for a telemetry reduction of a factor 11.

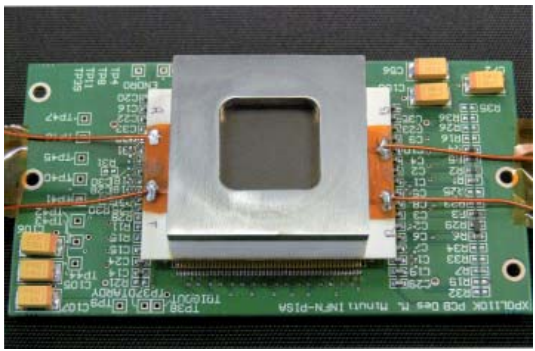

**Figure 4.26.** *The current prototype of the GPD.*

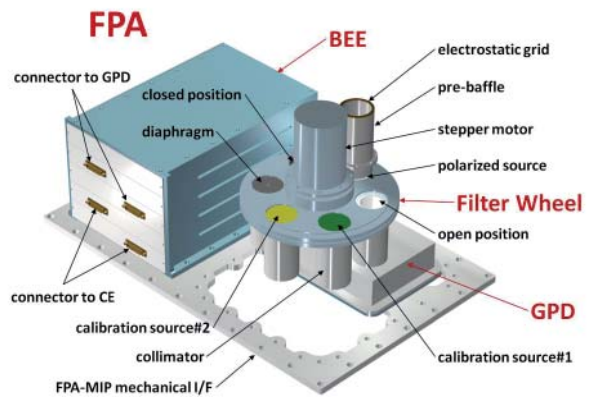

**Figure 4.27.** *Preliminary configuration for the XPOL FPA.*

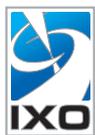

IXO-XPOL is a mature technology that offers the potential for breakthrough for the first time allowing polarimetric measurements at X-ray energies.

## 4.6   The X-ray Grating Spectrometer (XGS)

### 4.6.1   Performance Requirements and Specifications

The XGS is a wavelength-dispersive spectrometer for high-resolution spectroscopy of point sources in the energy band between 0.3 and 1.0 keV. The XGS is complementary to XMS in that it has higher spectral resolution at the longer wavelengths. Its top level requirements are listed in Table 4.8.

| Parameter | XGS requirement |
|---|---|
| Effective area | > 1000 cm² (0.3 – 1.0 keV) |
| Energy range | 0.3 – 1.0 keV |
| Spectral resolution | R = E/ΔE ≥ 3000 |

**Table 4.8.** *XGS Instrument Requirements.*

The instrument relies on a set of gratings placed between the X-ray mirror and a camera. Two concepts are being developed that will be detailed next. Both rely on the improved wavelength resolution obtained by sub-aperturing, whereby the gratings span a small fraction of the X-ray telescope in azimuth angle, exploiting the mirror modules' asymmetric PSF. The cameras are based on CCD arrays with very thin optical entrance filters to allow good efficiency at the lower X-ray energies. This, together with the rather large integration time, makes the XGS instrument most sensitive to optical straylight and sets the requirement on IXO at $<2\times10^8$ photons s$^{-1}$ cm$^2$. The intrinsic energy resolution of the CCDs is used to discriminate the different overlapping diffraction orders.

### 4.6.2   Critical Angle Transmission X-ray Grating Spectrometer (CAT-XGS)

#### 4.6.2.1   Instrument description and configuration

The CAT-XGS consists of a lightweight two-sector array of alignment-insensitive transmission gratings just aft of the flight mirror assembly (FMA), and a camera with a linear CCD array on the fixed instrument platform. The camera is laterally offset from the telescope focus, allowing soft X-ray spectroscopy at all times and simultaneously with the operation of any of the focal plane instruments on the MIP. The CAT-XGS has extensive heritage from the *Chandra* High Energy Transmission Grating Spectrometer, the *XMM-Newton* Reflection Grating Spectrometer, the *Chandra* Advanced CCD Imaging Spectrometer and the *Suzaku* X-ray Imaging Spectrometer. It is being developed by a consortium led by MIT that includes the Pennsylvania State University, Osaka University and NASA GSFC. The recently developed blazed CAT grating facets (Figure 4.28) disperse soft X-rays with high efficiency into a limited angular range and onto the diffraction-order-sorting camera. The expected grating dispersion efficiency in the various orders is shown in Figure 4.29.

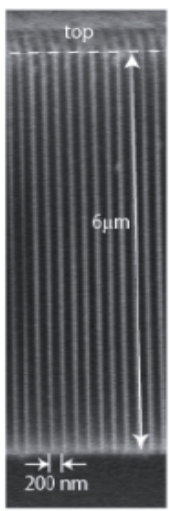

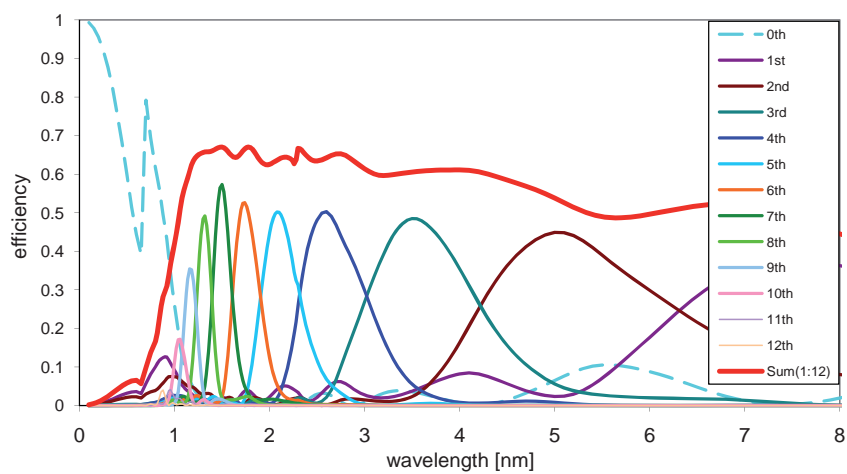

**Figure 4.28.** *Scanning electron micrograph of a cleaved cross section through a 200 nm-period CAT grating.*

**Figure 4.29.** *Theoretical diffraction efficiency of a 200 nm-period silicon CAT grating as a function of wavelength for diffraction orders 0 (contributes to effective area at the telescope imaging focus) through 12, and for the sum of orders 1-12.*

Each of the two azimuthally opposed grating sectors covers 30 degrees in azimuth and the outermost 22 MMs in the radial direction see Figure 4.30. The innermost MMs that provide most of the effective mirror area for imaging and spectroscopy at higher energies are left uncovered. Each sector consists of a grating array structure (GAS) that places 244 flat 50x50mm² grating facets on the surface of a Rowland torus that also contains the telescope focus and the camera CCD array. At higher energies CAT gratings are highly transparent (the silicon grating bars are only 6 μm thick) and simply transmit most X-rays towards the imaging focus.

The CAT design takes advantage of the combination of anisotropic scattering from the grazing-incidence mirrors and the azimuthally limited mirror coverage ("sub-aperturing"), which leads to a narrowing of the projected telescope PSF along the spectrometer dispersion axis. This, together with the long dispersion distance available with the CAT grating location just behind the FMA (Figure 4.30), provides the required resolving power with ample margin as validated by ray-tracing the full FMA/CAT-XGS system.

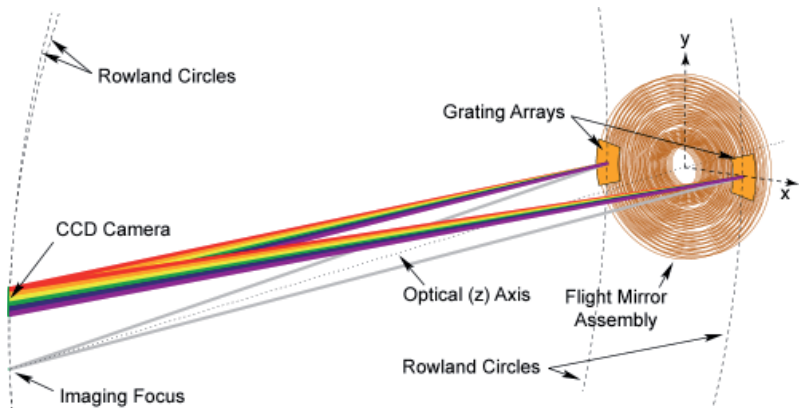

**Figure 4.30.** *Schematic of the optical design of the CAT-XGS.*

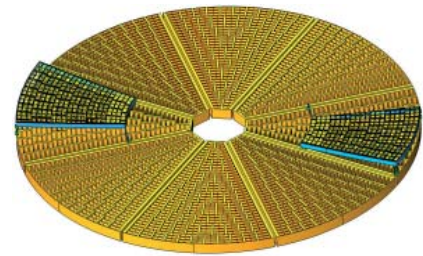

**Figure 4.31.** *CAD model of grating arrays mounted to the SPO FMA.*

The camera assembly holds the CCD array, and is connected to the separate detector electronics assembly (DEA). The CCDs are modified 25x25 mm² MIT Lincoln Laboratory CCID41 back-illuminated frame-transfer CCDs with 24x24 μm pixels (see Figure 4.32). The CCD focal plane is passively cooled to ~100 °C via heat pipes connected to a radiator. The camera housing provides shielding and accommodates fiducial lights for aspect reconstruction. The DEA holds two power boards, and each of its 16 detector processing boards services two CCDs. Digital data from the DEA is routed to the digital processing assembly (DPA), which processes the image data output from the DEA to extract X-ray events.

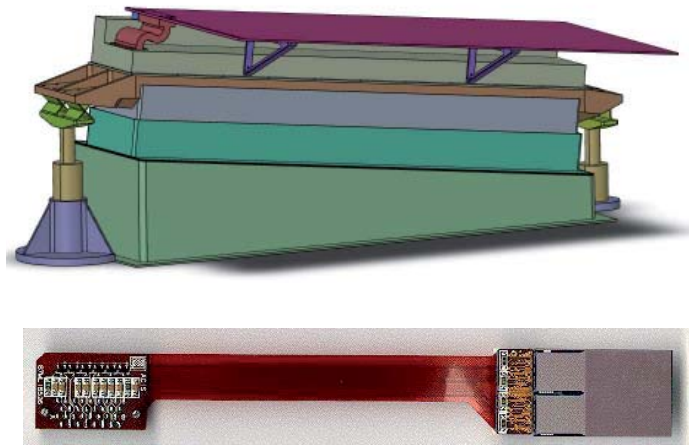

**Figure 4.32.** *Top*: *CAD model of CAT-XGS camera assembly. Bottom*: *CAT-XGS CCD packaging concept.*

### 4.6.3   Off-Plane X-ray Grating Spectrometer (OP-XGS)

The OP-XGS has been studied by a consortium including the Open University (UK) and in the US by the Universities of Iowa and Colorado.

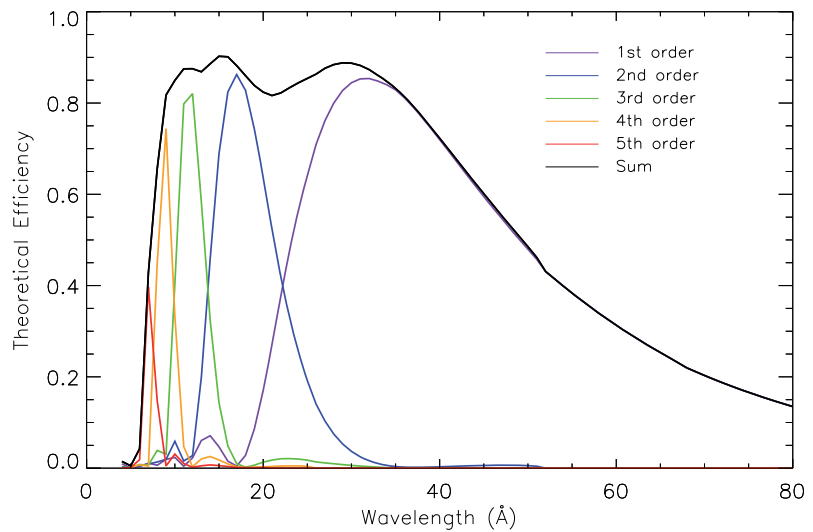

**Figure 4.33.** *Theoretical grating dispersion efficiency as a function of wavelength. Individual orders are shown in colour with the sum in orders as the black solid line. The modelled groove profile has been blazed to optimise 35 Angstroms in first order.*

#### 4.6.3.1   Instrument description and configuration

An example layout is shown in Figure 4.34. The OP-XGS comprises a grating array mounted upon a rigid lightweight tower which itself is mounted on the fixed instrument platform. The grating array diffracts a portion of the beam (approximately 10%) into several arcs, or spectra, into a fixed CCD camera, also mounted on the FIP. The camera consists of an array of CCDs with associated electronics, thermal control and radiation, stray light and contamination shielding.

The grating system consists of a grating array made from 6 separate, yet identical, modules as visible on the left detail of Figure 4.34 (only 4 shown). These grating modules are mounted to the top of the grating tower along with an independent thermal control system. Each of the 6 modules contains 23 gratings that differ only by their width and are co-aligned to form a single spectrum per module. The grooves on these gratings lie nearly parallel to the direction of the incoming X-rays (the off-plane mount) and exhibit a radial, blazed, high density profile that allows them to obtain high throughput and high resolution. The fabrication of the gratings is achieved through an industrial process which has been well established and therefore represents a low risk and manageable technology development/procurement.

The CCD camera draws upon the significant heritage of the X-ray cameras which are successfully employed on *XMM-Newton*/RGS and EPIC, and *Chandra*/ACIS. Due to the spectra having very high resolution, it is neither practical nor desirable to superimpose the outputs of the 6 grating modules onto a single spectrum. Instead, 6 separate spectra are projected onto the CCD camera. The overlapping spectra provide a high degree of redundancy in the design, where individual CCDs or their drive electronics can be lost without significantly impacting the science data return.

A number of advantages arise from the Off-Plane geometry in this configuration: It particularly has extended performance out to 1500 eV due to the efficiency of the gratings and CCDs at the higher energies, potentially provides resolving powers up to 7000, does not scatter into the focal plane instruments, and utilises a compact camera design.

The CCDs are mounted on an optical bench controlled at -80 °C via a radiator and heaters. 14 CCDs are arranged in two arcs to collect three spatially resolved spectra. Four further CCDs monitor four of the six zero order reflections from the gratings to provide wavelength calibration; this enables the instrument to self-calibrate the wavelength scale and takes out the effects of pointing errors. All CCDs are custom variants based upon proven e2v technologies processes; they are back-illuminated, 33x22 mm$^2$ devices operated in frame transfer mode and read out at a rate almost 2 orders of magnitude faster than RGS/XMM. The design is underpinned by relatively mature technology for the key components of gratings and CCDs.

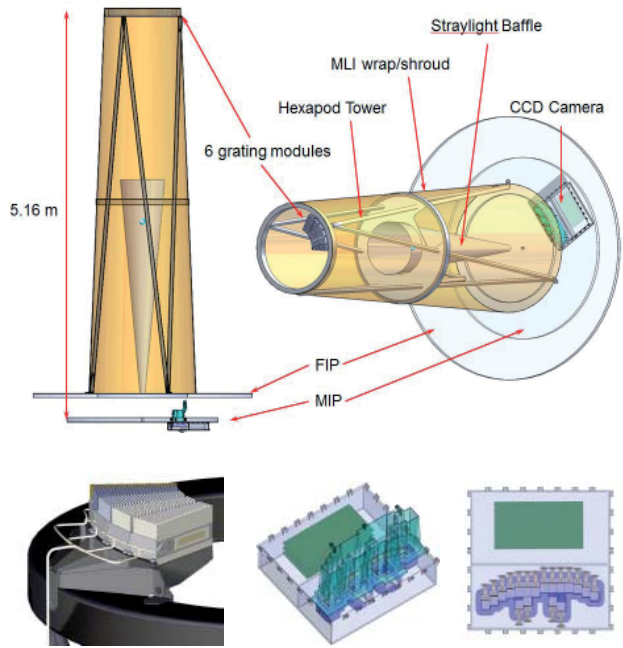

**Figure 4.34.** *Top: Illustrative layout of the OP-XGS instrument. 6 identical grating modules on a lightweight tower structure intercept ~10% of the beam and project 6 spectra onto a CCD Camera. The tower structure might incorporate system elements such as common straylight baffle, particle diverters, MLI wrap/shroud etc. to achieve system level mass savings. Bottom: Grating module detail (only 4 modules shown) and CCD Camera detail.*

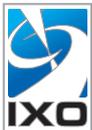

IXO-XGS can be implemented by either of two novel designs and offers two orders of magnitude performance improvement (Area x resolving power) compared with *XMM-Newton* and *Chandra*. XGS will be the most important instrument for detecting the hottest phase of the missing baryons.

## 4.7    Instrument Accommodation

The baseline payload includes 4 instruments locatable at the telescope focus: the XMS, the combination of WFI/HXI, the HTRS and XPOL. In addition there is the fixed dispersive grating spectrometer. All instrument boxes (except the grating camera) are mounted on a movable instrument platform. Only power and data (Spacewire) cables need to be routed from MIP to the spacecraft instrument module (IM). Spacewire routers and a power control & distribution unit (PCDU) are also foreseen on the IM to minimize the data and power harnesses to the service module (SVM). The instrument module includes a sunshield, a common straylight baffle and a particle diverter. A fixed straylight baffle protects the instruments from optical as well as X-ray straylight up to 70 keV. The structure of the baffle will also be used to fix the magnetic particle diverters. These magnets will deviate electrons and protons up to 25 and 75 keV respectively from the field of view of the instrument detectors. Higher energy particles, difficult to divert, will be discriminated in the instruments as they will deposit energies in their detectors in excess of the useful X-ray band-pass or be intercepted by anti-coincidence detectors.

The major issues related to the instrument accommodation that were identified and tackled during the study were:

- overall module mass,

- precision and stability of the pointing and alignment (deployment mechanism, thermo-elastic distortions) – the 1.5" astrometric pointing accuracy translates to 150μm at the focal plane, 20 m away from the mirror on a relatively flexible structure,

- space availability on MIP to accommodate all boxes,

- thermal control (large dissipation in small volume), cryogenics required for the microcalorimeter spectrometer (XMS),

- straylight rejection and

- contamination.

In conclusion one can say that the instrument complement fits the IXO mission and no show-stoppers for further design and implementation have been identified.

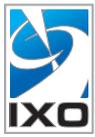
Most IXO instruments already meet the required TRL. The XMS is the most challenging, but the cryogenic systems are already being developed for *ASTRO-H*, while TES arrays with the required performance have been demonstrated in the lab and are being prepared for test flights.

## 4.8   Instrument resources summary

A summary of the instrument resources is given in Table 4.9. All power and mass figures include a 20% design maturity margin.

| Instrument/unit | Units [#] | Det. T [K] | Mass [kg] | Power [W] | | | Dim [cm³] | Telemetry [kbps] | | Location |
|---|---|---|---|---|---|---|---|---|---|---|
| | | | | OP | Ready | STBY | l×w×h | Nom | Peak | |
| **XMS** | | | | | | | | | | |
| Cryostat, FPA & coolers | 12 | 0.045 | 268.0 | 493.2 | 493.2 | 493.2 | Ø75×170 | | | MIP |
| Cooler Electronics | 5 | | 66.6 | 242.4 | 283.2 | 242.4 | 5× 30×32×13.5 | | | MIP |
| Instrument electronics | 6 | | 74.4 | 354.0 | 324.0 | 276.0 | 6× (avg.) 25×25×22 | | | MIP |
| *Total* | | | *409.0* | *1089.6* | *1100.4* | *1011.6* | | *64* | *840* | |
| **WFI** | | | | | | | | | | |
| Camera Head | 1 | 210 | 27.8 | 38.6 | 38.6 | 24.1 | 34×40×45 | | | MIP |
| HPPs | 2 | | 16.1 | 105.6 | 105.6 | 0.0 | 2× 20×25×35 | | | MIP |
| ICPCs | 2 | | 38.7 | 138.9 | 138.9 | 87.5 | 2× 25×30×20 | | | MIP |
| *Total* | | | *82.6* | *283.1* | *283.1* | *111.6* | | *45* | *450* | |
| **HXI** | | | | | | | | | | |
| HXI-S | 1 | 233 | 15.6 | 29.9 | 29.9 | 79.9 | 32×32×20 | | | WFI-CH |
| HXI-D | 1 | | 11.7 | 26.0 | 26.0 | 26.0 | 27×20×14 | | | MIP |
| *Total* | | | *27.3* | *55.9* | *55.9* | *105.9* | | *11* | *256* | |
| **HTRS** | | | | | | | | | | |
| FPA | 1 | 233 | 12.5 | 24.0 | 12.0 | 0.0 | 31×23×16.5 | | | MIP |
| Electronic box | 1 | | 17.8 | 121.0 | 48.0 | 0.0 | 36×23×17.5 | | | MIP |
| *Total* | | | *30.3* | *145.0* | *60.0* | *0.0* | | *840* | *840* | |
| **XPOL** | | | | | | | | | | |
| FPA | 1 | 283 | 7.4 | 20.0 | 11.0 | 0.0 | 17×19×18 | | | MIP |
| CE | 1 | | 7.2 | 41.1 | 23.2 | 0.0 | 14×19×10 | | | MIP |
| *Total* | | | *14.6* | *61.1* | *34.2* | *0.0* | | *840* | *840* | |
| **CAT-XGS** | | | | | | | | | | |
| Gratings | 2 | | 11.5 | 0.0 | 0.0 | 0.0 | 2× 86×94×5 | | | FMA |
| Camera assembly | 1 | 183 | 67.9 | 8.6 | 8.6 | 0.0 | 106×38×30 | | | FIP |
| DEA | 1 | | 34.2 | 103.7 | 103.7 | 0.0 | 41×20×15 | | | FIP |
| DPA | 1 | | 8.1 | 24.0 | 24.0 | 0.0 | 15×11×11 | | | FIP |
| *Total* | | | *121.7* | *136.3* | *136.3* | *0.0* | | *128* | *128* | |
| **OP-XGS** | | | | | | | | | | |
| Grating assembly | 1 | | 38.4 | 22.8 | 22.8 | 0.0 | 51x19x15 | | | FIP |
| Camera assembly | 1 | 183 | 51.6 | 87.6 | 87.6 | 0.0 | 57x52x15 | | | FIP |
| *Total* | | | *90.0* | *110.4* | *110.4* | *0.0* | | *128* | *128* | |

**Table 4.9.** *Summary of the instrument resources.*

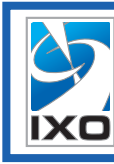 Following the assessment study, the definition of the instruments has been significantly advanced. The estimated mass and power resources have remained stable and are well within the allocated limits. No major impacts on the spacecraft systems have been identified.

# Appendix 1: Science Cases Flowdown

For each of the science objectives outlined, we now summarise the typical target information and the driving science requirements (see Table A.1, Table A.3, Table A.5 & Table A.7). In the tables of science performance (Table A.2, Table A.4, Table A.6 & Table A.8), the key driving requirements are highlighted in red, and N/A implies either there is no particular driving requirement or that the performance requirement is not applicable. For some science cases a secondary instrument is indicated where it is clear that the science investigations on different targets need more than one measurement type or range. There are of course similar arguments applying to most areas even where not explicitly noted in the tables.

## A.1.1 Co-Evolution of galaxies and their SMBH

| Science Topic | Typical Target | # Pointings | Total Obs Time (Ms) | Source Size (arcmin) | Typical Flux (ergs cm$^{-2}$ s$^{-1}$) | Science Analysis | Instrument (Secondary) |
|---|---|---|---|---|---|---|---|
| First SMBH | Chandra Deep Field | 38 | 10 | Point | 2 x 10$^{-17}$ | Spectro-imaging | WFI/HXI |
| Obscured Growth | NGC 4051 | 200 | 10 | Point | 10$^{-12}$ | Spectra | XMS (WFI/HXI) |
| Cosmic Feedback | MS0735+7421 | 50 | 9 | ~4 | 10$^{-13}$ | Spectra | XMS (WFI/HXI) |
| Black Hole Spin | NGC 4051 | 200 | 10 | Point | 10$^{-12}$ | Spectra | WFI/HXI (XMS) |

**Table A.1.** *Summary of the observing programme for the science topic "co-evolution of galaxies and their SMBH".*

Table A.2 summarises the various performance parameters that are driven by the science of co-evolution of galaxies and their supermassive black holes. The effective area at 1.25 keV and angular resolution are particularly important for these cases.

Instantaneously available sky coverage is not an issue, but the ability to meet exposure duration goals for a deep field in the same season implies a field should be visible for >500 ksec consecutively.

| Science Topic | FOV arcmin | Bandpass keV | PSF (HEW) arcsec | Effective Area | | | Energy Resolution | |
|---|---|---|---|---|---|---|---|---|
| | | | | 1.25 keV | 6 keV | 30 keV | FWHM eV | @E (keV) |
| First SMBH | 18dia. | 0.3-2 | 5 | 2.5 | 0.65 | 0.015 | 50 | 1 |
| Obscured Growth | 2 x 2 | 0.3-12 | 5 | 2.5 | 0.65 | 0.015 | 50 | 6 |
| Cosmic Feedback | 5 x 5 | 0.3-12 | 5 | 2.5 | 0.65 | 0.015 | 2.5 | 6 |
| Black Hole Spin | N/A | 1-40 | N/A | 2.5 | 0.65 | 0.015 | 1000 | 30 |

**Table A.2.** *Performance parameters driven by the science of "co-evolution of galaxies and their SMBH". Values highlighted in red are key driving requirements.*

# A.1.2    Large Scale Structure and the creation of chemical elements

| Science Topic | Typical Target | # Pointings | Total Obs Time *(Ms)* | Source Size *(arcmin)* | Typical Flux *(ergs cm$^{-2}$ s$^{-1}$)* | Science Analysis | Instrument (Secondary) |
|---|---|---|---|---|---|---|---|
| **Missing baryons & WHIM** | QSO B1426+428 | 30 | 15 | Point | 10$^{-11}$ | Spectra | XGS (XMS) |
| **Cluster Physics & evolution** | z = 1 cluster | 80 | 12 | 2- 10 | 5 x10$^{-15}$ | Spectra | XMS |
| **Galaxy Cluster Cosmology** | z = 1-2 cluster | 1000 | 15 | 2- 10 | 10$^{-15}$ | Spectra | XMS (WFI) |
| **Chemical Evolution** | M86 | 20 | 1 | 5 | 5 x10$^{-13}$ | Spectro-imaging | XMS |

**Table A.3.** *Summary of the observing programme for the science topic "large scale structure and the creation of chemical elements".*

Table A.4 summarises the performance drivers for this topic. The driving performance requirements for the XMS instrument come here from the combination of the good energy resolution, simultaneously with as large a field as possible.

Whereas some science can be performed with expected energy resolution of large FOV instruments, the study of dynamics requires high spectral resolution (~2 eV at 2 keV) over an extended FOV. As this resolution is incompatible with that expected from the large FOV instruments, mosaicing may be necessary to map interesting source regions.

| Science Topic | FOV *arcmin* | Bandpass *keV* | PSF (HEW) *arc sec* | Effective Area | | | Energy Resolution | |
|---|---|---|---|---|---|---|---|---|
| | | | | *1.25 keV* | *6 keV* | *30 keV* | *FWHM eV* | *@E (keV)* |
| **Missing baryons & WHIM** | N/A. | 0.3-1 | 5 | N/A | N/A | N/A | 0.1 | 0.3 |
| **Cluster Physics & evolution** | 2 x 2 | 0.3-7 | 5 | 2.5 | 0.65 | 0.015 | 2.5 | 6 |
| **Galaxy Cluster Cosmology** | 5 x 5 | 0.3-7 | 10 | 2.5 | 0.65 | N/A | 10 | 6.4 |
| **Chemical Evolution** | 5 x 5 | 0.3-7 | 5 | 2.5 | 0.65 | N/A | 2.5 | 6 |

**Table A.4.** *Performance parameters driven by the science of "large scale structure and the creation of chemical elements". Values highlighted in red are key driving requirements.*

# A.1.3   Matter under extreme conditions

| Science Topic | Typical Target | # Pointings | Total Obs Time (Ms) | Source Size (arcmin) | Typical Flux (ergs cm⁻² s⁻¹) | Science Analysis | Instrument (Secondary) |
|---|---|---|---|---|---|---|---|
| **Strong gravity and accretion** | MCG 6-30-15 | 60 | 10 | Point | 5 x10⁻¹¹ | Spectra | XMS (WFI/HXI) |
| **Black Hole Spin** | MCG 6-30-15 | 10 | 1 | Point | 5 x10⁻¹¹ | Polarisation | XPOL |
| **Neutron Star EOS** | 4U1636-536 | 30 | 5 | Point | 10⁻⁸ | Spectra | HTRS (XMS,XPOL) |
| **Stellar Mass Black Holes** | Cyg X-1 GX339-4 | 30 | 3 | Point | 10⁻⁸ | Spectra | HTRS |

**Table A.5.** *Summary of the observing programme for the science topic "matter under extreme conditions".*

Table A.6 summarises the performance requirements for this topic. Most observations can be satisfied with modest energy resolution. The topic drives the requirements for effective area at 6 keV and higher energies.

Timing requirement includes an ability to accurately phase fold, that demands a reconstructed barycentre correction for 100 μsec.

| Science Topic | FOV arcmin | Bandpass keV | PSF (HEW) arcsec | Effective Area | | | Energy Resolution | |
|---|---|---|---|---|---|---|---|---|
| | | | | 1.25 keV | 6 keV | 30 keV | FWHM eV | @E (keV) |
| **Strong gravity and accretion** | N/A. | 1-10 | N/A | 2.5 | 0.65 | 0.015 | 10 | 6 |
| **Black Hole Spin** | N/A | 2 - 10 | N/A | 2.5 | 0.65 | N/A | 1200 | 6 |
| **Neutron Star EOS** | N/A | 0.3-10 | N/A | 2.5 | 0.65 | 0.015 | 200 | 0.3-6 |
| **Stellar Mass Black Holes** | N/A | 03-10 | N/A | 2.5 | 0.65 | 0.015 | 200 | 0.3-6 |

**Table A.6.** *Performance parameters driven by the science of "matter under extreme conditions". Values highlighted in red are key driving requirements.*

## A.1.4  Life cycles of matter and energy

| Science Topic | Typical Target | # Pointings | Total Obs Time (Ms) | Source Size (arcmin) | Typical Flux (ergs cm⁻² s⁻¹) | Science Analysis | Instrument (Secondary) |
|---|---|---|---|---|---|---|---|
| SNR formation of elements | Tycho | 50 | 5 | 0.1-5 | $10^{-13}$ | Spectra | XMS |
| Shocks and particle acceleration | Galactic SNRs | 20 | 2 | 0.2-5 | $10^{-12}$ | Spectra | WFI/HXI (XMS) |
| Shocks and particle acceleration | Galactic SNRs | 15 | 1.5 | 0.2-5 | $10^{-12}$ | Polarisation | XPOL |
| ISM | 4U 1724-307 | serendipitous on other science | N/A | Point | $5 \times 10^{-11}$ | Spectra | XGS |
| Galactic Centre | Sgr-A* | 9 | 1 | 30 dia | $10^{-12}$ | Spectro-imaging | WFI/HXI (XPOL) |
| Stars and Planets | TW Hya Jupiter | 150 | 3 | Point | $10^{-12}$ | Spectra | XGS (XMS) |

**Table A.7.**  *Summary of the observing programme for the science topic "life cycles of matter and energy".*

For ISM studies a very large number of lines of sight are required, but the essential science of analysing the absorption edge energy shifts and fine structure can rely on XGS measurements serendipitously on normal science targets observed with the on-axis focal plane instruments.

Table A.8 describes performance requirements for the topic "life cycles of matter and energy". As befits general observatory science, no performance features are seen as critically driving the design.

| Science Topic | FOV arcmin | Bandpass keV | PSF (HEW) arcsec | Effective Area | | | Energy Resolution | |
|---|---|---|---|---|---|---|---|---|
| | | | | 1.25 keV | 6 keV | 30 keV | FWHM eV | @E (keV) |
| SNR formation of elements | 5 x 5 | 0.3-10 | N/A | 2.5 | 0.65 | N/A | 2.5 | 6 |
| Shocks and particle acceleration | 5 x 5 | 1 – 40 | N/A | 2.5 | 0.65 | 0.015 | 1000 | 30 |
| Shocks and particle acceleration | 2 x 2 | 2 – 10 | N/A | 2.5 | 0.65 | N/A | 1200 | 6 |
| ISM | N/A | 0.3-10 | N/A | N/A | N/A | N/A | 0.1 | 0.3 |
| Galactic Centre | >15 x 15 | 0.3-10 | N/A | N/A | N/A | N/A | 150 | 1 |
| Stars and Planets | 5 x 5 (for YSO) | 0.3-2 | 5 | N/A | N/A | N/A | 0.3 | 1 |

**Table A.8.**  *Performance parameters driven by the science of "life cycles of matter and energy".*

# Acronym List

| | | | | |
|---|---|---|---|---|
| 2MASS | 2 micron All Sky Survey | | JWST | James Webb Space Telescope |
| ACT | Atacama Cosmology Telescope | | L2 | Sun-Earth-Lagrange Point 2 |
| ADR | Adiabatic Demagnetization Refrigerator | | LISA | Laser Interferometer Space Antenna |
| | | | LSST | Large Synoptic Survey Telescope |
| AGN | Active Galactic Nuclei | | M1,M2…. | Mirror 1 (primary mirror), Mirror 2 (secondary mirror), etc |
| ALMA | Atacama Large Millimeter/ Submillimeter Array | | MA(M) | Mirror Assembly (Module) |
| AOCS | Attitude and Orbit Control Subsystem | | MIP | Movable Instrument Platform |
| APS | Active Pixel Sensors | | MIR | Medium Infrared |
| BEE | Back End Electronics | | MIS | Metal-Insulator Sensor |
| BH | Black Hole | | MM | Mirror Module |
| CAT | Critical Angle Transmission | | NASA | National Aeronautics and Space Agency |
| CCD | Charge Coupled Device | | | |
| CDF | Chandra Deep Field | | NuSTAR | Nuclear Spectroscopic Telescope Array |
| CE | Control Electronics | | | |
| CFRP | Carbon Fiber Reinforced Plastic | | PanSTARRS | Panoramic Survey Telescope Rapid Response System |
| CTA | Cherenkov Telescope Array | | | |
| CTE | Coefficient of Thermal Expansion | | PCA | Proportional Counter Array (RXTE) |
| CV | Cosmic Vision | | PCDU | Power Control and Distribution Unit |
| DEA | Detector Electronics Assembly | | PSF | Point Spread Function |
| DEPFET | Depleted P-channel Field-Effect Transistor | | QM | Qualification Model |
| | | | RF | Radio Frequency |
| DES | Dark Energy Survey | | ROSAT | Röntgen Satellite |
| DPA | Digital Processing Assembly | | RXTE | Rossi X-ray Timing Explorer |
| DSN | Deep Space Network | | SDD | Silicon Drift Detector |
| DSSD | Double-sided Si Strip Detector | | SGO | Segmented Glass Optics |
| E-ELT | European Extremely Large Telescope | | SKA | Square Kilometer Array |
| eROSITA | Extended Röntgen Survey Imaging Telescope Array | | SLI | Single layer Insulation |
| | | | SMBH | Super massive Black Hole |
| FIR | Far Infrared | | SNe | Supernovae |
| FM | Flight Model | | SNR | Supernova Remnant |
| FMA | Flight Mirror Assembly | | SPICA | Space IR Telescope for Cosmology and Astrophysics |
| FoV | Field of View | | | |
| FPA | Focal Plane Assembly | | SPO | Silicon Pore Optics |
| FPIs | Focal Plane Instruments | | SPT | South Pole Telescope |
| FW | Filter Wheel | | SVM | Service Module |
| GAS | Grating Array Structure | | SWG | Science Working Group |
| GBH | Galactic Black Hole | | TDA | Technology Development Activity |
| GEM | Gas Electron Multiplier | | TDM | Time Domain Multiplexing |
| GPD | Gas Pixel Detector | | TRL | Technology Readiness Level |
| HEW | Half Energy Width | | TT&C | Tracking Telemetry and Command |
| HGA | High Gain Antenna | | VISTA | Visible and Infrared Survey Telescope for Astronomy |
| HPD | Half Power Diameter | | | |
| HST | Hubble Space Telescope | | WFI | Wide Field Imager (IXO) |
| HTRS | High Time Resolution Spectrometer (IXO) | | WFIRST | Wide Field Infrared Survey Telescope |
| | | | WHIM | Warm-Hot Intergalactic Medium |
| HXI | Hard X-ray Imager (IXO) | | WMAP | Wilkinson Microwave Anisotropy Probe |
| HW | Hardware | | XEUS | X-ray Evolving Universe Explorer |
| IM | Instrument Module | | XGS | X-ray Grating Spectrometer (IXO) |
| IXO | International X-ray Observatory | | XMS | X-ray Microcalorimeter Spectrometer |
| JAXA | Japan Aerospace Exploration Agency | | XPOL | X-ray Polarimeter (IXO) |
| JDEM | Joint Dark Energy Mission | | XPBF | X-ray Pencil Beam Facility |